\documentclass[aip, amsmath,amssymb, citeautoscript, floatfix,
preprint,%
nofootinbib, tightenlines,cha,
]{revtex4-1}

\usepackage[pdftex,bookmarks, colorlinks, citecolor =blue, breaklinks]{hyperref}

\usepackage{graphicx}
\usepackage{multirow}
\usepackage{dcolumn}
\usepackage{caption}
\usepackage[multidot]{grffile} 
\usepackage{bm}

\usepackage{enumitem}   
\setenumerate[1]{label=(\alph*)}

\usepackage[utf8]{inputenc}
\usepackage[T1]{fontenc}
\usepackage{mathptmx}
\usepackage{float}
\usepackage[subrefformat=parens,labelformat=parens]{subfig}
\usepackage{xcolor}
\usepackage{xspace}
\usepackage{amsmath}
\usepackage{amssymb}
\usepackage{algorithm}
\usepackage{placeins}
\usepackage[noend]{algpseudocode} 

\newcommand{\bR}{{\mathbb{ R}}}

\newcommand{\citeInline}[1]{Ref.~[\onlinecite{#1}]}
\newcommand{\citesInline}[1]{Refs.~[\onlinecite{#1}]}

\begin{document}
	
	\preprint{Currently in review at Chaos: An Interdisciplinary Journal of Nonlinear Science}
	
	\title{Towards automated extraction and characterization of scaling regions in dynamical systems}
	\author{Varad Deshmukh}
	\affiliation{Department of Computer Science, University of Colorado, Boulder, CO, USA}
	\author{Elizabeth Bradley}
	\affiliation{Department of Computer Science, University of Colorado, Boulder, CO, USA}
	\affiliation{Santa Fe Institute, Santa Fe, NM, USA}
	\author{Joshua Garland}
	\affiliation{Santa Fe Institute, Santa Fe, NM, USA}
	\author{James D. Meiss}
	\affiliation{Department of Applied Mathematics, University of Colorado, Boulder, CO, USA}
	
	\date{\today}
\begin{abstract}
		
  Scaling regions---intervals on a graph where the dependent variable
  depends linearly on the independent variable---abound in dynamical
  systems, notably in calculations of invariants like the
  correlation dimension or a Lyapunov exponent.  In these
  applications, scaling regions are generally selected by hand, a
  process that is subjective and often challenging due to noise,
  algorithmic effects, and confirmation bias.  In this paper, we
  propose an automated technique for extracting and characterizing
  such regions.  Starting with a two-dimensional plot---e.g., the
  values of the correlation integral, calculated using the
  Grassberger-Procaccia algorithm over a range of scales---we
  create an ensemble of intervals by
  considering all possible combinations of endpoints,
  generating a distribution of slopes from least-squares fits
  weighted by the length of the fitting line and the
  inverse square of the fit error.  The mode of this
  distribution gives an estimate of the slope of the scaling region
  (if it exists).  The endpoints of the intervals that correspond to
  the mode
  provide an estimate for the extent of that region.  When there is \textbf{no}
  scaling region, the distributions will be wide and the resulting error
  estimates for the slope will be large.  We demonstrate this method for
  computations of dimension and Lyapunov exponent
  for several dynamical systems, and show that it
  can be useful in selecting values for the parameters in time-delay
  reconstructions.
  
\vspace*{1ex}
\noindent
\end{abstract}
\maketitle
\bigskip

{\bf Lead Paragraph:}
Many problems in nonlinear dynamics involve identification of regions
of constant slope in plots of some quantity as a function of a length or time scale.
The slopes of these \textit{scaling regions}, if they exist, allow one
to estimate quantities such as a fractal dimension or a Lyapunov
exponent.  In practice, identifying scaling regions is not always
straightforward.  Issues with the data and/or the algorithms often
cause the linear relationship to be valid only for a limited range in
such a plot, if it exists at all.  Noise, geometry and data quality
can disturb its shape in various ways, and even create multiple
scaling regions.  Often the presence of a scaling region, and the
endpoints of its range, are determined by eye,
\cite{kantz97,Holger-and-Liz} a process that is subjective and not
immune to confirmation
bias.\cite{Ji2011,Chen2019,clauset2009power,Zhou2018}
Worse yet, we know of no formal results about the relationship between
the width of the scaling region and the validity of the resulting
estimate; often practitioners simply use the heuristic notion that
``wider is better.'' In this work, we propose an ensemble-based method to objectively
identify and quantify scaling regions, thereby addressing some of the
issues raised above. 

\section{Introduction}\label{sec:Intro}

To identify a scaling region in a plot, our method begins by
generating multiple fits using intervals of different lengths and
positions along the entire range of the data set.  Penalizing the
lower-quality fits---those that are shorter or have higher error---
we obtain a weighted distribution of the slopes across all possible
intervals on the curve.  As we will demonstrate by example, the slope
of a scaling region that is broad and straight manifests as a dominant
mode in the weighted distribution, allowing easy identification of an
optimal slope.  Moreover, the extent of the scaling region is
represented by the modes of a similar distribution that assesses the
fit as a function of the interval endpoints. As we will show, for a
long, straight scaling region, markers closer to the endpoints of the
scaling region are more frequently sampled (due to the combinatorial
nature of sampling) and have a higher weighting (due to longer lengths
of the fits).  Thus, the method gives the largest reasonable scaling
interval for the optimal fit along with error bounds on the estimate
as computed from the distributions.

Section~\ref{sec:method} covers the details of this ensemble-based
method and illustrates its application with three examples.  In
\S\ref{sec:results}, we demonstrate the approach on several dynamical
systems problems, starting in \S\ref{sec:dcorr} with the estimation of
correlation dimension for two canonical examples.  We show that our
method can accurately detect spurious scaling regions due to noise and
other effects, and that it can be useful in systematically estimating
the embedding dimension for delay reconstruction.  We then demonstrate
that this method is also useful for calculating other dynamical
invariants, such as the Lyapunov exponent (\S\ref{sec:lyap}).  We
present the conclusions in \S\ref{sec:conclusions}.

\section{Characterizing scaling regions}\label{sec:method}

\begin{figure}[ht]
		\begin{center}
			\subfloat[$y(x) = s(x) + d(x)$]{
                  \includegraphics[width=0.5\linewidth]{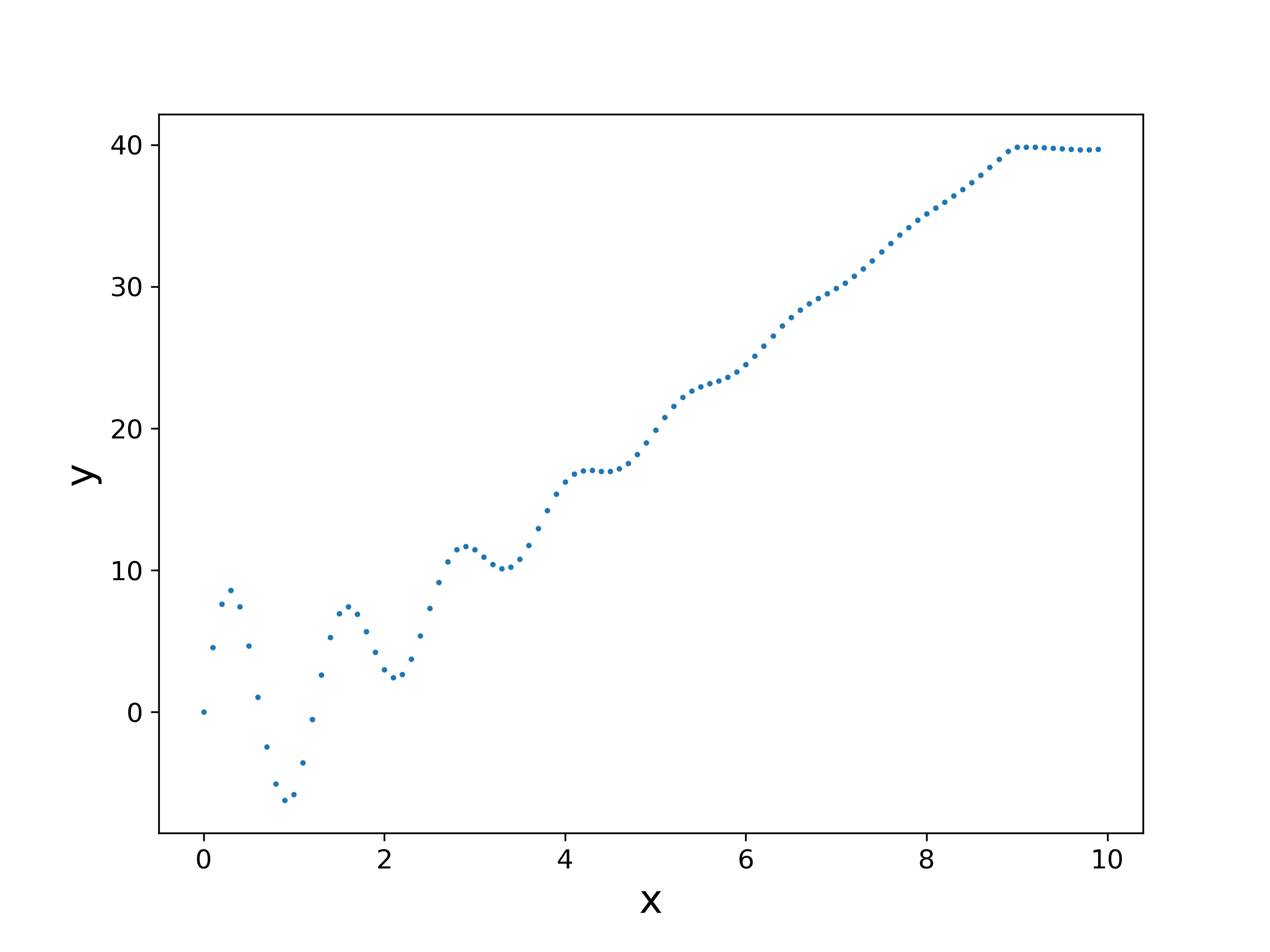}
                } \subfloat[Weighted slope histogram with the KD estimator]{
                  \includegraphics[width=0.5\linewidth]{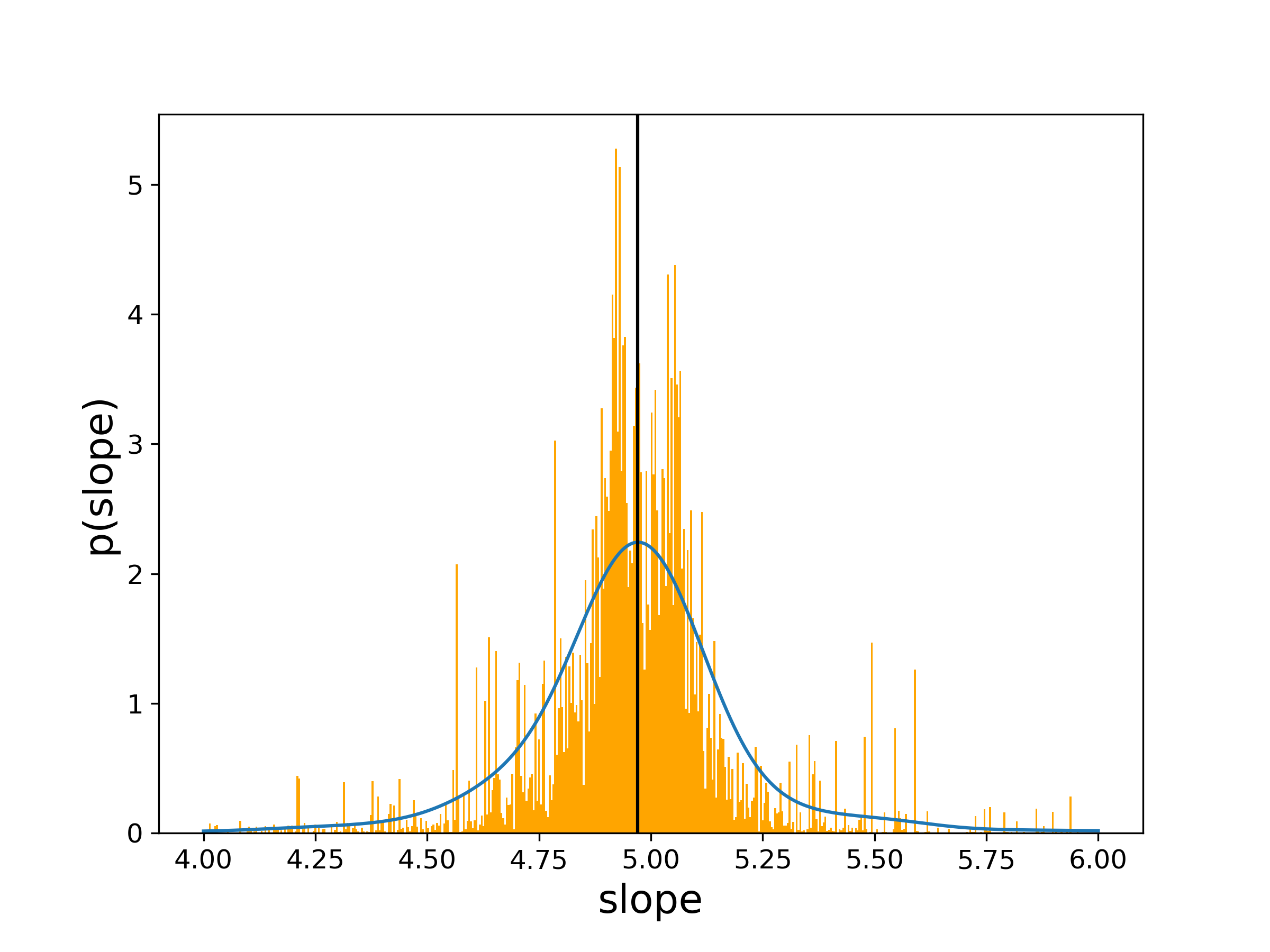}
                }\\ \subfloat[Weighted distributions of interval endpoints]{
                  \includegraphics[width=0.5\linewidth]{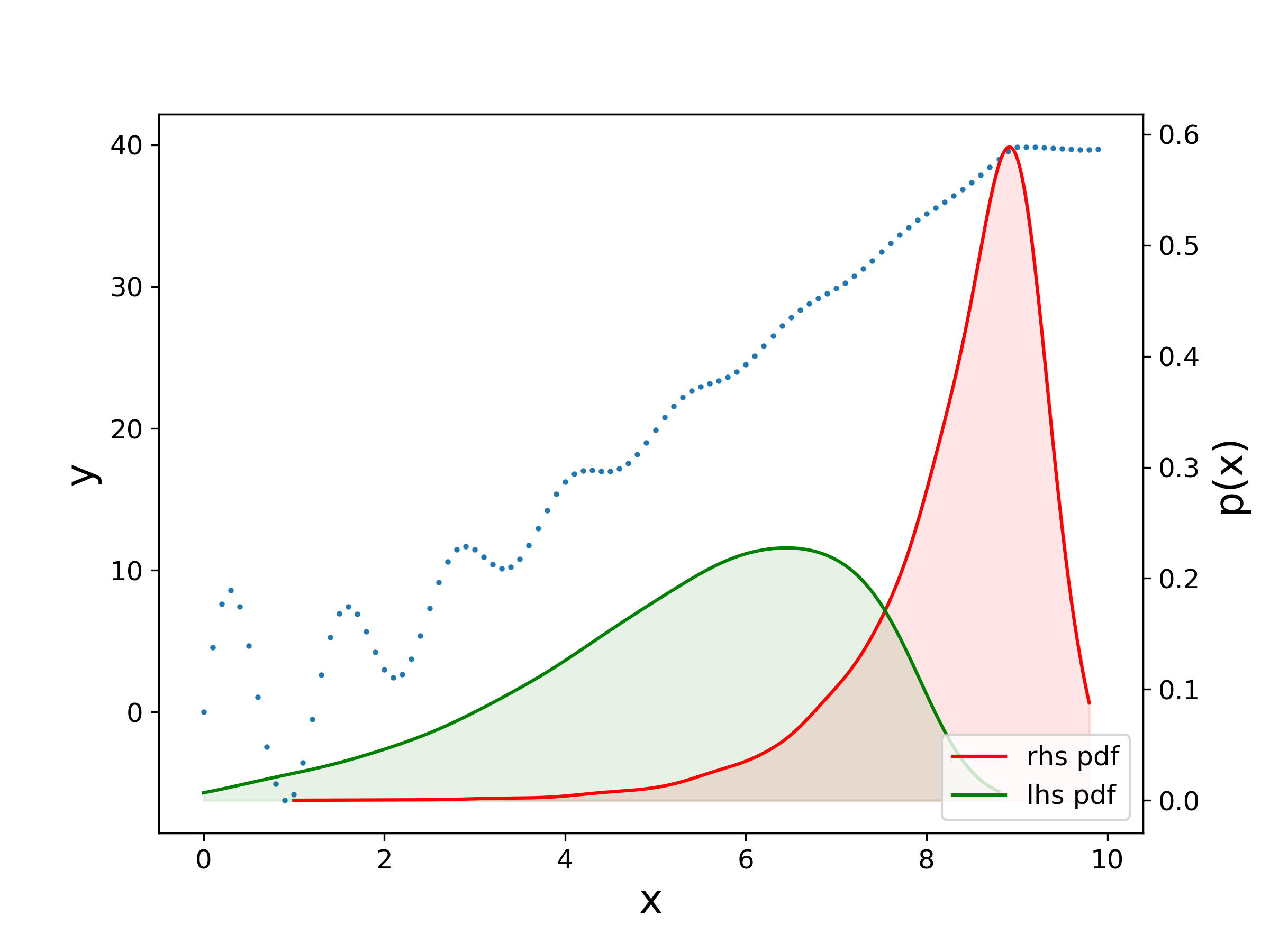}
                }
            \end{center}
\caption{Extracting a scaling region using an ensemble-based approach.
  (a) A synthetic example with a linear range that is bounded below by
  oscillations and above by a flat interval.  (b) Distribution of
  slopes (orange histogram) generated from an ensemble of fits in
  different intervals of this plot, weighted by the inverse square of
  the fitting error.  A kernel density estimation for the probability
  density is shown in blue, with its mode marked by the black line.
  The horizontal axis is clipped to [4,6].  (c) Probability
  distributions of the interval endpoints $[x_l,x_r]$ from the
  ensemble, with the same weighting as (b).}
\label{fig:scaling_region}
\end{figure}

A \textit{scaling law} ``...describe[s] the functional relationship
between two physical quantities that scale with each other over a
significant
interval.''\footnote{https://www.nature.com/subjects/scaling-laws}
Such a scaling law manifests as a straight segment on a
two-dimensional graph of these physical quantities, known as a
\textit{scaling region}.  In practical situations, the data will 
typically not lie exactly on a line or may do so only over a 
limited interval, see {\sl e.g.}, the synthetic example in
Fig.~\ref{fig:scaling_region}(a).
These data are obtained from a function $y(x) = s(x) + d(x)$, where
$s(x)$ is piecewise linear, with slope $5$ in the region $1 \le x \le
9$:
	\[
		s(x) =
		\begin{cases}
			0, & x < 1  \\
			5x - 5,& 1 \leq x \leq 9  \\
			40,& 9 < x 
		\end{cases} .
	\]
This function exhibits a scaling region in the range $[1,9]$ that is
bounded by intervals with zero slope: a shape that is similar to what
is often seen in computations of dynamical invariants even for
well-sampled, noise-free data.  To make the problem less trivial, we
add a damped oscillation,
	\[
		d(x) = 10\ e^{-0.5x} \sin(5x) ,
	\]
to $s(x)$.  The data set for this example is taken to be $y(x)$ on
$[0,10)$ with a sampling interval $\Delta x = 0.1$.

To find a scaling region in a diagram like
Fig.~\ref{fig:scaling_region}(a), one would normally choose
two points---markers $x_l$ and $x_r$ for the bounds of 
the linear region---then fit a line to the interval $[x_l,x_r]$.
A primary challenge is to identify the appropriate interval.  For
Fig.~\ref{fig:scaling_region}(a), the range $7\le x \le 9$ appears to
be linear, but should this range extend to smaller $x$?  One could
certainly argue that the lower boundary could be $x_l=6$ instead of
$7$, but it would be harder to defend a choice of $x_l \le 4$ because
of the growing oscillations.

Several approaches have been proposed for formalizing this procedure
in the context of the largest Lyapunov exponent (LLE) problem.  These
include what \citeInline{Zhou2018} calls the ``small data
method''---used, for example, in \citeInline{Rosenstein1993}---which
is efficient but suffers from the subjectivity problem described
above; an object searching method of the ``maximum nonfluctuant,"
which often reports a local optimal solution \cite{yongfeng-lyap}; 
and a fuzzy C-means clustering approach
\cite{shuang-lyap}, which also has limitations since the number of clusters
must be pre-specified by the user.

Our approach formalizes the selection of a scaling interval by first
choosing all possible ``sensible'' combinations of left- and
right-hand-side marker pairs.  Specifically, given a sequence of data
points at $\{x_j, \; j=1..N\}$, we choose $x_l$ and $x_r$ such that
\begin{equation}\label{eq:lrConstraint}
	r-l > n,
\end{equation}
so that the left marker $x_l$ must be below the right one ($x_r$) and
there must also be at least $n+1$ data points on which to perform the
linear fit.  For each pair $[x_l,x_r]$, we perform a least-squares fit
using the data bounded by the pair to compute both the slope and the
least-squares fit error. To suppress the influence of
  intervals with poor fits, we construct a histogram of their slopes,
  weighting the frequency of each directly by the length of the fit and inversely by the
  square of the associated fitting error. Finally, we use a kernel
density estimator to fit a smooth probability density function (PDF)
to the histogram.  The details are presented in
Appendix \ref{sec:appendix_algorithm}, along with a discussion of the 
choices for $n$ and the exponents associated with the fit width and the
fitting error (which are set to 1 and 2, respectively, in the
experiments reported here).

A central claim in this paper is that the resulting distribution
contains useful information about the existence, extent, and slope of
the scaling region.  The slopes for which the PDF is large represent
estimates that not only occur for more intervals in the graph, but also
that are longer and have lower fit errors.
In particular, we conjecture that the mode of this distribution gives
the best estimate of the slope of the scaling region: {\sl i.e.}, a
value with low error that persists across multiple choices for the
position of the interval.  More formally, most intervals $[x_l,x_r]$
that result in slopes near the mode of the PDF
(within full width half maximum \cite{Weik2001}) correspond to 
high-quality fits of regions in the graph
whose bounds lie within the interval of the scaling region.  If the
graph has a single scaling region, the PDF will be unimodal; a
narrower peak indicates that the majority of high quality fits lie
closer to the mode.  Broad peaks and multimodal distributions might 
arise for multiple reasons, including noise; we address
these issues in \S\ref{sec:results}.

For the example in Fig.~\ref{fig:scaling_region}, there are $N=100$
data points and we set $n=10$ so the ensemble of endpoints has $4005$
elements.  The resulting weighted slope distribution is unimodal, as
shown in panel (b), with mode $4.97$---not too far from the expected
slope of 5.0.  To provide a confidence interval around this estimate,
we compute three quantities:
\begin{enumerate}
	\item the full width half maximum ($FWHM$) of the PDF around the mode;
	\item the fraction, $p_{FWHM}$, of ensemble members that are
          within the $FWHM$;
	\item the standard deviation, $\sigma$, for the slopes of the ensemble members
          within the $FWHM$.     
\end{enumerate}
For the example in Fig.~\ref{fig:scaling_region}, we obtain
\[
	FWHM = 0.36, \quad  p_{FWHM} = 0.67, \quad \mbox{and }  \sigma = 0.11.
\]
That is, we estimate the slope of the scaling region as $4.97$ $\pm$
$0.11$, noting that fully $67\%$ of the estimated values are within
$\pm 0.18$.  These error estimates quantify the shape of the peak,
and give assurance that the diagram does indeed contain a scaling
region.

The method also can estimate the boundaries of the scaling region.
To do this, we compute secondary distributions using the positions of
the left-hand and right-hand side markers.  Since we choose all
possible pairs, subject to the constraint \eqref{eq:lrConstraint},
the probability densities of the resulting histograms will decrease
linearly for $x_l$ and grow linearly for $x_r$ from left to right. To
penalize poor fits, we again scale the frequency for each point in
these secondary histograms directly by the length and
inversely by the square of the error of the corresponding fit. The
resulting PDFs for the synthetic example of
Fig.~\ref{fig:scaling_region}(a) are shown in panel (c).
The modes of the LHS and RHS histograms fall at $x_l = 6.45$ and $x_r
= 8.92$, respectively.
These align neatly with what an observer might choose as the endpoints
of the putative scaling region in Fig.~\ref{fig:scaling_region}(a). 
Note that the $x_l$ distribution is wider with
a relatively lower peak, whereas the $x_r$ distribution is narrower
and taller.  This indicates that we can be more certain about the
position of the right endpoint than we are
about the left one.  There is 
more flexibility in choosing $x_l$ than $x_r$. 

\begin{figure}[ht]
	\begin{center}
		\subfloat[$y(x) = x^2$]{
			\includegraphics[width=0.5\linewidth]{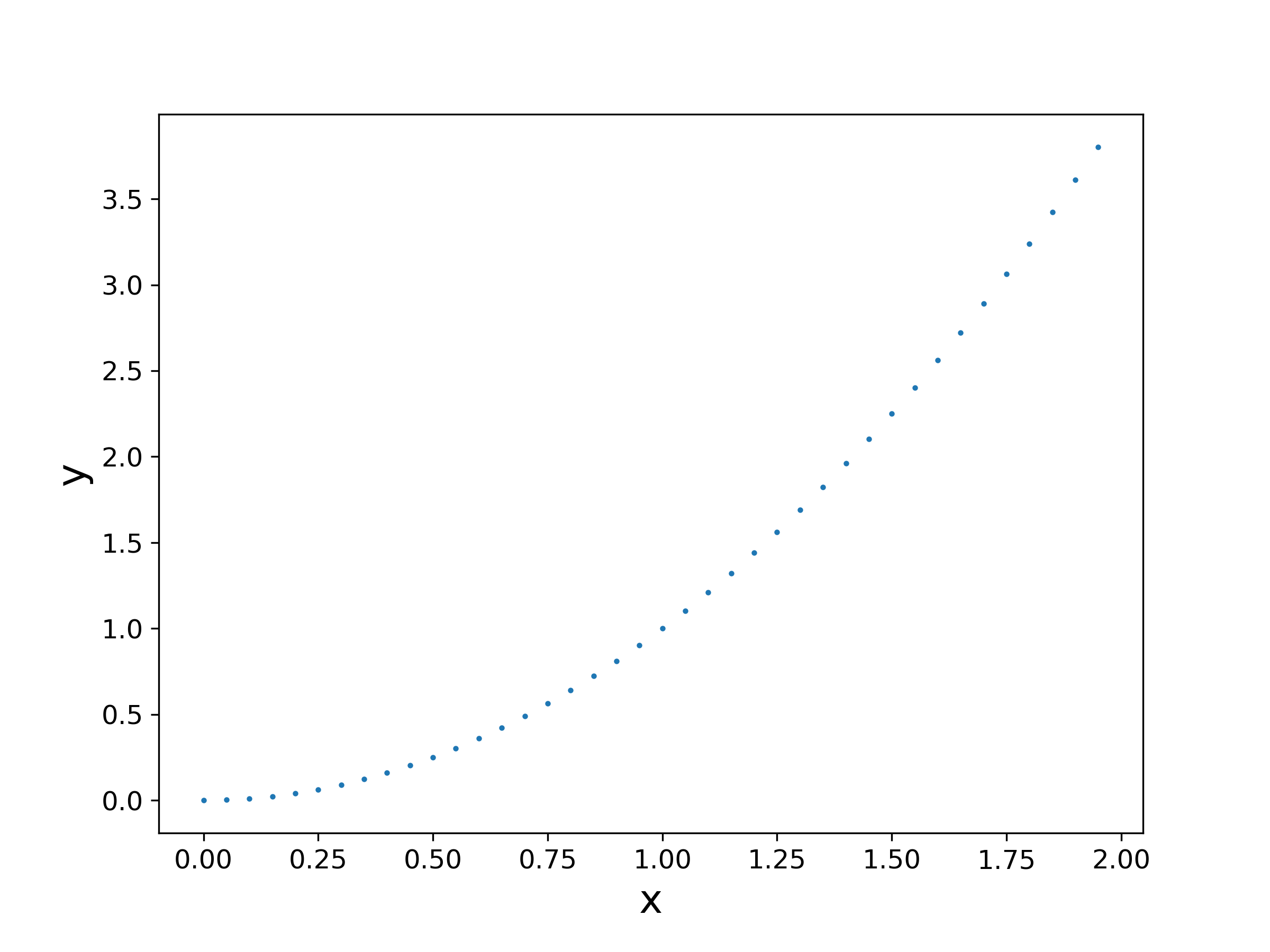}
		} \subfloat[Weighted slope histogram with the KD estimator]{
			\includegraphics[width=0.5\linewidth]{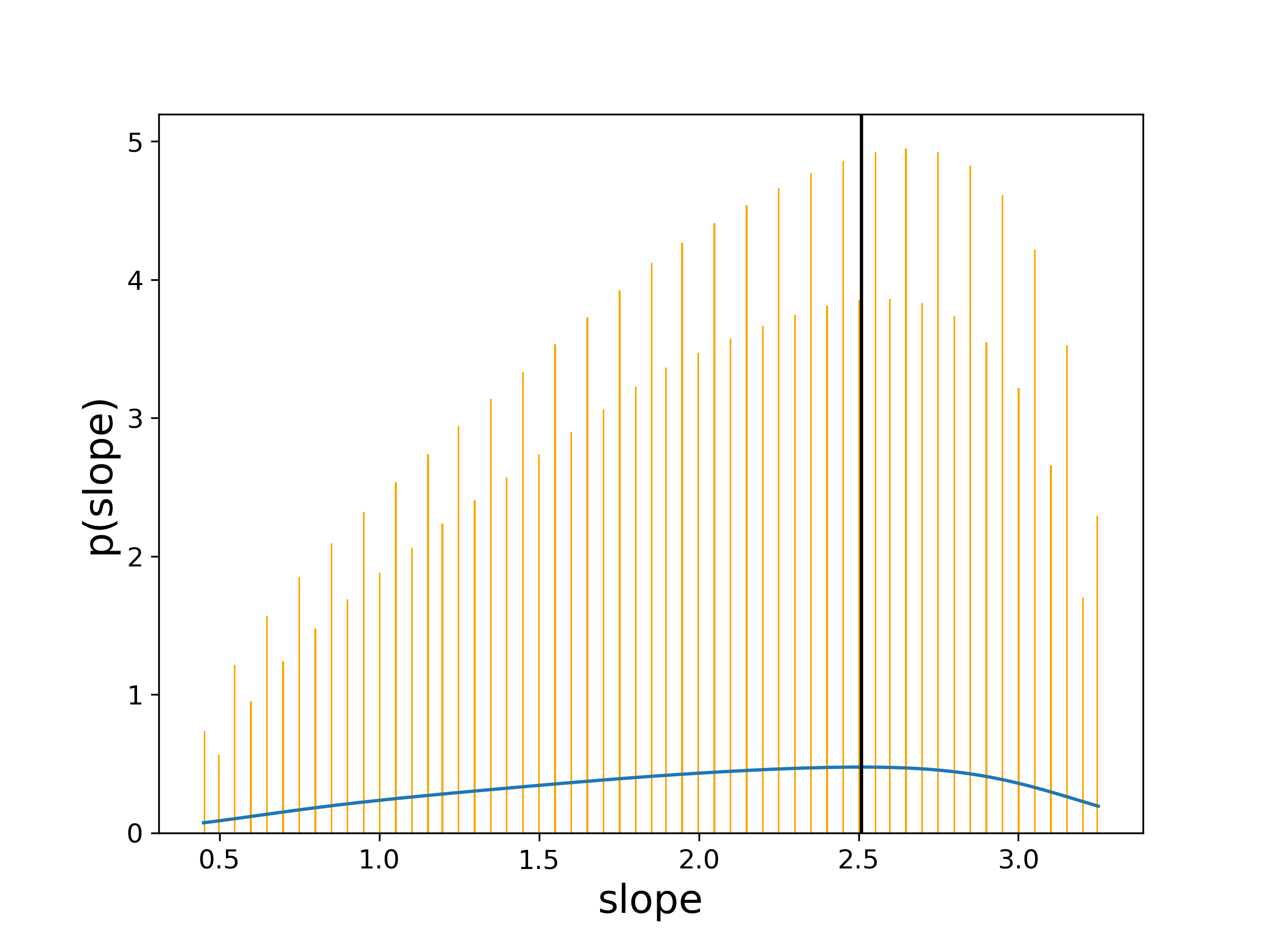}
		}\\ \subfloat[Weighted distributions of interval endpoints]{
			\includegraphics[width=0.5\linewidth]{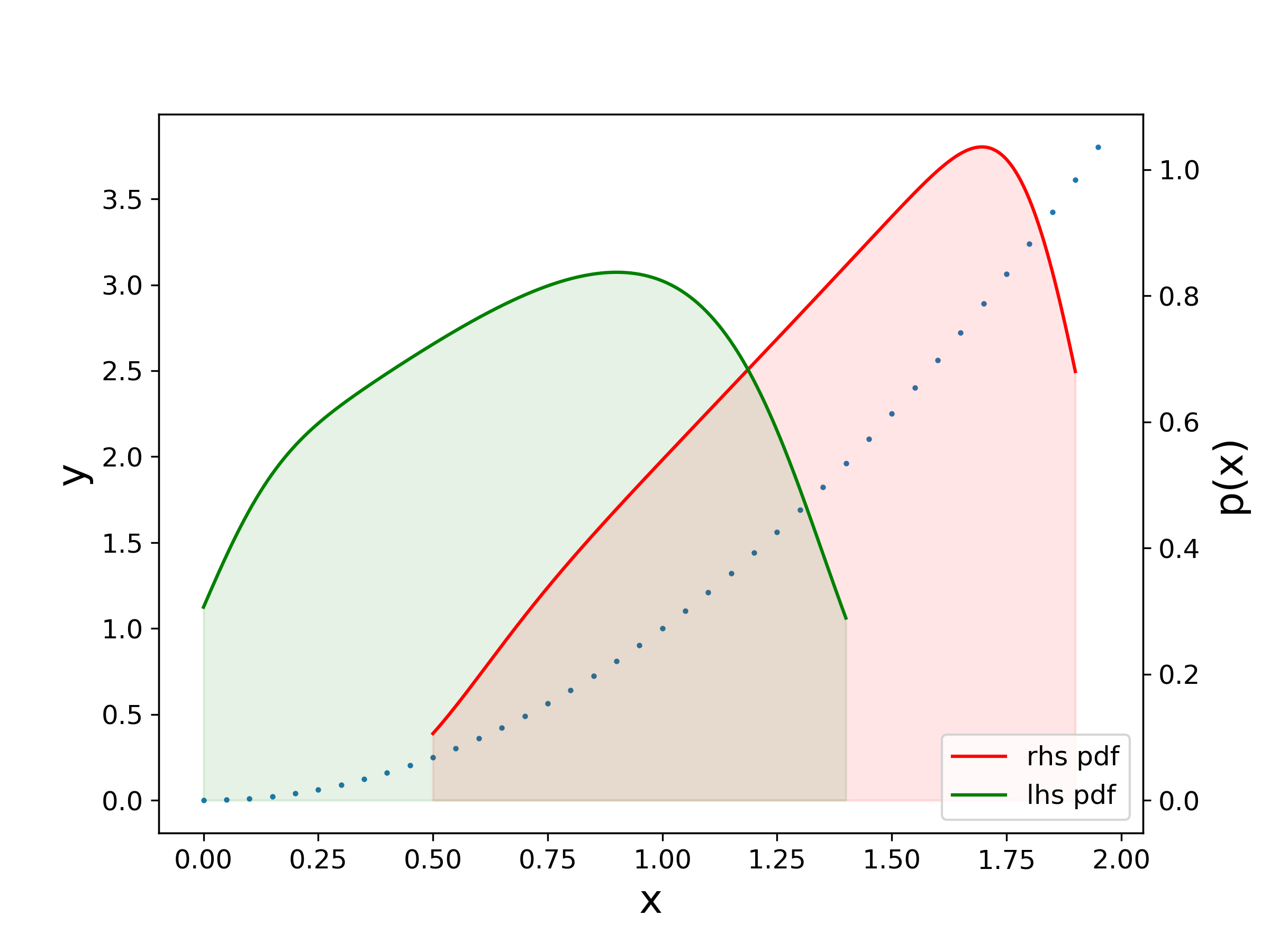}
		}
	\end{center}
	\caption{Estimating the scaling region of a quadratic
          function.}
	\label{fig:quadratic}
\end{figure}

To test our ensemble method on a dataset without an obvious scaling
region, consider the quadratic function
\[
y(x) = x^2\\.
\]
We choose the range $x \in [0,2]$ and a sampling interval $\Delta x =
0.05$.  A graph of resulting function, Fig.~\ref{fig:quadratic}(a), of
course has no straight regions; though, one might imagine that the
curve becomes nearly linear near the top of the range
even though the actual slope grows linearly with $x$.
The slope distribution, shown in Fig.~\ref{fig:quadratic}(b), has mode 
\[
	2.51 \pm 0.63 \quad (FWHM = 2.18 \mbox{ and  } p_{FWHM} =0.89) ,
\]
and as the endpoint distributions in panel (c) show, 
this corresponds to the interval $[x_l, x_r] =[0.90, 1.70]$.
However, the large width of the slope distribution and the
correspondingly high magnitude of $\sigma$ convey little 
confidence as to the existence of a scaling region.  This shows
that the error statistics and the $FWHM$ can be useful indicators as to whether or
not a scaling region is present.  A high error estimate and/or a high
FWHM contraindicates a scaling
region.

In the case of the two simple examples above, scaling regions imply
simple linear relationships between the independent and dependent
variables.  Our technique can just as well be applied to plots with 
suitably transformed axes---like the
log-linear or log-log axes involved in the computation of many
dynamical invariants---as we demonstrate next by
computing the dimension of a fractal.

The box-counting dimension, $d_{box}$, is the growth rate of the number
$N(\epsilon)$ of boxes of size $\epsilon$ that cover a set:
\begin{align}
	N(\epsilon) \propto \epsilon^{-d_{box}} .
\end{align}
In practice, $d_{box}$ is estimated by computing the slope of the
graph of $\ln\ N(\epsilon)$ versus $\ln \tfrac{1}{\epsilon}$.
\begin{figure}[ht]
	\centering
	\subfloat[100,000 points on the Sierpinski Triangle]{
		\includegraphics[width=0.5\linewidth]{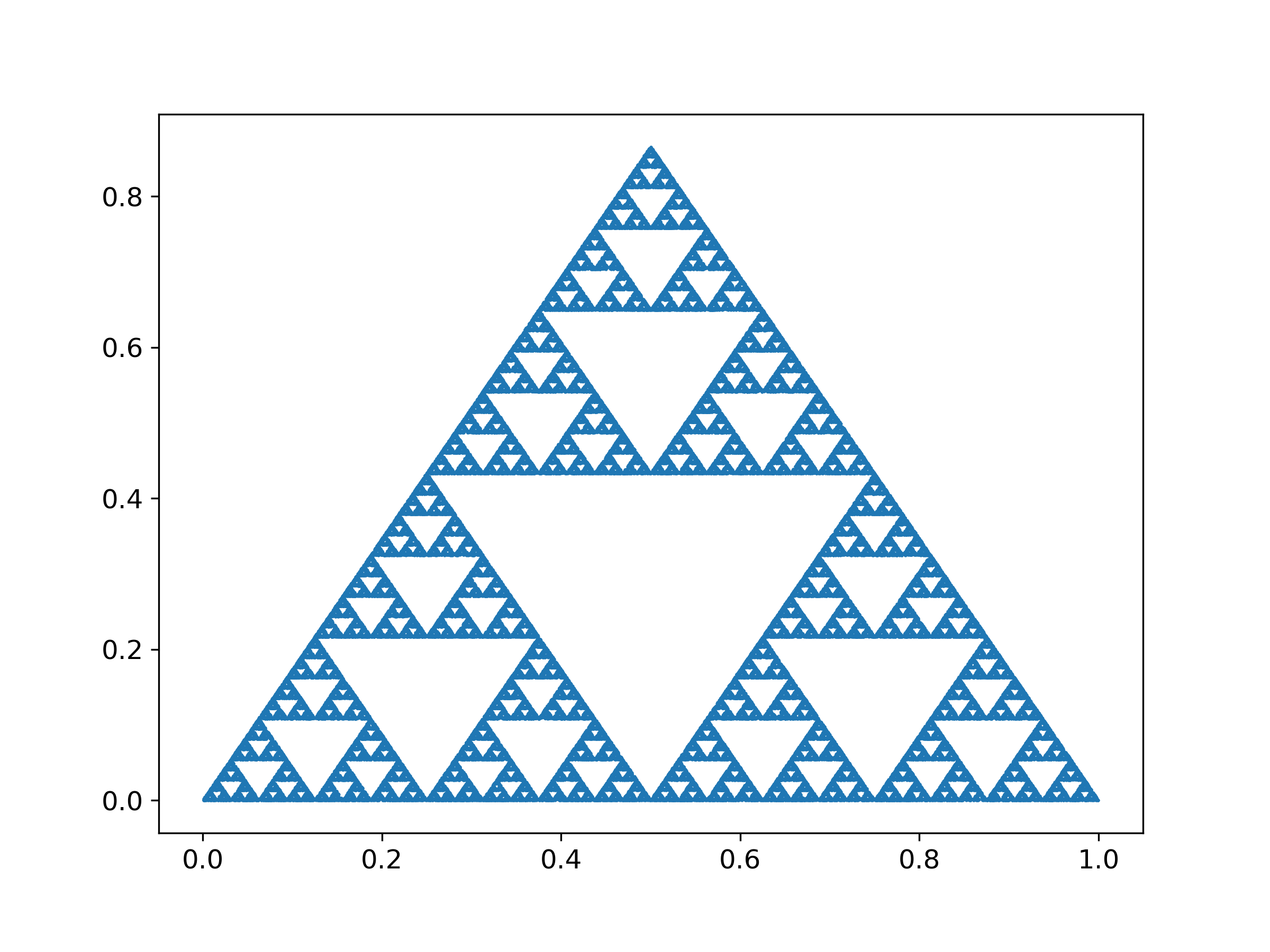}
	}\\
	\subfloat[Log-log plot of $N(\epsilon)$]{
		\includegraphics[width=0.5\linewidth]{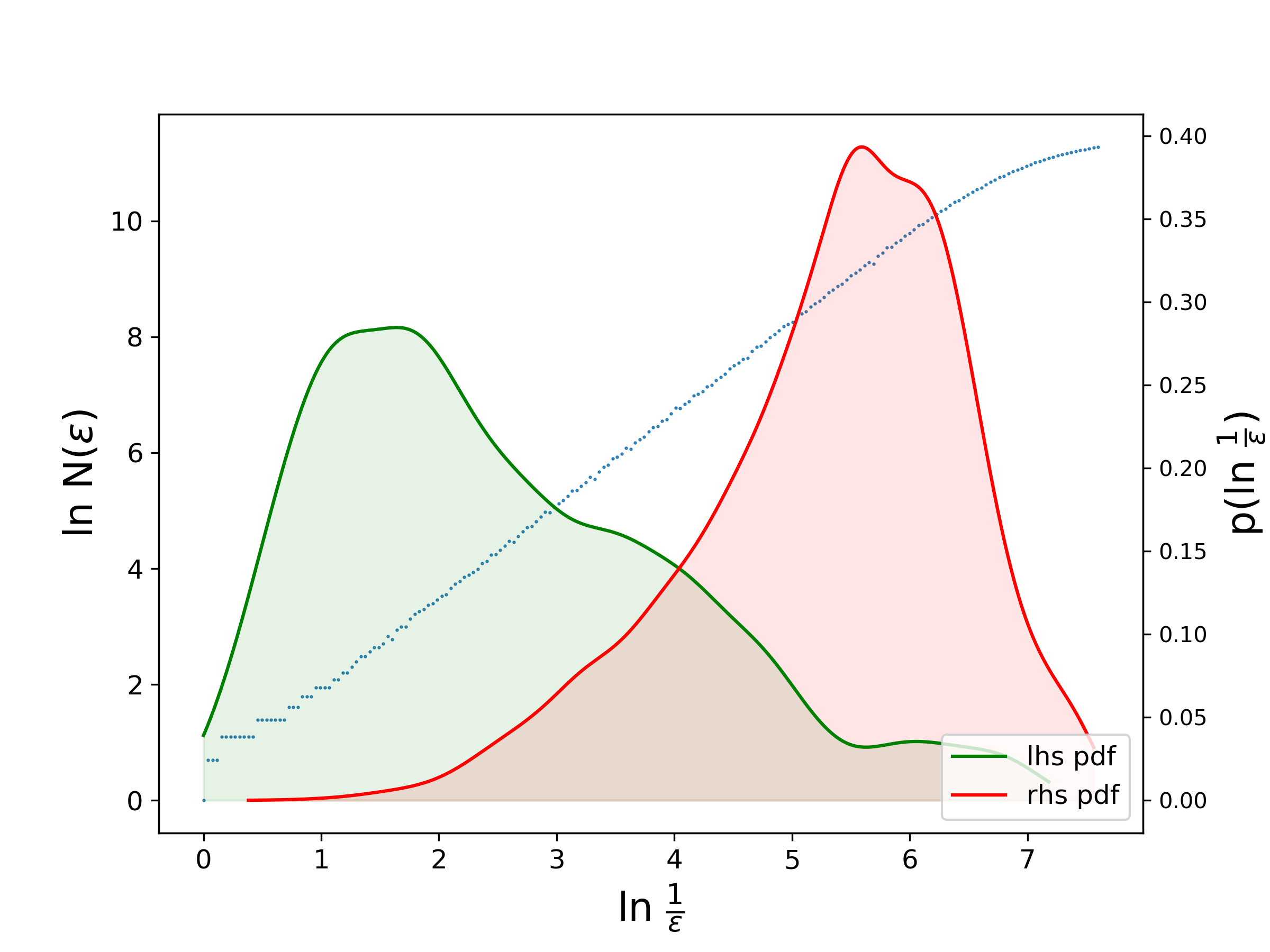}
	}
	\subfloat[Weighted slope distribution]{
		\includegraphics[width=0.5\linewidth]{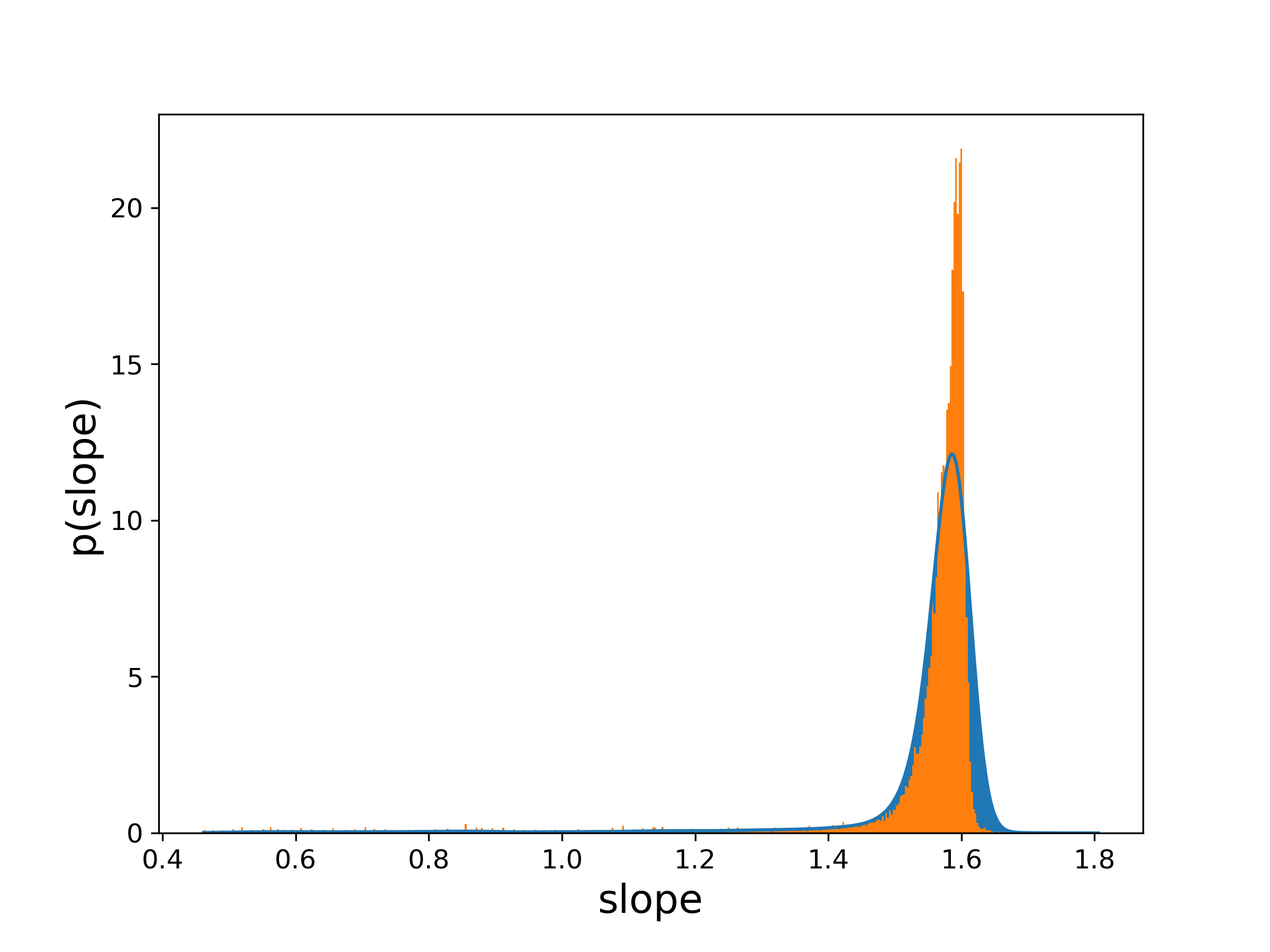}
	}
	\caption{Estimating the box-counting dimension of the
          Sierpinski Triangle. }
	\label{fig:sierpinski}	
\end{figure}
As an example, we consider the
equilateral Sierpinski triangle of side one, generating a set of $10^5$ 
points using an iterated function system \cite{Peitgen}
to obtain Fig.~\ref{fig:sierpinski}(a).  A log-log plot of
$N(\epsilon)$ for this data set is shown in
Fig.~\ref{fig:sierpinski}(b) and the resulting slope distribution is
shown in Fig.~\ref{fig:sierpinski}(c). The mode of this distribution
falls at $1.585$ $\pm$ $0.020$, where---as before---the error is the
estimate $\sigma$. This is close to the true dimension $\frac{\ln
  3}{\ln 2}$ $\approx$ $1.585$.  This peak is narrow, with $FWHM
=0.07$, though $p_{FWHM} = 0.68$, similar to the first example.  Thus
about $70\%$ of the weighted fits lie within
$\pm0.035$ of the estimated slope, strengthening confidence in the
estimate. The small periodic oscillations in the curve in
Fig.~\ref{fig:sierpinski}(b) are due to the self-similarity of the
fractal.  Note that the linear growth saturates when
$\ln\tfrac{1}{\epsilon} < 0.5$, the point at which the box size
becomes comparable to the diameter of the set.  Neither of these
effects significantly influences the mode of the slope
distribution. For this example, the LHS and RHS distributions in panel
(b) have similar widths, with modes $x_l = 1.64$ and $x_r =5.59$, respectively.

The examples of this section show that the ensemble-based method
effectively selects an appropriate scaling region---if one exists---and
is able to exclude artifacts near the edges of the data.

\section{Applications to Dynamical Systems}\label{sec:results}
	
In this section, we apply this scaling region identification and
characterization method to calculations of two dynamical
invariants---the correlation dimension (\S\ref{sec:dcorr}) and the
Lyapunov exponent (\S\ref{sec:lyap})---for two well-known examples,
the Lorenz system and the driven, damped pendulum.  We also explore the
effects of noise (\S\ref{sec:noise}) and the selection of embedding
parameters for delay reconstructions (\S\ref{sec:embed}).

\subsection{Correlation Dimension}
\label{sec:dcorr}
	
Correlation dimension, one of the most common and useful dynamical invariants, can be
calculated using the Grassberger-Procaccia algorithm
\cite{grassberger2004measuring,grassberger1983characterization}.  Given a set of points that
sample an object, such as an attractor of a dynamical system, this
algorithm estimates the average number of points in an
$\epsilon$-neighborhood of a given point, $C(\epsilon)$.
For an attractor of correlation dimension $d_2$, this 
scales with $\epsilon$ as
	\begin{equation}
		C(\epsilon) \propto \epsilon^{d_2}.
	\end{equation}
Estimating $d_2$ is therefore equivalent to finding a scaling region
in the log-log plot of $C(\epsilon)$ versus
$\epsilon$ \cite{GrassbergerPhysicaD}.
	
In this section, we use {\tt d2}, the {\tt TISEAN}
\cite{kantz97} implementation,
which outputs estimates of
$C(\epsilon)$ for a range of $\epsilon$ values.  We apply our
ensemble-based method to a graph of log $C(\epsilon)$ versus log
$\epsilon$ and identify the mode of the resulting slope distribution
to obtain an estimate of the correlation dimension of the data set.
When the slope distribution is multi-modal, our method can also reveal
the existence of potential scaling regions for different ranges of
$\epsilon$, some of which may not be obvious on first observation.
This analysis, then, not only yields values for the slope and extent
of the scaling region; it also provides insight into the geometry of
the dynamics, the details of the Grassberger-Procaccia algorithm, and
the interactions between the two.
	
\subsubsection{Lorenz}
\label{sec:lorenz}
	
\begin{figure}[ht]
		\centering \subfloat[Correlation sum calculations]{
                  \includegraphics[width=0.5\linewidth]{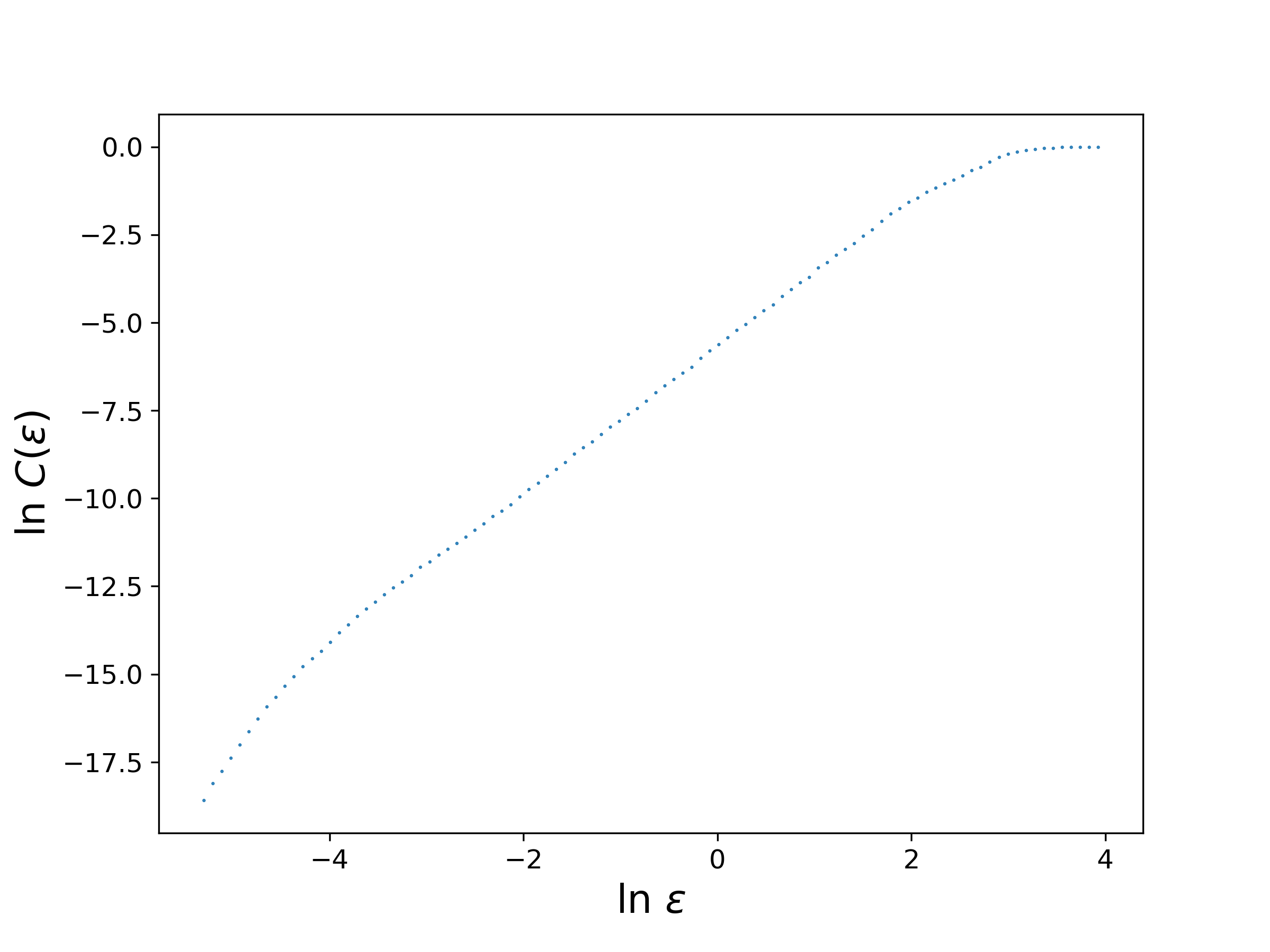}
                } \subfloat[Weighted slope distribution]{
                  \includegraphics[width=0.5\linewidth]{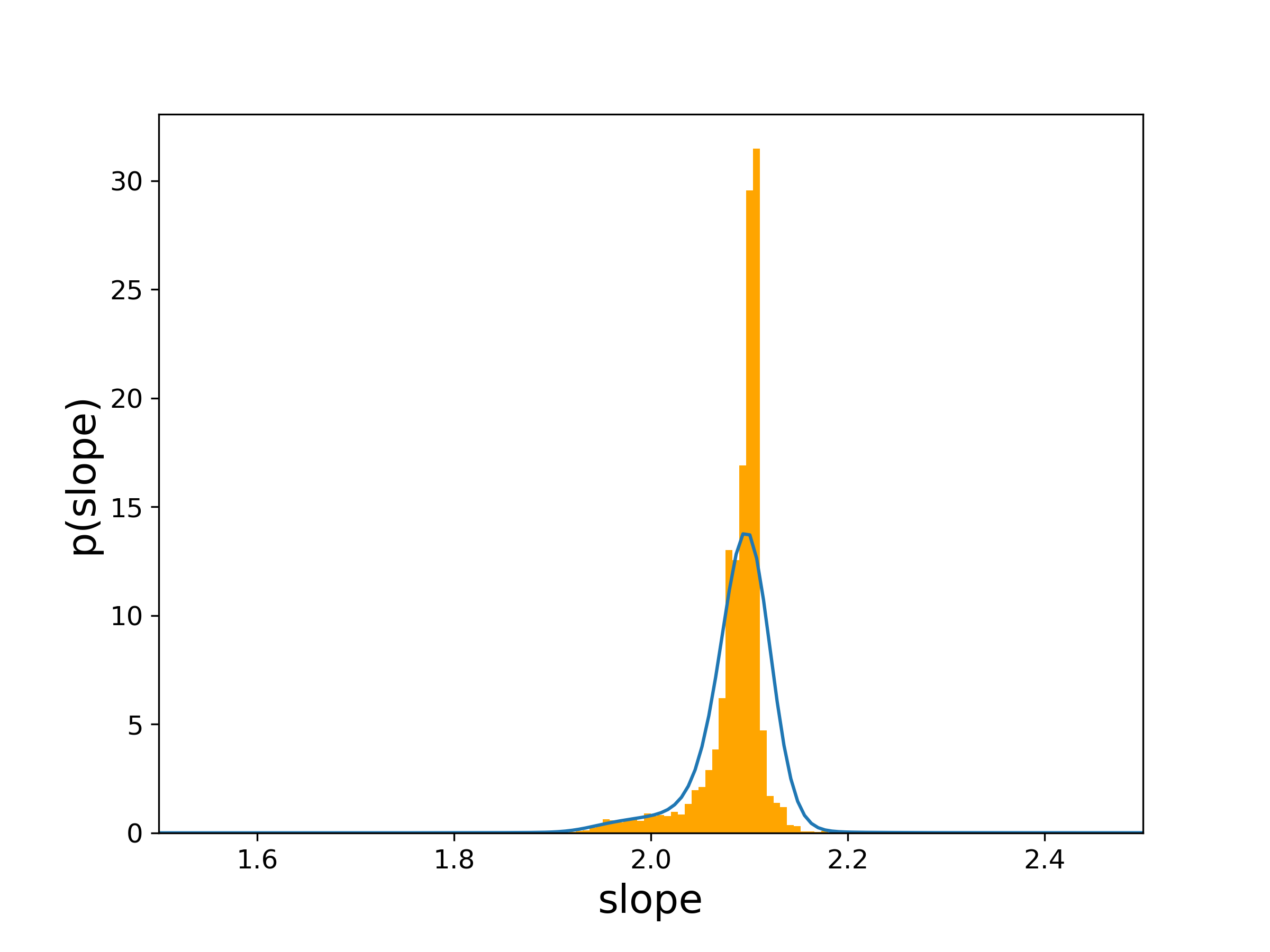}
                }\\ \subfloat[Weighted distributions of interval endpoints]{
                  \includegraphics[width=0.5\linewidth]{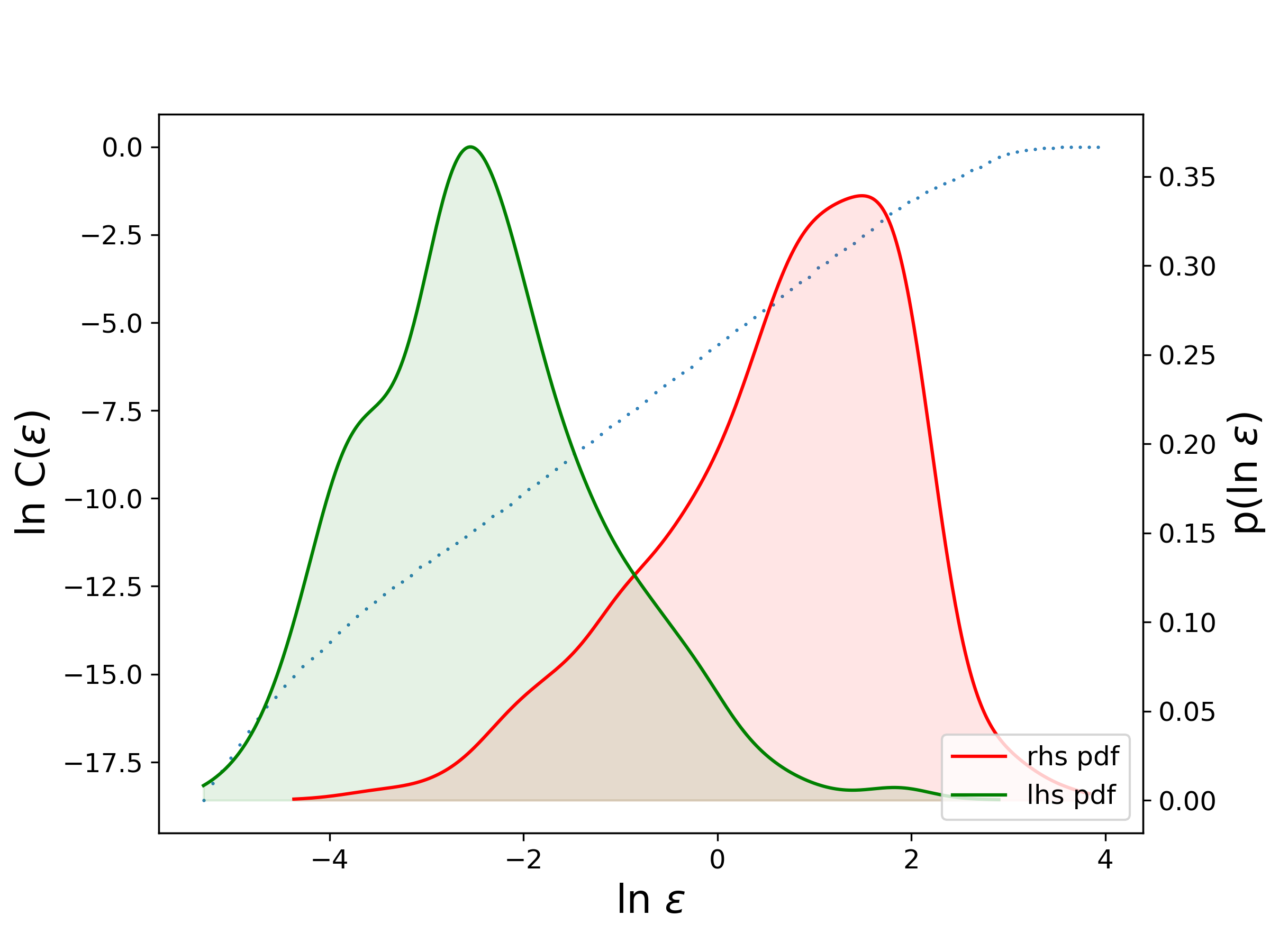}
                } \subfloat[Heatmap of the endpoint ensemble]{
                  \includegraphics[width=0.50\linewidth]{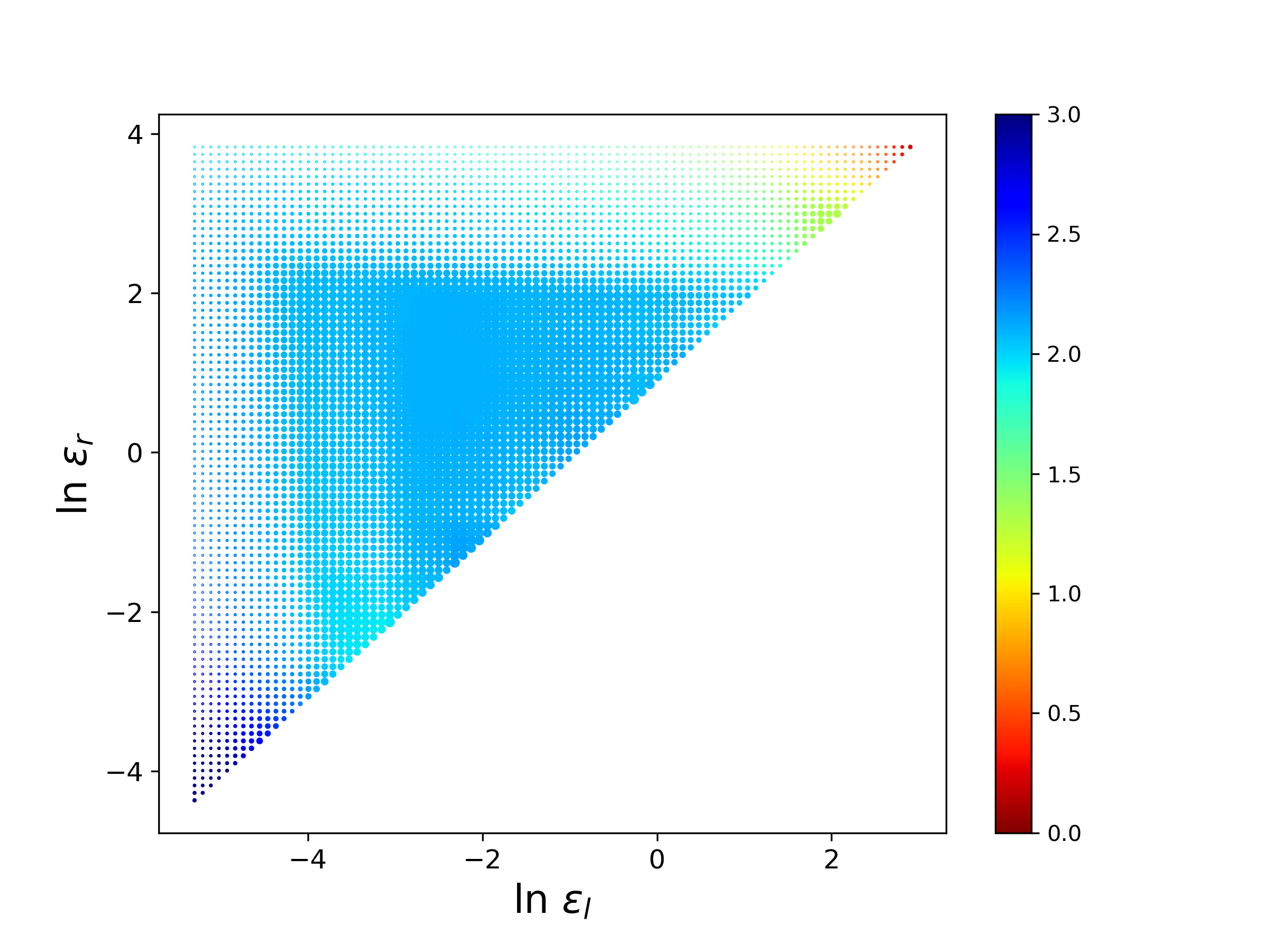}
                }
		\caption{Estimating the correlation dimension of the
                  Lorenz attractor}
		\label{fig:lorenz}
\end{figure}

We start with the canonical Lorenz system:
	\begin{align}
		\dot{x} &= \sigma(y - x), \nonumber \\
		\dot{y} &= x (\rho - z) - y, \nonumber \\
		\dot{z} &= xy - \beta z, \nonumber
	\end{align}
with $\sigma = 10$, $\beta = \frac{8}{3}$, and $\rho = 28$
\cite{Lorenz63}.  We generate a trajectory from the initial condition
$(0, 1, 1.05)$ using a fourth-order Runge-Kutta algorithm for $10^5$
points with a fixed time step $\Delta t = 0.01$, discarding the first
$10^4$ points to remove the transient, to get a total trajectory length
of 90,000 points. The chosen initial condition lies in the basin of
the chaotic attractor and the discarded length is sufficient to remove
any transient effects.  Figure \ref{fig:lorenz}(a) shows a log-log plot of $C(\epsilon)$ versus $\epsilon$
produced by the {\tt d2} algorithm.
To the eye, this graph appears to contain a scaling region in the
approximate range $[\ln \epsilon_l,\ln \epsilon_r] = [ -3,2]$.  As in
the box-counting dimension example in \S\ref{sec:method}, the curve
flattens when $\epsilon$ is larger than the diameter of the attractor,
since $C(\epsilon)$ will not change when $\epsilon$
increases beyond this point.  Since the diameter of the Lorenz
attractor is $25.5$, this flattening occurs for $\ln \epsilon >
\ln(25.5) \approx 3.2$.

Figures~\ref{fig:lorenz}(b) and~(c) show the results of our
ensemble-based approach. The mode of the slope PDF is 
\[
	2.09 \pm 0.02 \quad (FWHM = 0.06 \mbox{ and } p_{FWHM} = 0.69);
\]
this gives an estimate for the correlation dimension for the Lorenz
attractor that agrees with the accepted range, $2.06-2.16$
\cite{Viswanath04, Martinez93}.  The PDFs of the LHS and RHS
markers in Fig.~\ref{fig:lorenz}(c) have modes $\ln \epsilon_l = -2.6$
and $\ln \epsilon_r = 1.5$, respectively.
Estimates of both endpoints by our algorithm are close 
to the informal ones, although slightly on the conversative side.
In other words, our method can indeed effectively formalize 
the task of identifying both the existence and the extent 
of scaling regions.

The endpoint distributions can be broken
down further in the heatmap-style scatter plot of
Fig.~\ref{fig:lorenz}(d) to reveal additional details.
Each dot represents a
single element of the ensemble: {\sl i.e.}, a fit for a particular
interval $[\ln \epsilon_l, \ln \epsilon_r]$.  Its color encodes the
slope and its radius is scaled by the fitting weight.  
If samples come from intervals that are close, the associated
dots will be nearby; if these have low error, the dots will be large
and their effective density will be high.  Note that the domain of
panel (d) is a triangle since $\epsilon_l < \epsilon_r$ by
\eqref{eq:lrConstraint}.

This visualization provides an effective way to identify scaling
region ranges that produce similar slope estimates: long scaling regions
will manifest as large regions of
similarly colored points.  Fig.~\ref{fig:lorenz}(d) shows such a
triangular high-density region in blue (slope $\approx$ 2.0),
above the diagonal, bounded from the left by $\ln
\epsilon_l \approx -4$, and from above by $\ln \epsilon_r \approx 2$.
This clearly corresponds to the primary scaling region in
Fig.~\ref{fig:lorenz}(a) and the corresponding mode in panel (b).
This heatmap can reveal other, shorter scaling
regions.  Indeed, panel (d) shows other clusters of similarly colored dots
that are somewhat larger than the dots from neighboring regions: {\sl
  e.g.}, the green cluster for $\ln\epsilon_l \approx 2$ and $\ln
\epsilon_r \approx 3$, with a slope near $1.5$.  Close
examination of the original curve in Fig.~\ref{fig:lorenz}(a) reveals
a small straight segment in the interval $[2,3]$.  This, not immediately apparent, feature
stands out quite clearly in the distribution visualization. Two much smaller scaling regions
at the two ends of panel (a) are
also brought out by this representation: the interval
of slope $\approx$ 3.0 for small $\epsilon$ 
(the dark blue cluster at the lower left corner of the scatter plot)
and the zero slope for large $\epsilon$ (the red cluster at the upper
right corner).
	
\subsubsection{Pendulum}
\label{sec:pend}

\begin{figure}[ht]
\includegraphics[width=0.8\linewidth]{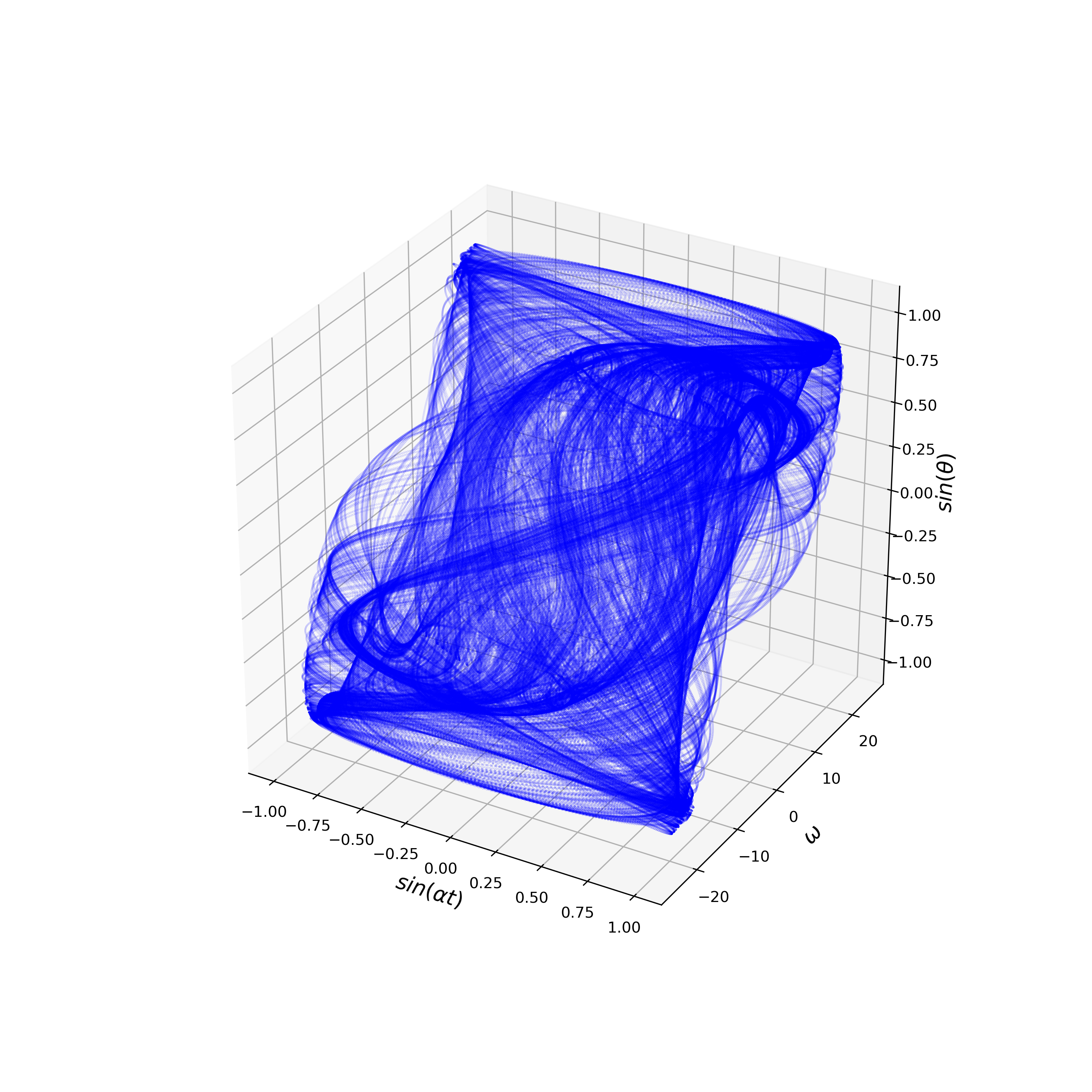}
\caption{A trajectory of the driven, damped pendulum \eqref{eq:Pendulum}.}
\label{fig:pendulum-trajectory}
\end{figure}

As a second example, we study the driven, damped pendulum:
	\begin{equation}\begin{split}\label{eq:Pendulum}
			\dot{\theta} &= \omega, \\ \dot{\omega} &= -\beta\omega
			- \nu_0^2 \sin(\theta) +A \cos(\alpha\ t),\\
	\end{split}\end{equation}
where the natural frequency is $\nu_0 = 98$, the damping coefficient
is $\beta = 2.5$, and the driving force has amplitude $A = 91$
and frequency $\alpha = 0.75\nu_0$.
The coordinates of the three-dimensional extended phase space
$\mathbb{T}^2 \times \mathbb{R}$ are the angle
$\theta$, the time $t$, and the angular velocity $\omega$. We generate a
trajectory for this system using a fourth-order Runge-Kutta algorithm
with initial condition $(3.14,0, 50)$ for $1.1\times10^6$ points with
fixed time step $0.001$, discarding the first $10^5$ points to remove
the transient, resulting in a final time series of length one million.
To avoid issues with periodicity in $\theta$ and $t$, we project the
time series from $\mathbb{T}^2 \times \bR$ onto
$(\sin(\theta), \sin(\alpha t),\omega) \in \mathbb{\bR}^3$.
The resulting trajectory is shown in
Fig.~\ref{fig:pendulum-trajectory}.  Note that the attractor has the
bound $|\omega| \le 24.88$, but that the range of the other two
variables is $[-1,1]$ due to the sinusoidal transformation.

\begin{figure}[ht]
		\centering \subfloat[Correlation sum calculations]{
                  \includegraphics[width=0.5\linewidth]{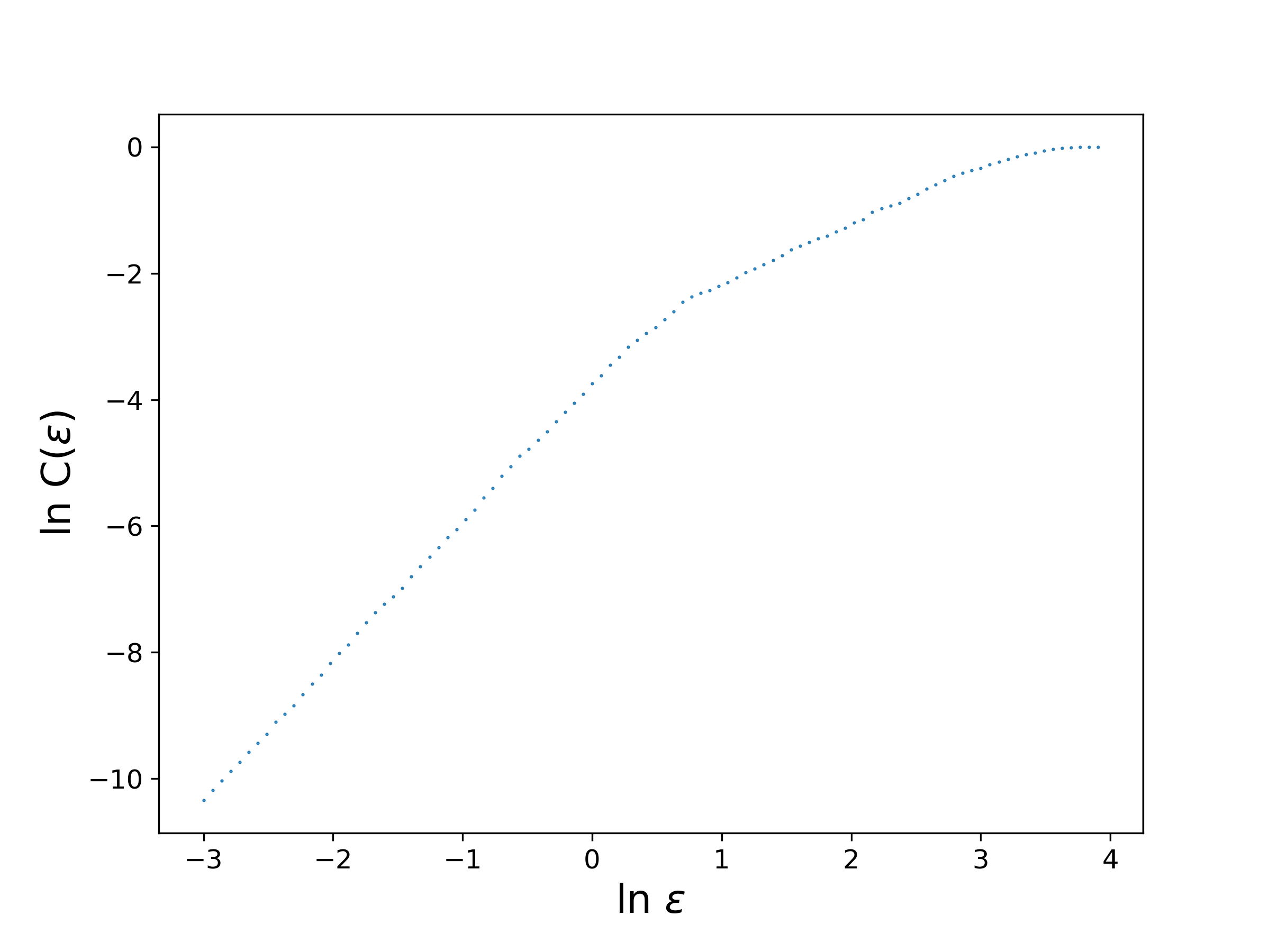}}\\ \subfloat[Weighted
                  slope distribution]{
                  \includegraphics[width=0.5\linewidth]{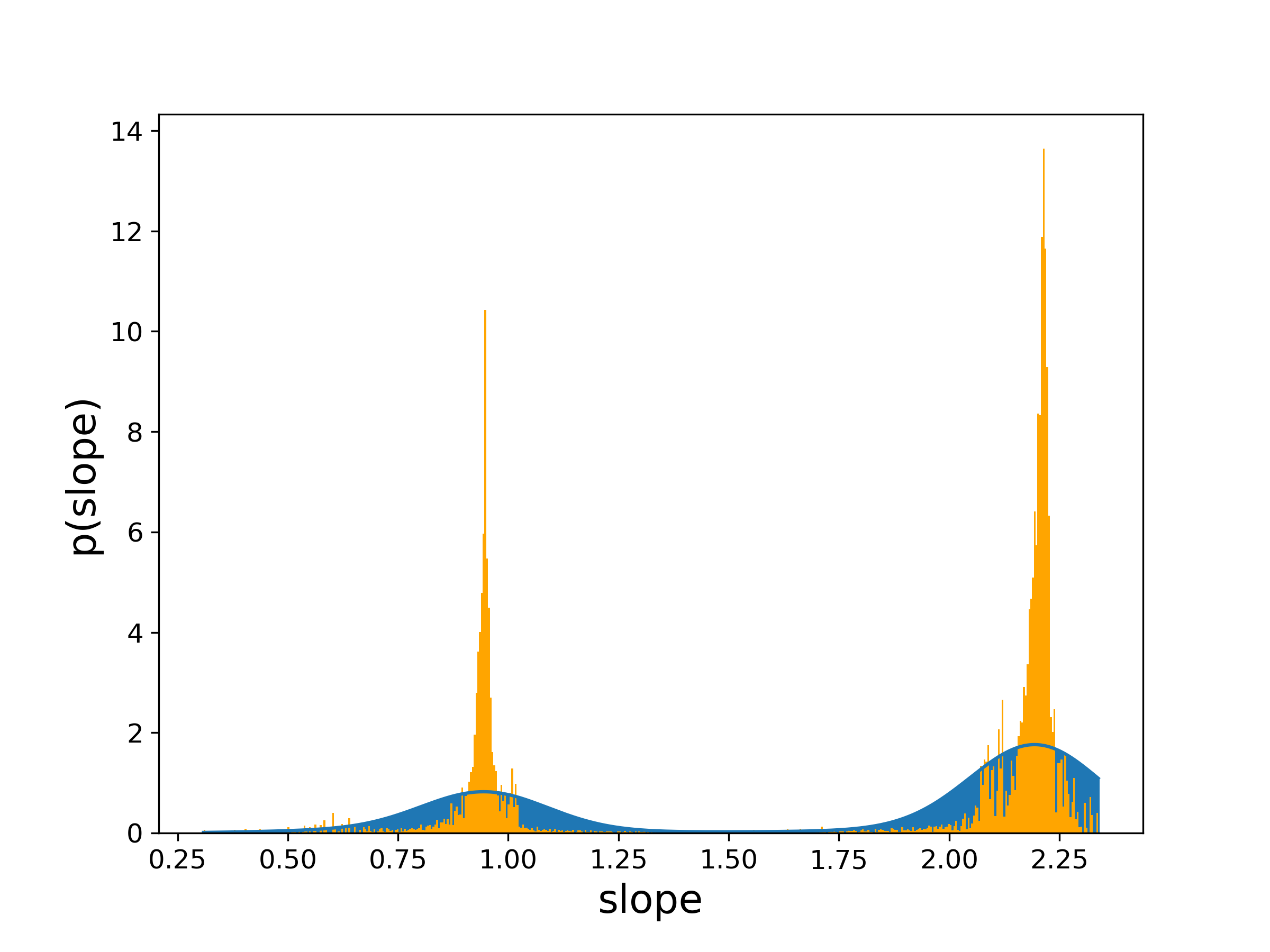}}
                \subfloat[Weighted distributions of interval endpoints
                  with the correlation sum graph from (a) superimposed]{
                  \includegraphics[width=0.5\linewidth]{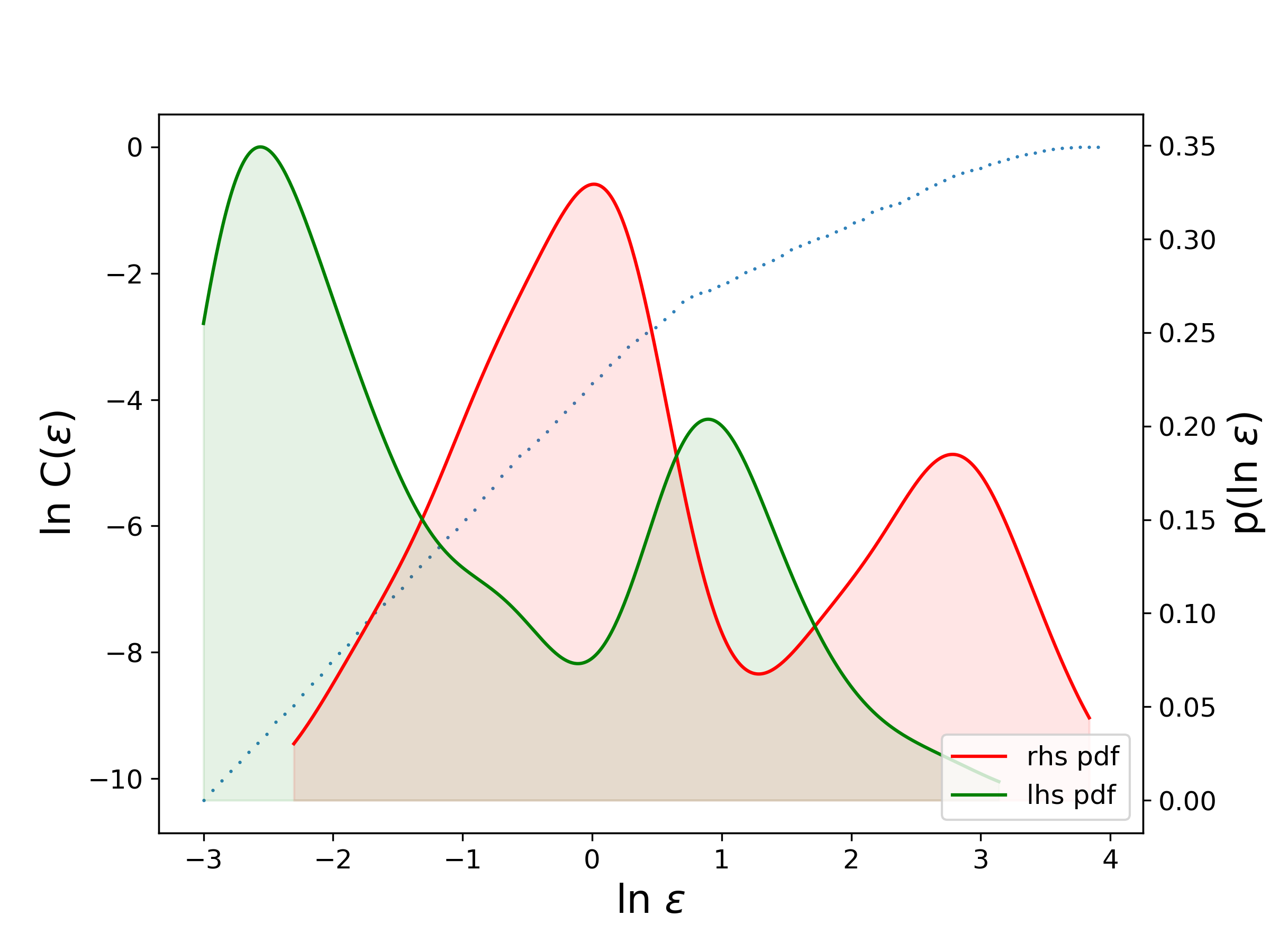}}
\caption{Estimating the correlation dimension of the driven, damped
  pendulum trajectory shown in Fig.~\ref{fig:pendulum-trajectory}.}
\label{fig:pendulum}
\end{figure}

To the eye, results of a {\tt d2} calculation on this trajectory,
shown in Fig.~\ref{fig:pendulum}(a), exhibit two apparent scaling
regions above and below $\ln \epsilon \approx 1$.  The slope
distributions produced by our method not only confirm, but also
formalize, this observation.  The larger peak of the distribution in
panel (b) of the figure falls at $d_2 = 2.19 \pm 0.09$
($FWHM$ = 0.32 and $p_{FWHM} = 0.53$), which
is equivalent to the correlation dimension of the attractor as
reported in~\citeInline{Deshmukh2020}. Note that, to account for the
fact that the distribution is clipped on the right before the density
falls below half the peak value, the computation of $FWHM$ uses this
as the upper bound.
The interval endpoint distributions in panel (c) indicate that the
extent of this scaling region is $-2.6 < \ln \epsilon < 0$. To the
eye, $\epsilon_r \sim 0.8$ would seem a more-appropriate choice
for the RHS endpoint of the scaling region; however, minor
oscillations in the interval $\epsilon \in [0,1]$ cause the
ensemble-based algorithm to be more conservative and choose
$\epsilon_r = 0$.
The presence of a second scaling region in panel (a) for $\ln \epsilon > 1$ gives a
second mode at the slope $0.94 \pm 0.10$ ($FWHM = 0.35$ and $p_{FWHM}
= 0.26$).  The secondary peaks in the endpoint distributions in
panel~(c) suggest an extent of $0.8 < \ln \epsilon < 2.8$
for this second scaling region, which is consistent with visual inspection of
panel~(a).

This second scaling region is an
artifact of the large aspect ratio of the attractor: only one of the
variables, $\omega$, has a range larger than $1$, so when $\epsilon >
1$, the effective dimension is one.  To test this hypothesis, we
rescaled the components of the phase space vectors so that the
attractor has equal extent of $[0,1]$ in all three directions (using
the {\tt -E} flag in the {\tt TISEAN d2} command) and repeated the
{\tt d2} calculations.  This leads to rescaling of the axis of the
$d_2$ plot---$\ln \epsilon$ now has the range $[-7,0]$---and causes
the second scaling region in the previous results to disappear,
leaving a single scaling region $-7 < \ln \epsilon < -2$
with slope
\[
	d_2 = 2.22 \pm 0.02 \quad (FWHM = 0.05, \mbox{ and } p_{FWHM} = 0.70).
\]
Note that, in addition to resolving the artifact of the second scaling
region, this rescaling also reduces the width of the mode and gives a
much tighter error bound.

By revealing the existence, the endpoints, and the slopes of different
scaling regions in the data, our ensemble-based approach has not only
produced an objective estimate for the correlation dimension, but also
yielded some insight into the interaction between the {\tt d2}
algorithm and the attractor geometry. 
	
\subsubsection{Noise}
\label{sec:noise}

\begin{figure}[ht]
		\centering
		\subfloat[$\delta  = 0.01$]{
			\includegraphics[width=0.4\linewidth]{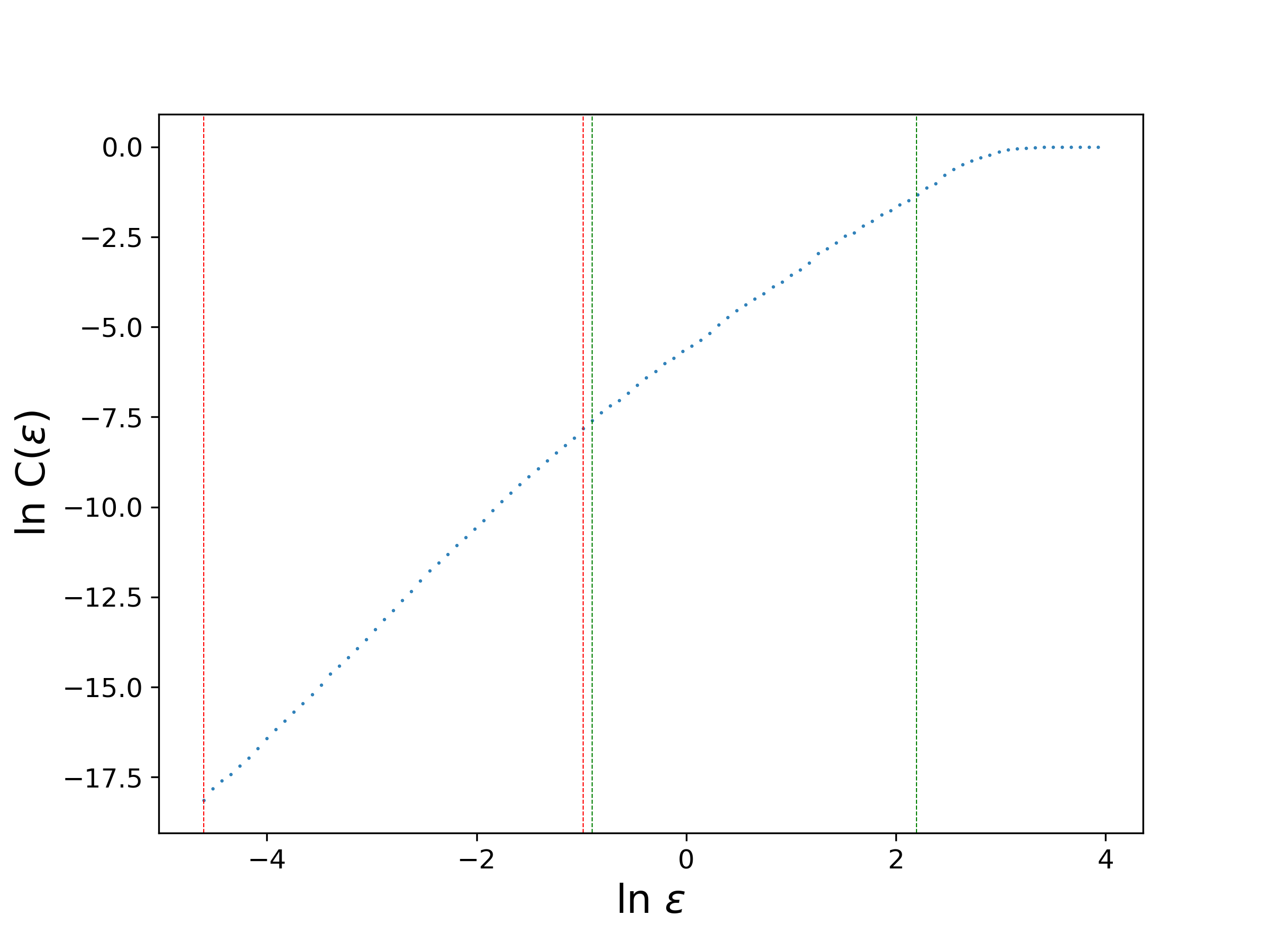}
		}
		\subfloat[Weighted slope distribution for (a)]{
			\includegraphics[width=0.4\linewidth]{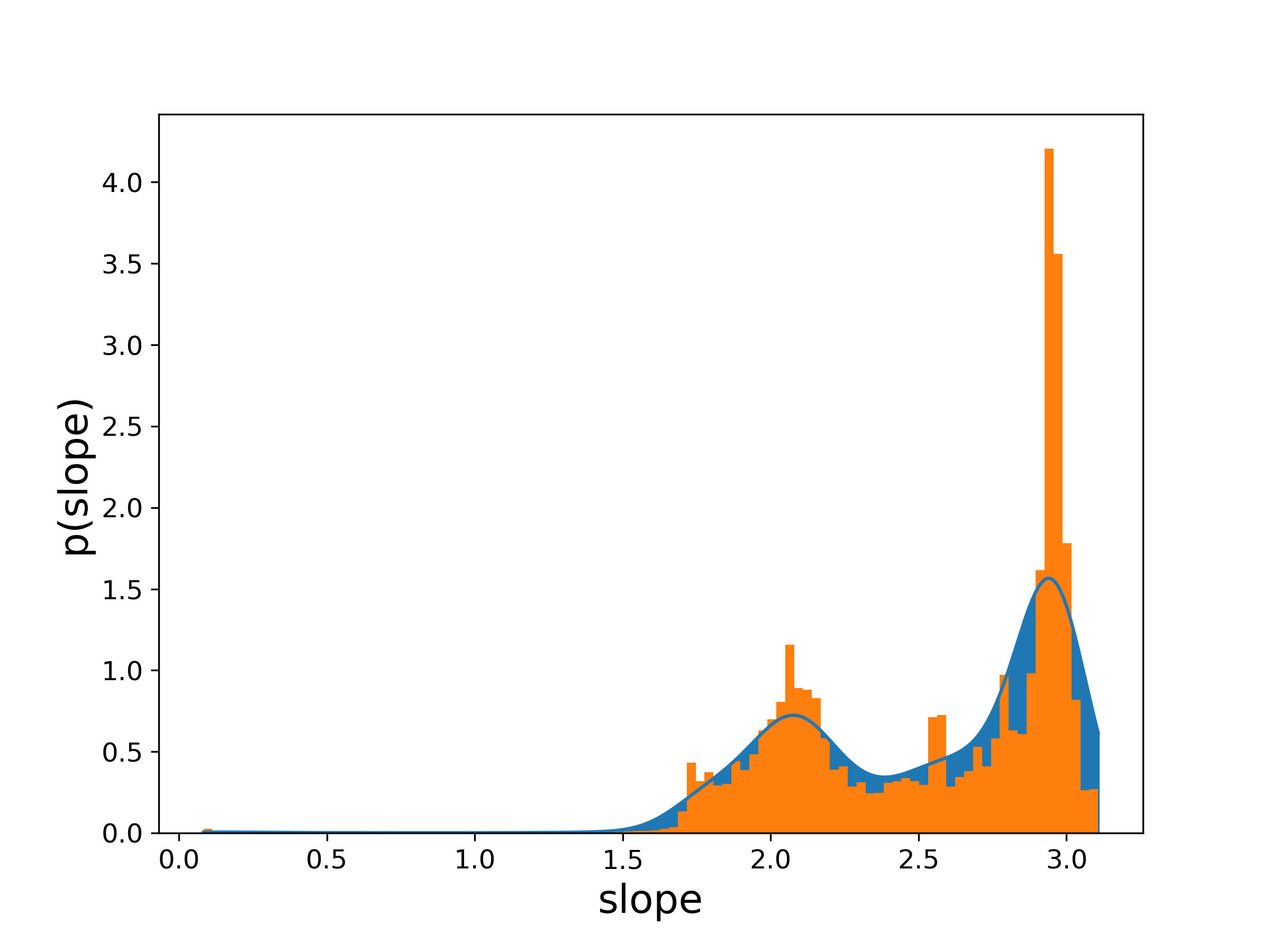}
		}\\
		\subfloat[$\delta  = 0.1 $]{
			\includegraphics[width=0.4\linewidth]{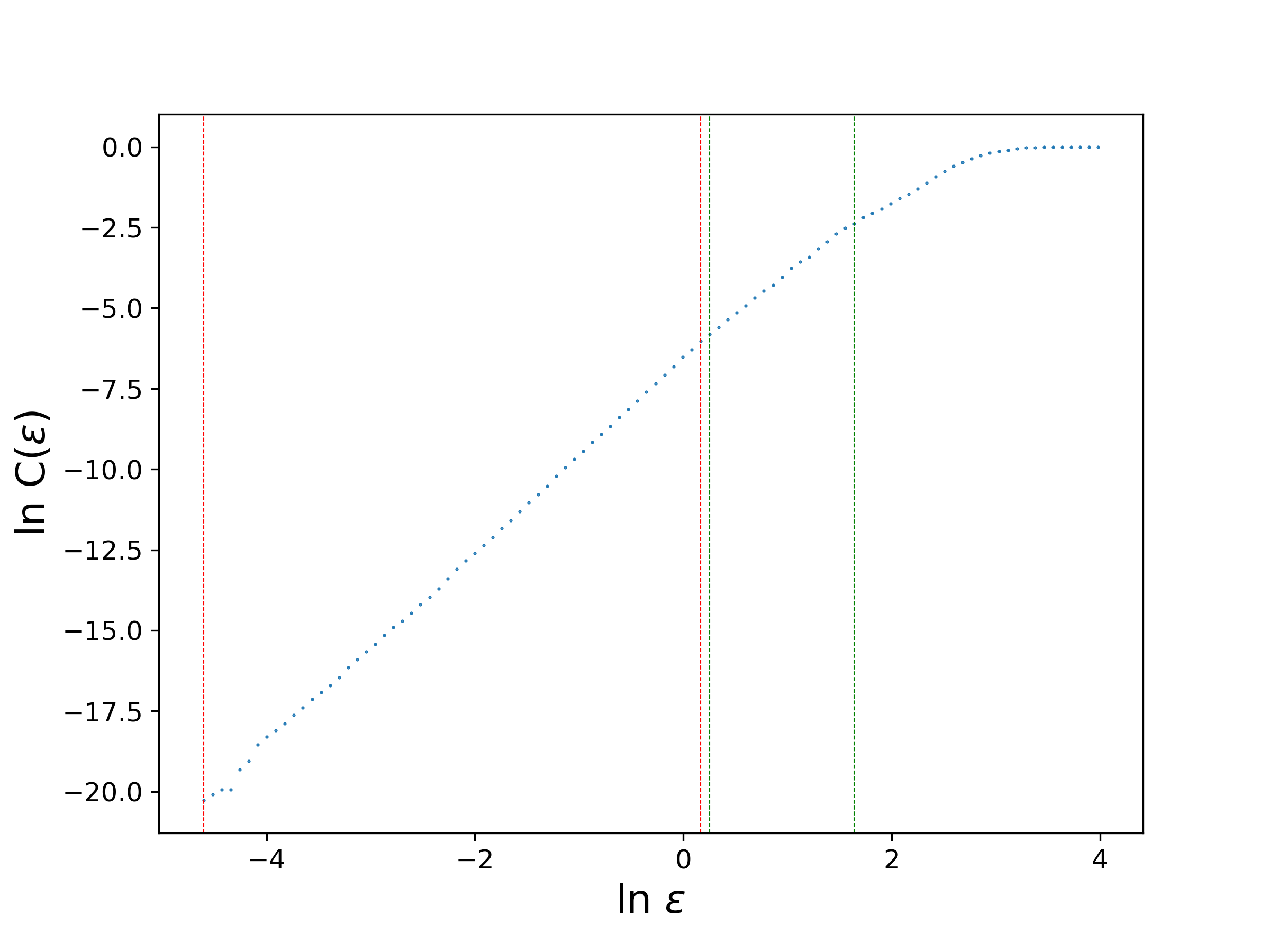}
		}
		\subfloat[Weighted slope distribution for (c)]{
			\includegraphics[width=0.4\linewidth]{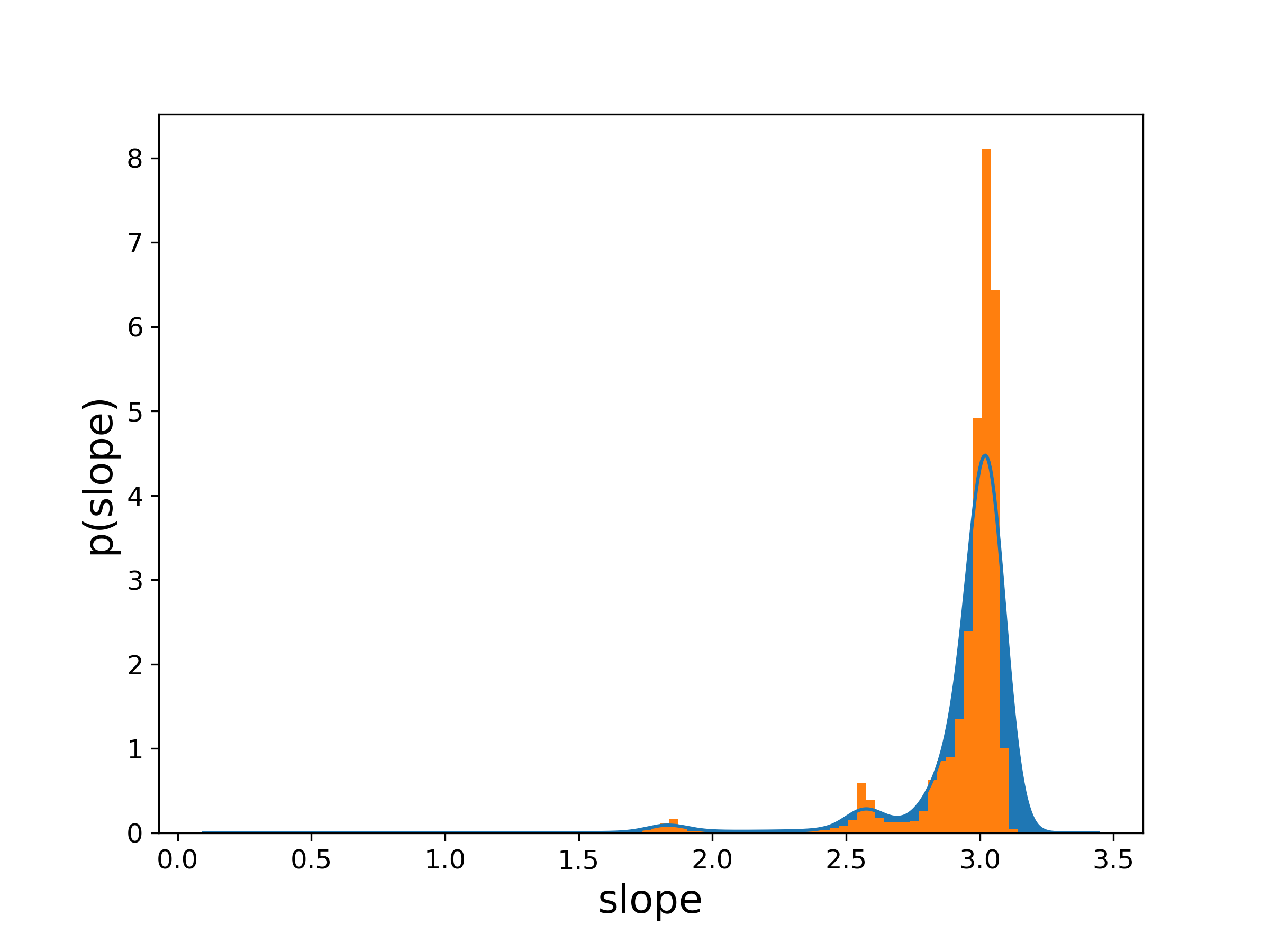}
		}\\
		\subfloat[$\delta  =  0.1$]{
			\includegraphics[width=0.4\linewidth]{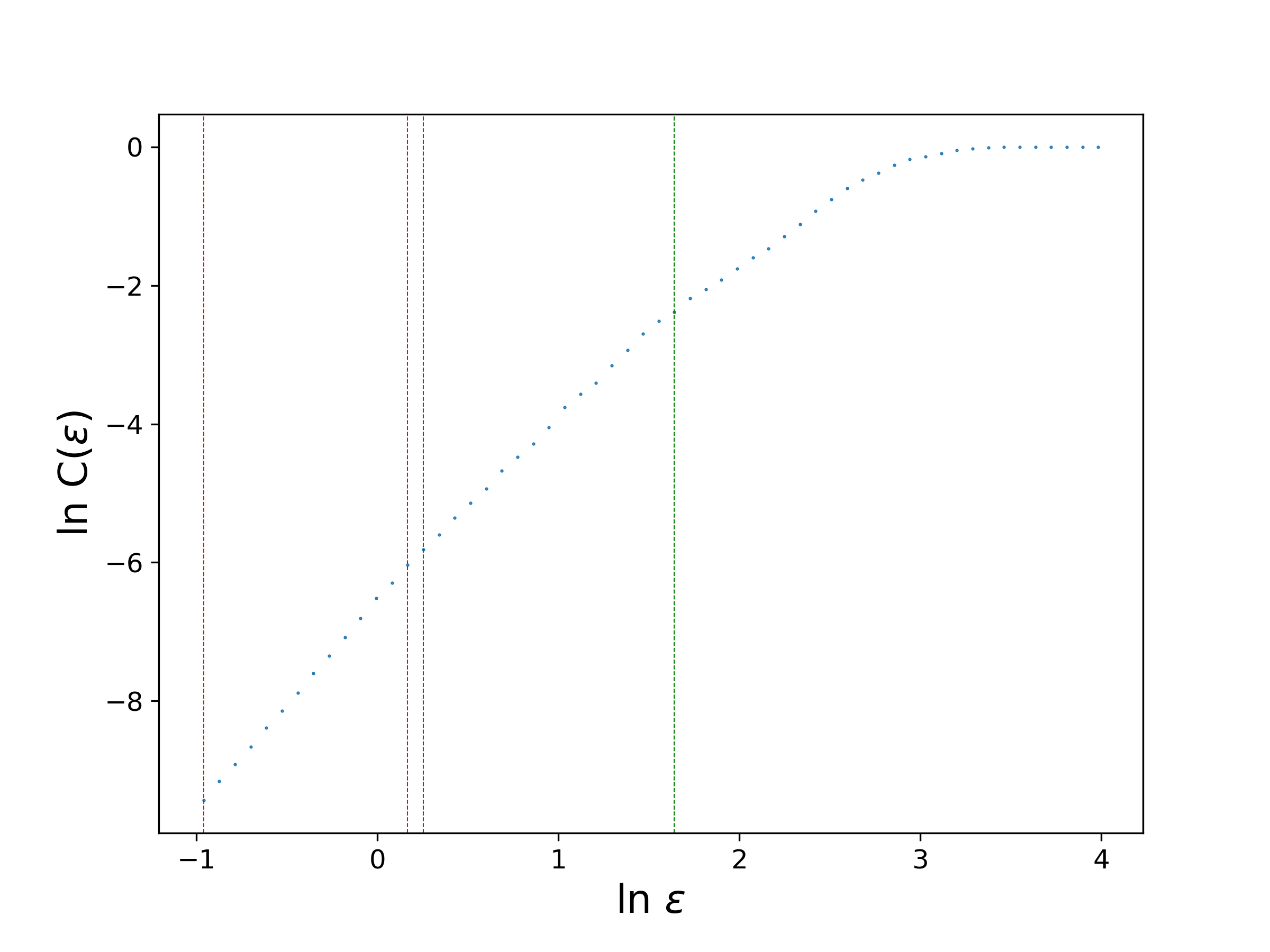}
		}
		\subfloat[Weighted slope distribution for (e)]{
			\includegraphics[width=0.4\linewidth]{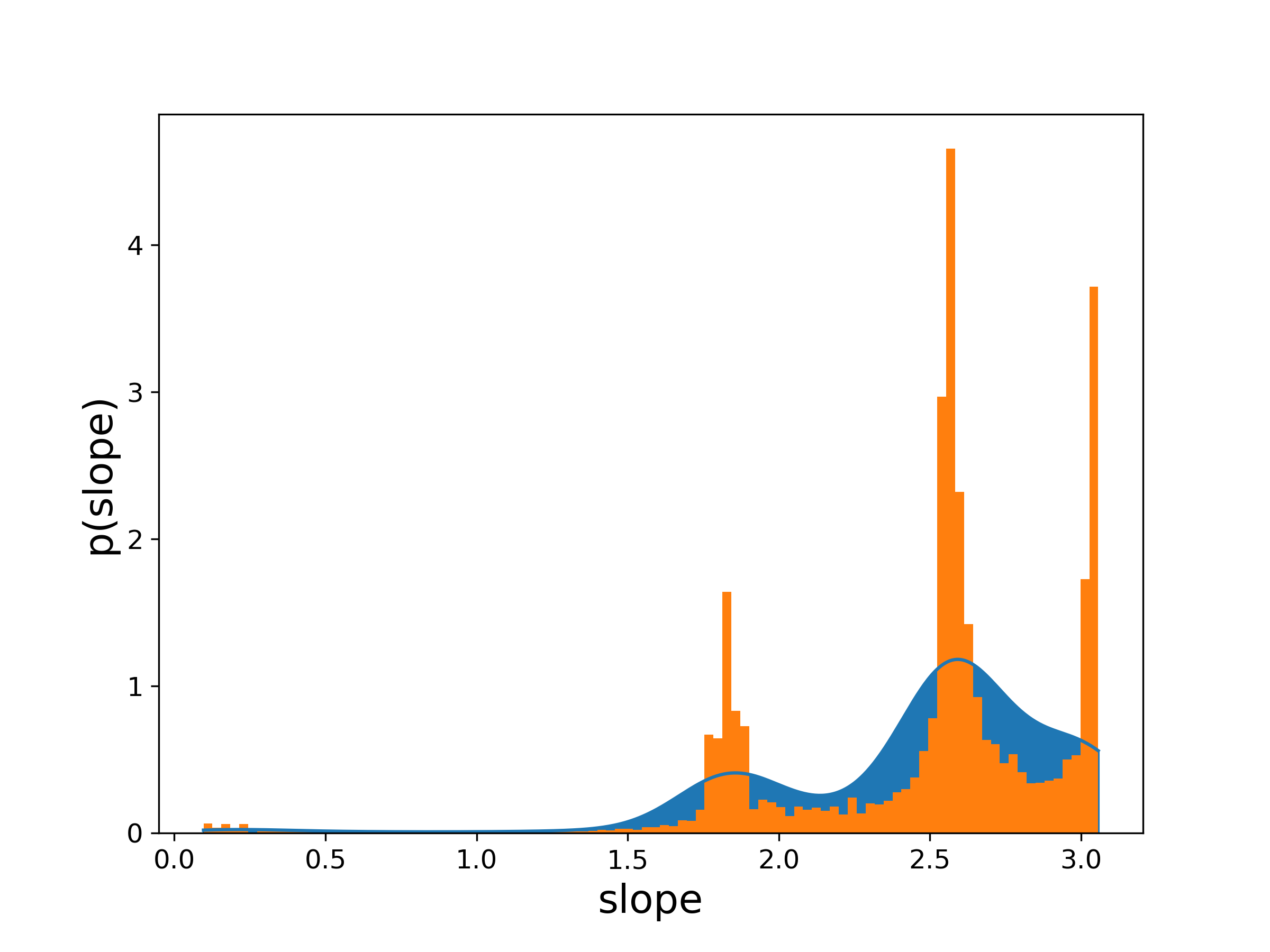}
		}
		\caption{Correlation dimension of
                  the Lorenz attractor with added noise
                  $\delta = 0.01$ for (a)-(b), and $\delta = 0.1$
                  for (c)-(f).  
                  Panels (e)-(f) exclude $\ln \epsilon < -1$.
                  The red markers in (a), (c) and (e) denote
                  scaling regions corresponding to noise while the
                  green markers indicate that for the dynamics.}
		\label{fig:lorenz_noise_0.01}
\end{figure}

Noise is a key challenge for any practical nonlinear time-series
analysis method.  We explore the effects of measurement noise on our
method using the Lorenz trajectory of \S\ref{sec:lorenz} by adding
uniform noise \textit{post facto} to each of the three state variables.
The noise amplitude in each dimension is proportional to the radius of
the attractor in that dimension.  The correlation sum
$C(\epsilon)$ for a noise amplitude $\delta = 0.01$---{\sl i.e.},
$1\%$ of the radius of the attractor in each dimension---is shown in
Fig.~\ref{fig:lorenz_noise_0.01}(a).
There appear to be two distinct scaling regions in this plot, above
and below a slight knee at $\ln \epsilon \ \approx\ -1$.  The slope
distribution produced by our method is shown in
Fig.~\ref{fig:lorenz_noise_0.01}(b).  As in the pendulum example, this
distribution is bimodal, indicating the presence of two scaling
regions with slopes
\begin{align*}
	2.94 \pm 0.09 &\quad(FWHM = 0.33 \mbox{ and } p_{FWHM} = 0.42), \mbox{ and }\\ 
	2.08 \pm 0.15 &\quad (FWHM = 0.51 \mbox{ and } p_{FWHM} = 0.30),
\end{align*} 
respectively. We claim that these results are consistent with
the geometry of the noise and of the dynamics, respectively.  The
taller peak corresponds to the interval $\ln \epsilon \in [-4.60,-1]$,
delineated by red markers in panel (a).  Note that the right endpoint
of this region is close to $\epsilon_r \approx 0.01 \times 26$, the
maximum extent of the noise. The computed slope of $2.94$ in this
interval captures the geometry of a noise cloud in three-dimensional
state space.  The smaller, secondary peak at $2.08$
reflects the dimension of the attractor for scales larger than the noise, the
interval $\ln \epsilon \in [-1,2.2]$ that is bounded by green markers in panel (a).

As the noise level increases, the geometry of the attractor is
increasingly obscured (see Fig.~\ref{fig:lorenz_noise_attractor}
in Appendix~\ref{sec:appendix_lorenz_noise}). 
Figures \ref{fig:lorenz_noise_0.01}(c) and~(d) show results for $\delta = 0.1$.
As one would expect, the right-hand
boundary of the lower scaling region is increased, shown as the red markers in
panel~(c). We observe that $\ln\ \epsilon_r \approx 0.16$, 
near the the noise cloud radius of $1.3$.
With this noise level, the slope distribution is nearly unimodal with mode
\[
	3.02 \pm 0.05 \quad (FWHM = 0.18 \mbox{ and } p_{FWHM} = 0.65), 
\]
again reflecting the geometry of the noise.  While there appears to be
a secondary peak in the slope distribution in panel (d), it is nearly
obscured.

Interestingly, the identification of the scaling region due to the
noise can be used to refine the calculation and retrieve some
information about the dynamical scaling: one simply repeats the
slope-distribution calculations using only the data for larger $\epsilon$
values; that is, discarding the values below the noise. 
Restricting to the interval $[-1.0, 4.0]$,
gives the curve  and the slope distribution
shown in panels (e) and (f). In addition to the noise peak near $3.0$, this
reveals two additional peaks. The first,
\[
	2.59\pm 0.20 \quad (FWHM = 0.68 \mbox{ and } p_{FWHM} = 0.67),
\]
was hinted at by the tail of the distribution in panel (d).  This, we
conjecture, corresponds to the correlation dimension of the noisy
dynamics. The final peak, near $1.8$, corresponds to the smaller
scaling region $[2.0,3.0]$ that can be seen in a heatmap-style scatter
plot (not shown).  The validity of this region can be easily confirmed
by further limiting the ensemble to the interval $\ln \epsilon \in
[1.0,4.0]$ which then gives a single mode at $1.83$.  Note that this
region corresponds to a smaller scaling region that was also seen in the
noiseless case: the small green cluster in Fig.~\ref{fig:lorenz}(d).

The ability of the ensemble-based slope distribution method to reveal
secondary scaling regions and suggesting refinements allows one to identify
noise scales and even extract information that might be hidden by the noise.

\subsubsection{Time Series Reconstructions}
\label{sec:embed}

The previous examples assumed that all of the state variables are
known. This is rarely the case in practice; rather, one often has data
only from a single sensor. Delay-coordinate embedding
\cite{takens,packard80}, the foundation of nonlinear time-series
analysis, allows one to reconstruct a diffeomorphic representation of
the actual dynamics from observations of a scalar time series $x(t)$
provided that a few requirements are met: $x(t)$ must represent a
smooth, generic observation function of the dynamical
variables\cite{takens,Whitney36} and the measurements should be evenly
spaced in time.  This method embeds the time series $x(t)$ into
$\mathbb{R}^m$ as a set of delay vectors of the form
$[x(t),x(t-\tau),\dots,x(t-(m-1)\tau)]$ given a time delay $\tau$.
Theoretically the only requirements are $\tau >0$\cite{takens} and
$m>2d_{cap}$, where $d_{cap}$ is the capacity dimension of the
attractor on which the orbit is dense \cite{sauer91}.  Many heuristics
have been developed for estimating these two free
parameters\cite{Olbrich97,josh-tdAIS,fraser-swinney,kantz97,Buzug92Comp,liebert-wavering,
  Gibson92,Buzugfilldeform,Liebert89,joshua-pnp,rosenstein94,Cao97Embed,Kugi96,KBA92,Hegger:1999yq,Deshmukh2020,kennel92},
notably the first minimum of the average mutual information
for selecting $\tau$ \cite{fraser-swinney} and the
false near neighbors algorithm for selecting $m$\cite{kennel92}.  See
\citesInline{Holger-and-Liz, tisean-website} for more details.

\begin{figure}[ht]
		\centering
		\subfloat[Correlation sum calculations]{
			\includegraphics[width=0.5\linewidth]{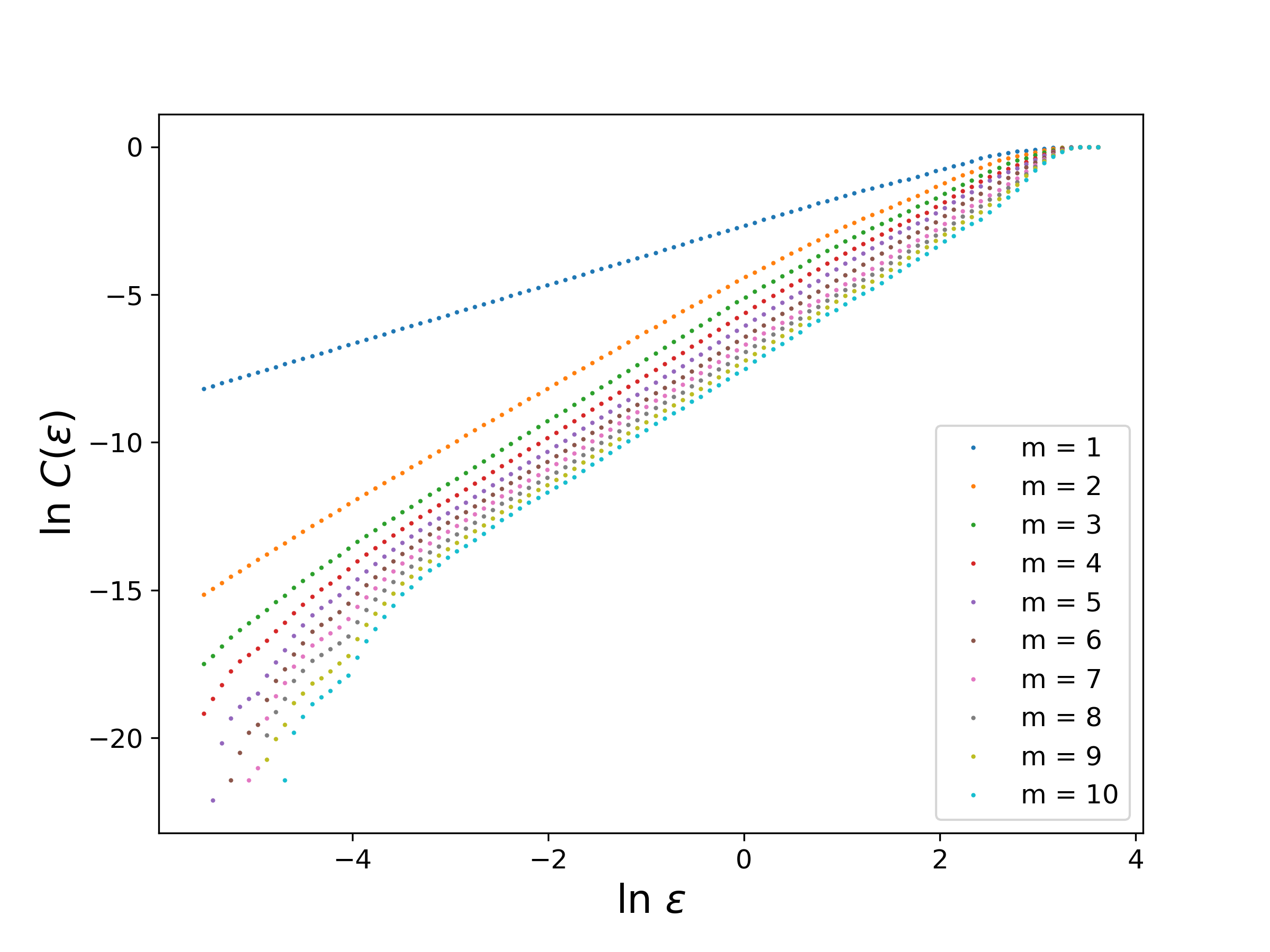}
		}
		\subfloat[Weighted slope distributions]{
			\includegraphics[width=0.5\linewidth]{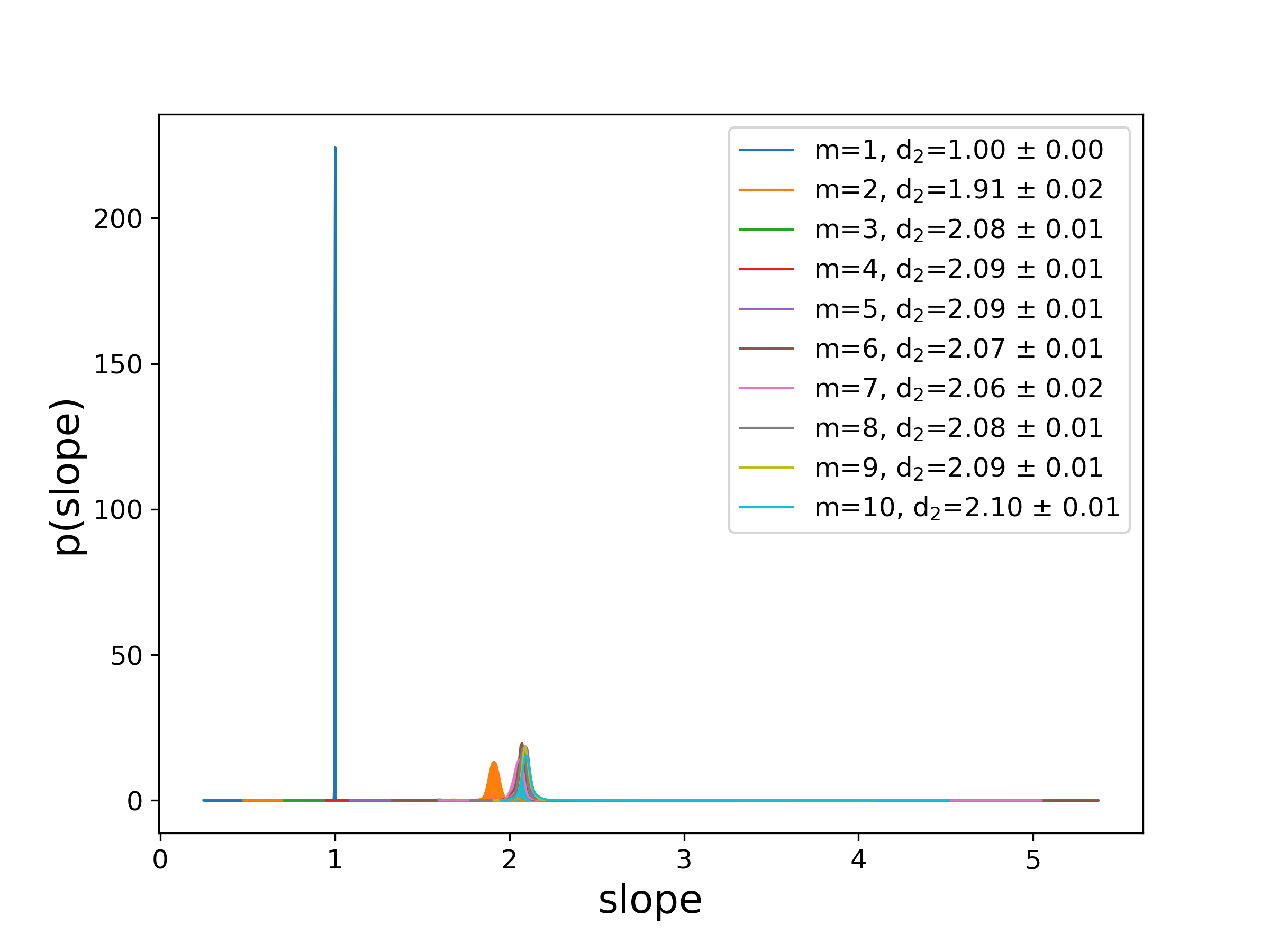}
		}\\
		\subfloat[Wasserstein distance]{
			\includegraphics[width=0.5\linewidth]{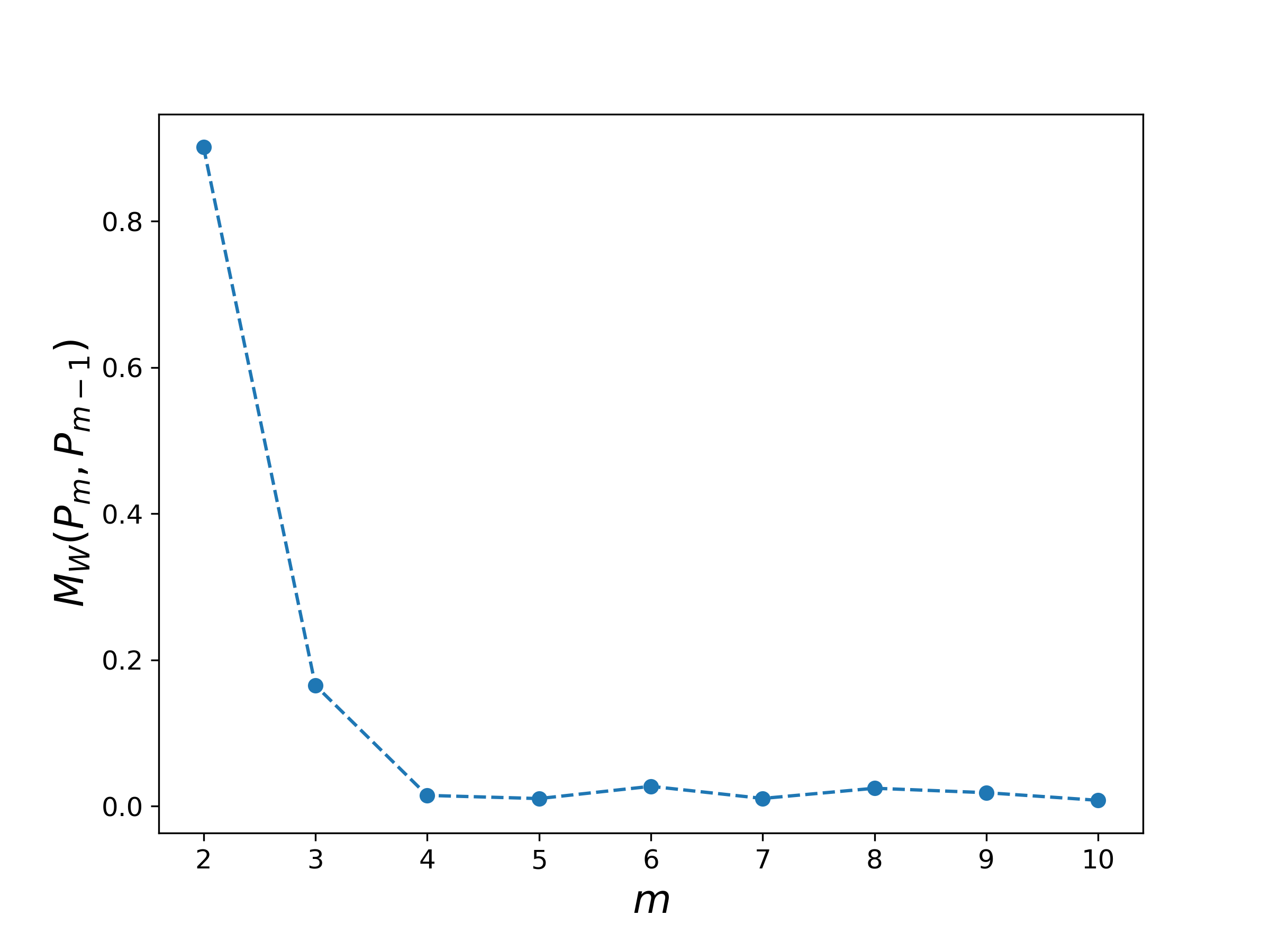}
	}
\caption{Correlation dimension from time-delay reconstructions of the Lorenz
    system for $x(t)$, $\tau = 18$ and $m \in [1,10]$. Each $m$ from (a) gives a slope distribution (b),
    and these are pairwise compared in (c) using the 
    Wasserstein distance. }
\label{fig:lorenz_reconstruction}
\end{figure}

Since the correlation dimension is preserved under a diffeomorphism,
it can be computed with the {\tt d2} algorithm on properly
reconstructed dynamics.  Moreover, the calculations of the correlation
dimension for different values of $m$ can help diagnose the
correctness of an embedding.  To explore how our method can contribute
in this context, we embed the $x(t)$ time series of the Lorenz
trajectory from \S\ref{sec:lorenz} using $\tau = 18$, which was chosen
using the curvature heuristic described in ~\citesInline{Deshmukh2020}
and corroborated using the first minimum of the average mutual
information. Using {\tt TISEAN}, we then create a
series of embeddings for a range of $m$ values and apply the
Grassberger-Procaccia algorithm to each one.  The results are shown in
Fig.~\ref{fig:lorenz_reconstruction}(a) for $m \in [1,10]$.

As is well known, we can assess the convergence of $d_2$ by reasoning
about such a sequence of curves.  When $m$ is too small, the
reconstructed attractor is improperly unfolded, giving an incorrect
dimension, while for large enough $m$, $d_2$ typically
converges---modulo data limitation issues---to the nominally correct
value.  
Since our  ensemble methodology automates the calculation 
of scaling regions, it can simplify this calculation for multiple curves.
To determine the
value of $m$ for which the slopes converge, one simply computes the
modes of each slope distribution and looks for the $m$ value at which
the positions of those modes stabilize.  For
Fig.~\ref{fig:lorenz_reconstruction}(a), the mode for $m=1$ 
is, of course, $d_2 = 1$, but when $m$ reaches $3$, the
distributions begin to overlap and for $3 \le m \le 5$ the modes are $2.09 \pm
0.01$, essentially the estimate from the full dynamics in \S\ref{sec:lorenz}.

To formalize this procedure, we use the Wasserstein metric $M_{W}$ to
compare the two distributions\cite{vallender1974}.  Informally, this
metric---called the ``earth mover's distance''---treats each
distribution as a pile of soil, with the distance between the two
distributions defined as the minimum ``effort'' required to turn one
pile into another.
Thus $M_{W} = 0$ only when the distributions are identical.
We use the {\tt python scipy.stats.wasserstein\_distance}
package to compute this metric.\footnote{https://docs.scipy.org/doc/scipy/reference/generated/scipy.stats.wasserstein\_distance.html}

Given slope distributions $P_m$ and $P_{m-1}$ for successive embedding
dimensions from Fig.~\ref{fig:lorenz_reconstruction}(b), the distance $M_W(P_m,P_{m-1})$ is shown in
panel (c) of the figure.
Note that $M_W(P_m,P_{m-1})$ initially
decreases rapidly with increasing $m$, approaching zero at
$m=4$.  Up to fluctuations, $M_W(P_m,P_{m-1})$ remains approximately
constant thereafter.  Through experiments across multiple examples, we
heuristically propose a threshold $M_W(P_m,P_{m-1}) \sim 0.1$ to
estimate convergence of $d_2$ as $m$ grows.
Such a threshold has two benefits, giving both an estimate of the
embedding dimension and of $d_2$.

To assess the generality of these ideas, we applied the ensemble-based
method to four additional scalar time series: the noisy Lorenz data of
\S\ref{sec:noise} with $\delta = 0.1$ using $x(t)$, the pendulum trajectory of
\S\ref{sec:pend} using the $\omega(t)$, a shorter,
$200,000$ point segment of this orbit, and finally, a trajectory from
the Lorenz-96 model \cite{lorenz96Model}.  In each case, we again used
the curvature-based heuristic of \citeInline{Deshmukh2020} to estimate
$\tau$.
This set of examples represents a range of problems that can arise
in practice: measurements affected by noise, and a trajectory that is
too short to fully cover
an attractor, and a high-dimensional attractor.
\setlength{\tabcolsep}{12pt}
\begin{table}
	\centering
	\begin{tabular}{l|r|rrr}
		\multicolumn{1}{c|}{System} & \multicolumn{1}{c|}{Full Dynamics} & \multicolumn{3}{c}{Reconstructed}\\
		& \multicolumn{1}{c|}{$d_2$} & \multicolumn{1}{c}{$d_2$} & $\tau$ & $m$ \\
		\hline 
		Lorenz (\ref{sec:lorenz})      & $2.09\pm 0.02$  & $2.09\pm 0.01$  & 18 & 4\\
		Noisy Lorenz ($\delta = 0.1$)  & $2.59\pm 0.20$  & $2.27\pm 0.14$  & 21 & 7\\
		Pendulum (\ref{sec:pend})      & $2.22\pm 0.02$  & $2.16\pm 0.02$  &120 & 4\\  
		Truncated Pendulum             & $2.23\pm 0.07$  & $2.20\pm 0.03$  &120 & 4\\
		Lorenz-96                      & $5.63 \pm 0.09$  & $5.80\pm 0.07$  & 24 & 10\\  
	\end{tabular}
	\caption{Correlation dimensions calculated using the
          ensemble-based method with Wasserstein distance for five
          cases. The histograms for each of these examples are shown
          in Appendix \ref{sec:appendix_reconstruction}.}
	\label{tab:d2_results}
\end{table}
The results, presented in Table~\ref{tab:d2_results} show that $d_2$ 
for the reconstructed dynamics is close to
that of the full dynamics as well as that given previously by manually
fitting scaling regions \cite{Deshmukh2020}.  However, the
embedding dimensions are often significantly
smaller that those suggested by other work.  For the first four
examples in Table~\ref{tab:d2_results}, for example,
\citeInline{Deshmukh2020} used $m = 7 = 2d+1$, the dimension required
by the Takens theorem for a three-dimensional state space
\cite{takens}; for the Lorenz-96 example, that paper used $m = 12$,
obtained from the heuristic $m > 2d_{cap}$ \cite{sauer91}.
Both $m \geq 2d +1$ and $m > 2d_{cap}$ are \textit{sufficient}
conditions, of course.  The Wasserstein test used in
Table~\ref{tab:d2_results} shows that accurate estimates of $d_2$ can
often be obtained with a lower embedding dimension and without
prior knowledge of the original dynamics.

\subsection{Lyapunov Exponent}
\label{sec:lyap}

Computing a Lyapunov exponent also often involves identification and
characterization of scaling regions. Here we will
use the Kantz algorithm
\cite{Kantz1994} to estimate the maximal Lyapunov exponent $\lambda_1$
for a reconstructed trajectories of the Lorenz and chaotic
pendulum examples from \S\ref{sec:embed}.
We embed the scalar data
using a delay $\tau$ found by standard heuristics, and experiment with
various embedding dimensions. 
The Kantz algorithm starts by finding all of the points in
a ball of radius $\epsilon_s$ (also called the ``scale'') around
randomly chosen reference points on an attractor.  By marching
through time, the algorithm then computes the divergence between the
forward image of the reference points and the forward images of the
other points in the $\epsilon_s$ ball.  The average divergence across
all reference points at a given time is the \textit{stretching
factor}. To estimate $\lambda_1$, one
identifies a scaling region in the log-linear plot of the stretching
factor as a function of time.

\begin{figure}[ht]
	\centering
	\subfloat[Stretching factor calculations (Lorenz)]{
		\includegraphics[width=0.5\linewidth]{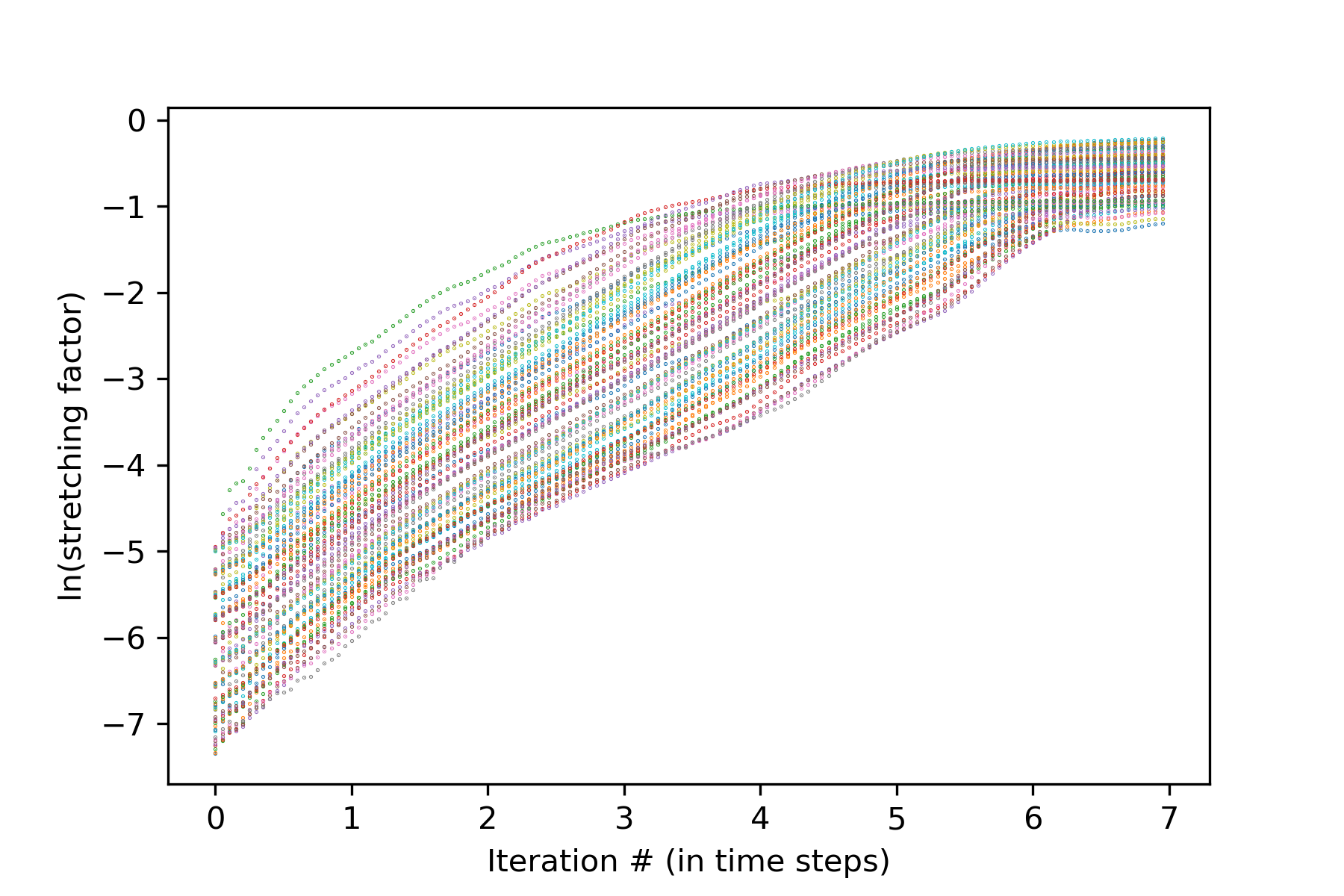}
	}
	\subfloat[$\lambda_1$ estimate (Lorenz)]{
		\includegraphics[width=0.5\linewidth]{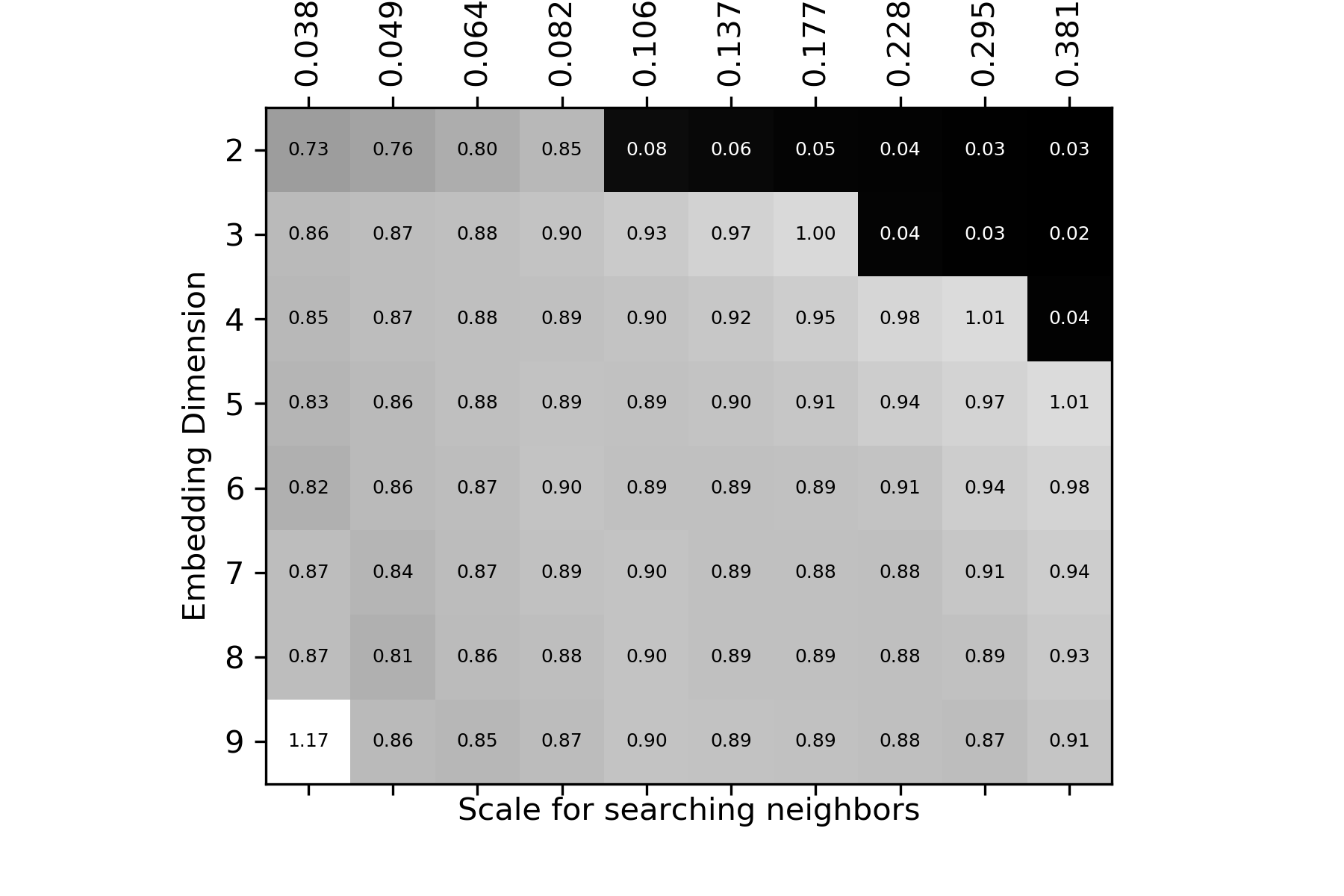} 
	}	\\
	\subfloat[Stretching factor calculations (pendulum)]{
		\includegraphics[width=0.5\linewidth]{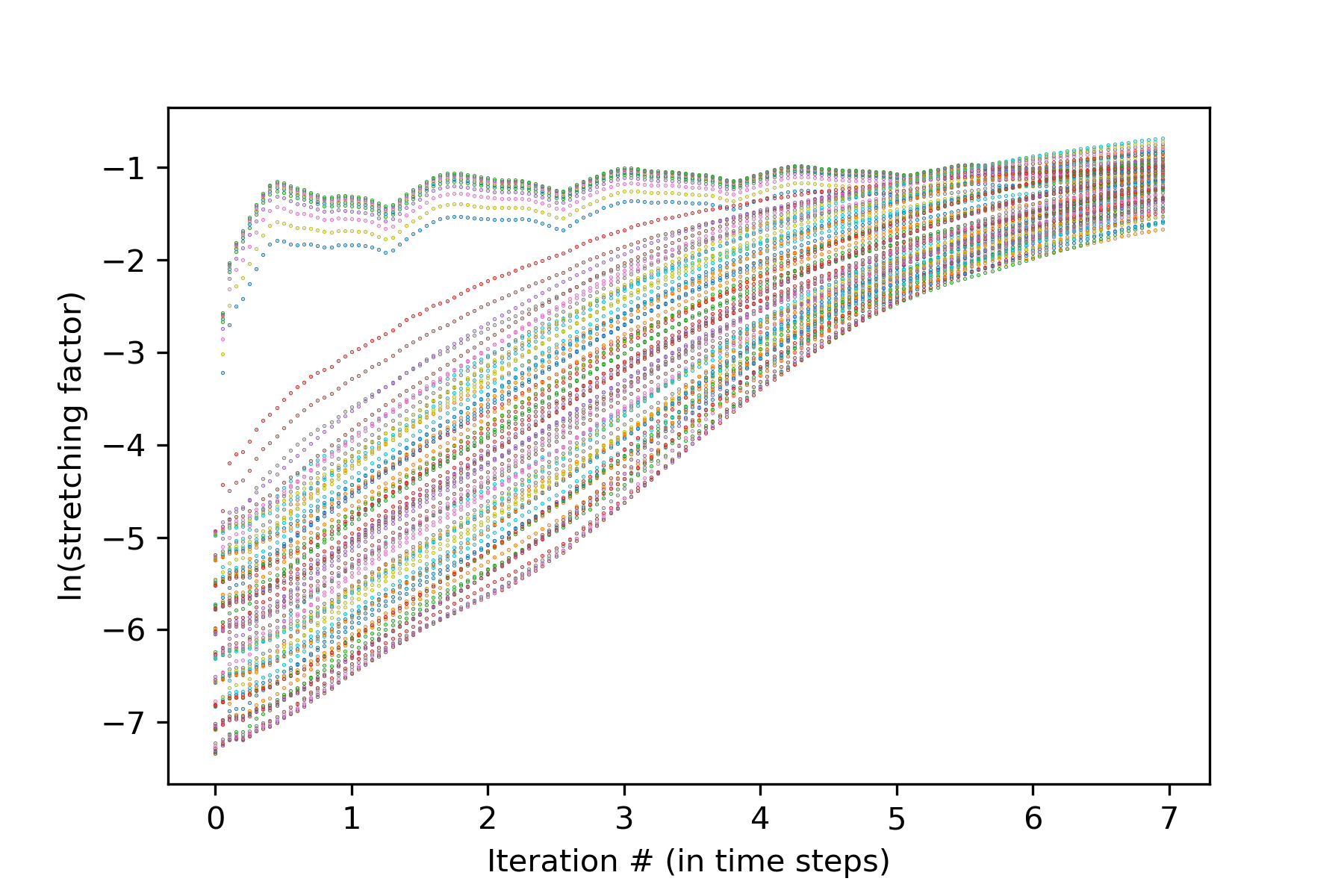}
	}
	\subfloat[$\lambda_1$ estimate (pendulum)]{
		\includegraphics[width=0.5\linewidth]{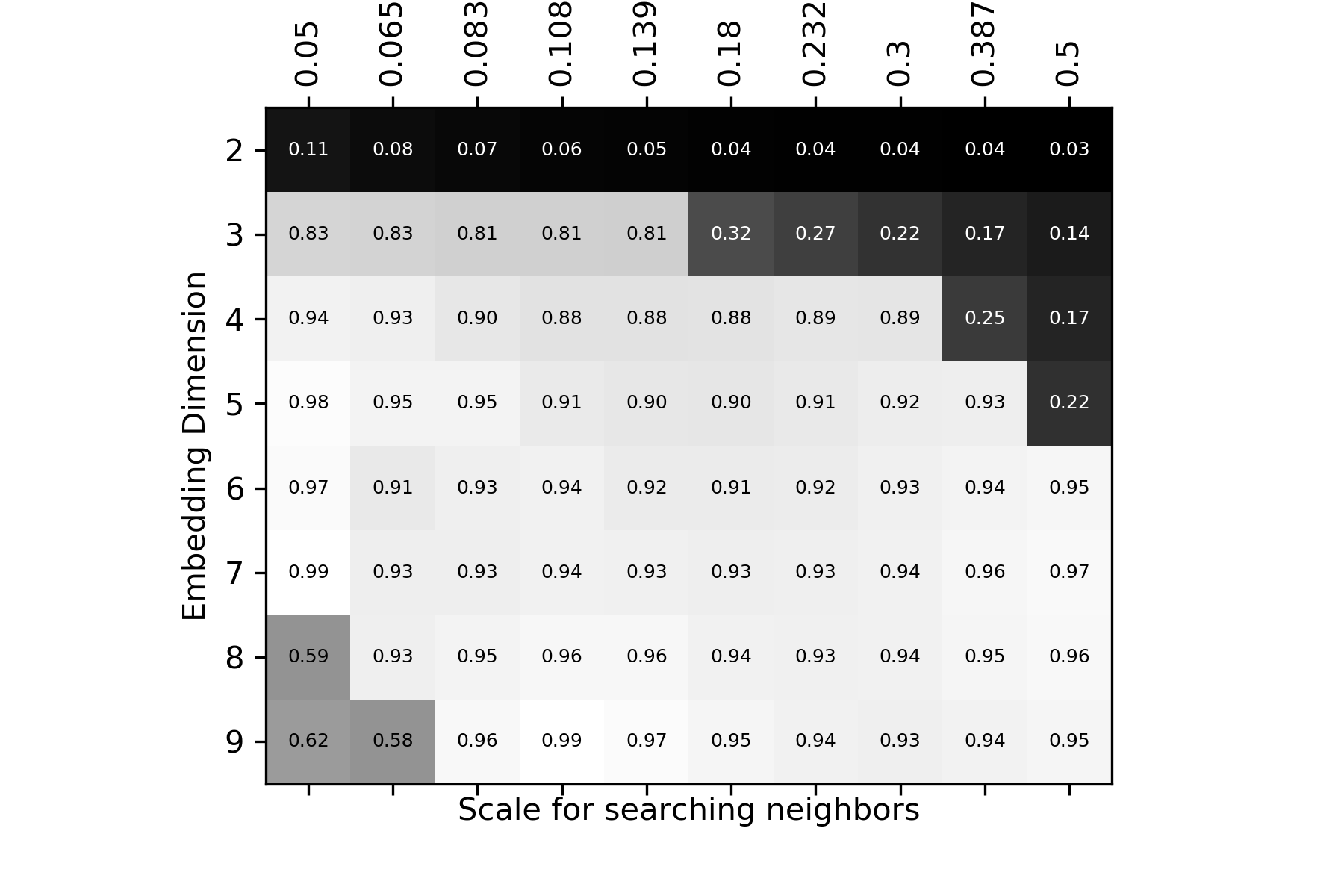}
	}
	\caption{Estimating the largest Lyapunov exponent $\lambda_1$
          of the Lorenz and driven, damped pendulum examples. Panels
          (a) and (c) show stretching-factor calculations for the
          eight embedding dimensions $m$ and ten search scales
          $\epsilon_s$.  The estimated exponents for all the
          dimensions and scales are shown in grid form in panels (b)
          and (d).  The gray scale in each grid cell represents the
          magnitude of the exponent.
          }
	\label{fig:lyap}
\end{figure}

Note that this procedure involves three free parameters: the
embedding parameters $\tau$ and $m$ and the scale of the balls
used in the algorithm.\footnote
{Other parameters---the number of reference points, Theiler window,
and length of the temporal march, etc.---were
set to TISEAN's default values.}
To obtain accurate results one is confronted by an onerous multivariate parameter tuning
problem, and an automated approach can be extremely
advantageous.  If, for example, one embeds the $x$ coordinate of the
Lorenz data from \S\ref{sec:lorenz} with $m$ ranging from 2 to 9 and
chooses 10 values of $\epsilon_s$ ({\sl e.g.,} on a logarithmic grid
with 10 values between 0.038 and 0.381), then there will be 80
different curves, as seen in Fig.~\ref{fig:lyap}(a).
Manually fitting scaling regions to each curve would be a demanding task.
The ensemble method gives the automated results shown in panel (b) of the figure.

Each row of this 2D grid
corresponds to fixed $m$ and each column to fixed $\epsilon_s$.  The
estimated $\lambda_1$---the location
of the mode in the slope PDF---is the value shown in the cell.
To help detect convergence, each cell is shaded according to the corresponding
$\lambda_1$ value.  Note that the majority of the
configurations (48 out of 80, colored in intermediate gray) give an
estimate of $\lambda_1 = 0.89\pm 0.02$, which is consistent with the
known value \cite{Grigorenko2003}.

The pattern in the grid make sense.  The cells in its center
correspond to midrange combinations of the free parameters: $m \in
[5,9]$ and $\epsilon_s \in [0.064,0.177]$.  Values outside these
ranges create well-known issues for the Kantz algorithm in the context
of finite data.  If $\epsilon_s$ is too small, the ball will not
contain enough trajectory points to allow the algorithm to effectively
track the divergence; if $\epsilon_s$ is too large, the ball will
straddle a large portion of the attractor, thereby conflating
state-space deformation with divergence.  
For $m$ too small, the attractor is insufficiently unfolded, so the
values in the top few rows of the grid are suspect.  In the upper
right corner, these two effects combine:
the attractor is both insufficiently unfolded and also
spanned by the $\epsilon_s$ ball.  Also, the
zero slope portion on the right ends of the stretching factor curves
becomes dominant
(since since the $\epsilon_s$ ball reaches the edge of a flattened
attractor more quickly) thus causing the ensemble-based algorithm to return
a slope close to zero. Finally, we see a slightly higher estimate of 
$\lambda_1 = 1.17$ for $m = 9$ and $\epsilon_s = 0.038$. On close inspection
of the stretching factor plot for this configuration, we see significant oscillations
distorting most of the curve. A relatively straighter part of these oscillations
is chosen by the algorithm as a scaling region, leading to that specific estimate.
Seeing how visually distorted this curve is from the curves generated for 
other parameter combinations, it is clear that this parameter combination should
be avoided when computing the $\lambda_1$ estimate.

The process is repeated for the driven damped pendulum example in
Fig.~\ref{fig:lyap}(c,d). Here, we use a slightly different trajectory
than in section \ref{sec:pend}: $500,000$ points (after removing a
transient of $50,000$ points) with a time step of $0.01$. For this
trajectory, the heuristic of \citeInline{Deshmukh2020} suggests an
embedding delay of $\tau = 21$.  From the grid, we observe similar
patterns as in the Lorenz example: 46 out of 80 combinations give
$\lambda = 0.93 \pm 0.04$ (for $m \in [4,9]$ and $\epsilon_s \in
[0.083, 0.3]$). We observe
significantly lower estimates when the embedding dimension is 
low and $\epsilon_s$ is high. Additionally, an anomalous behavior
similar to the Lorenz example is observed for higher embedding
dimensions ($m \in [8,9]$) and smaller scales ($\epsilon_s \in [0.05, 0.065]$). 
Here, we see lower estimates ($0.60 \pm 0.02$) for these 
parameter configurations. The underlying cause in this case is low 
frequency oscillations in the stretching factor plots, with a relatively
straight region towards the right end of the plots where they 
start to flatten. As for the Lorenz example, these parameter combinations
should be avoided for the $\lambda_1$ estimate.

The ensemble approach allows us to easily automate computation of scaling regions 
for various values of the hyperparameters, thus
sparing significant manual labor.  Moreover, the
grid visualization allows us to find the ``sweet spot'' in the
free-parameter space.  While this is only a relatively crude
convergence criterion, one could instead use a more rigorous test such
as the Wasserstein distance of \S~\ref{sec:embed}. 

\section{Conclusions}
\label{sec:conclusions}

The technique described above is intended to formalize a
subjective task that arises in the calculation of dynamical
invariants, the identification and characterization of a scaling
region: an interval in which one quantity grows linearly with some
other quantity.  By manipulating the axes, one can extend this to
detect exponential or power-law relationships: e.g., using log-log
plots for fractal dimensions and log-linear plots for Lyapunov
exponents, as shown in Sections~\ref{sec:dcorr} and~\ref{sec:lyap}.
Moreover, linearity is not a  limitation in the
applicability of our approach: it can be used to identify regions
in a dataset with any hypothesized functional relationship
(e.g. higher order polynomial, logarthmic, exponential, etc.).
One would simply compute the least squares fit over the data
using the hypothesized function instead.

A strength of our method is that the scaling
region is chosen automatically as the optimal family of intervals over
which the fits have similar slopes with small errors.  To do this, we
create an ensemble of fits over different intervals to build PDFs, to identify
the boundaries and slopes of any scaling regions, and to
obtain error estimates.
When the scaling region is clear, our
ensemble-based approach computes a slope similar to that which would
be manually determined. We showed that a convergence test on the
slope distributions helps choose the delay reconstruction dimension.
This is a challenging task: even though
there are powerful theoretical guidelines, the information
needed to apply them is not known in advance from a given scalar
time-series.

This kind of objective, formalized method is useful not only because
it reveals scaling regions that may not be visible, but also because
it provides confidence in its existence by providing error estimates. 
Moreover, since computing a dynamical invariant, such as the Lyapunov
exponent, can involve finding scaling regions in many curves, an
automated method provides a clear advantage over hand selection of
scaling regions.  Our method could be potentially useful in areas
outside dynamical systems as well: {\sl e.g.}, estimating the scaling
exponent for data drawn from a power-law distribution.  This is an active area
of research~\cite{alstott2014powerlaw,clauset2009power}. 
Standard techniques involve inferring
which power-law distribution (if any) the data is most likely drawn
from, and---if so---which exponent is most likely.
Our method could potentially help
narrow down the range of exponents to begin such a search.

There are a number of interesting avenues for further work,
beginning with how to further refine the confidence intervals for
slope estimates. We have used the standard deviation of the slopes
within the $FWHM$ around the mode of the distribution.  Alternatively,
one could use the average of the least squares fit
error across all the samples within the $FWHM$ as the confidence
interval.  Another possibility is to first extract a single scaling
region (by determining the modes for the left- and right-hand side
markers) and computing the error statistics over all the sub-intervals
that are sampled from this scaling region ({\sl e.g.,} standard
deviation of the slopes or the average of the fit errors).
We used the Wasserstein distance to assess the convergence of a sequence
of slope distributions to help choose an
embedding dimension.  This may be too strict since it requires that
the \emph{entire} distributions are close.  One could instead target
the positions of the modes, quantifying closeness using some relevant
summary statistic like their confidence interval.
Of course, in cases of multimodal distributions, the Wasserstein
distance test is more appropriate since we cannot choose a single 
mode for computing the intervals.

\begin{acknowledgments}
The authors acknowledge support from NSF grants EAGER-1807478 (JG), AGS-2001670 (VD, EB, and JDM) and DMS-1812481
(JDM).  JG was also supported by Omidyar and Applied Complexity Fellowships
at the Santa Fe Institute.  Helpful conversations with Aaron Clauset
and Holger Kantz are gratefully acknowledged.
\end{acknowledgments}
\section*{Data Availability}
The data that supports the findings of this study are available within the article.  

\appendix
\section{Algorithm}
\label{sec:appendix_algorithm}

The pseudocode for the proposed ensemble-based approach is 
described in Algorithm~\ref{alg:ensemble_algorithm}.
There are four important choices in this algorithm.  The first is the
parameter $n$, the minimum spacing between the left-hand and
right-hand side markers $LHS$ and $RHS$.
This choice limits what is deemed to be a ``significant interval''
for a scaling region.
We set $n = 10$ in this paper, which we chose using a persistence
analysis, that is varying $n$ and seeing if the results change. One 
could argue for increasing $n$ for a data set with more points; 
however, pathologies in the data set could violate that argument. 
The second 
is the choice of a kernel density estimator (KDE) for the histograms.
We used the Gaussian KDE with the {\tt python} implementation {\tt
	scipy.stats.gaussian\_kde}\cite{2020SciPy-NMeth} of Scott's
method\cite{scott1979optimal} to automatically select the kernel
bandwidth. Alternatively, one might choose other bandwidth selection methods,
such as Silverman's method\cite{Silverman86}, or simply
manually specify the bandwidth. 
Scott's method, which is the default method used by the package,
worked well for all our examples. 

\begin{algorithm}[H]
	\caption{Ensemble approach for estimating the slope of scaling regions} 
	\begin{algorithmic}[1]
		\State Assume a plot $(x[0:length-1],y[0:length-1])$ that potentially contains a scaling region.
		\State Initialize empty arrays $slopes$, $lengths$, $errors$, $x_l$ and $x_r$.
		\For {$lhs=0,1,2,\ldots,length-1$}
		\For {$rhs=0,1,2,\ldots,length-1$}
		\If {$rhs - lhs > n $} 
		\State Fit a line using least squares to data $x[lhs:rhs], y[lhs:rhs]$.
		\State Obtain an estimate for slope $m$ and intercept $c$.
		\State Compute the least squares fit length and fitting error, 
		\begin{gather}
			\text{fit length} = \sqrt{1+m^2} \left|x[rhs] - x[lhs] \right| , \\
			\text{fit error} = \sqrt{\frac{\sum_{i = lhs}^{rhs} (y[i] - m x[i] - c) ^ {2}}{rhs - lhs}} .
			\label{eqn:length_error}
		\end{gather}
		\State Append the slope, length and error to the arrays $slopes$, $lengths$ and $errors$, respectively.
		\State Append $x[lhs]$ and $x[rhs]$ to endpoint arrays $x_l$ and $x_r$, respectively.
		\EndIf
		\EndFor
		\EndFor
		\State Generate a histogram $H$ from the $slopes$ array, weighting each point by
		\begin{equation}
			\text{weight[i]} \propto \frac{\text{(fit length[i])}^{p}}{\text{(fit error[i])}^{q}} ,
			\label{eqn:scaling_term}
		\end{equation}
		for suitable powers $p$ and $q$.
		\State Using a kernel density estimator, generate a probability distribution function $P$, from $H$ 
		(see e.g. Fig.~\ref{fig:scaling_region}(b)).
		\State Compute the mode(s) of $P$ and the error estimates as described in Sec.~\ref{sec:method}.
		\State Generate histogram $H_l$ and $H_r$ for $x_l$ and $x_r$ from Step 10,
		weighting the frequency with \eqref{eqn:scaling_term}. Generate 
		PDFs and evaluate them on the $x$ interval, as in Step 12, to generate distributions $P_l$ and $P_r$ 
		(see e.g. Fig.~\ref{fig:scaling_region}(c)).
	\end{algorithmic} 
	\label{alg:ensemble_algorithm}
\end{algorithm}

The two other important parameters are the powers $p$ and $q$ used for
the weights of the length and the error of the fit in \eqref{eqn:scaling_term}.
We explored ranges for these parameters, constraining $p$ and $q$ to
nonnegative integers, and
determined suitable values that worked well for all of the examples
considered here.  Some interesting patterns emerged in these
explorations.  Firstly, we found that $p>0$ helps to reduce the error
estimate for the slope $\sigma$ and improve $p_{FWHM}$.
Note that $p>0$ penalizes the shorter fits near the
edges of the $FWHM$, suppressing their influence on the histogram.
The $FWHM$ therefore narrows, in turn reducing the error estimate
$\sigma$.  Secondly, setting $p \le q$ was found to be advantageous in
all cases, ensuring that the algorithm does not prioritize
unnecessarily longer fits of poor quality.  Enforcing these two
conditions, we experimented with a few choices for $p$ and $q$ in the
context of the first example of Fig.~\ref{fig:scaling_region}, 
and found that lower $p,q$ values
generally work better.  Higher powers tend to magnify effects of 
small errors and
small lengths, making the algorithm very conservative in terms of 
what constitutes a good fit.  With this
in mind, we settled on $p = 1$ and $q= 2$ for the correlation dimension
estimation, finding that those values
generalized well across all the examples. For the
Lyapunov exponent examples, on the other hand, we found that
$p = 1$ and $q = 1$ does better, generating tighter
confidence interval bounds around the estimate across the various
parameter combinations.

This algorithm always uses the full data set to compute the ensemble
but note that it does not require {\sl even} spacing of the data.
Given a much larger data set, it might make sense to downsample to
speed up the algorithm. In its current implementation, the run-time
complexity of the algorithm is $O(N^2)$, for a relationship curve with
$N$ points. In the future, it might be useful to develop faster algorithms
for sampling and generating the slope distributions.

\section{Effects of noise on the Lorenz attractor} 
\label{sec:appendix_lorenz_noise}

\begin{figure}[ht]
	\centering
	\subfloat[Noiseless Lorenz trajectory]{
		\includegraphics[width=0.5\linewidth]{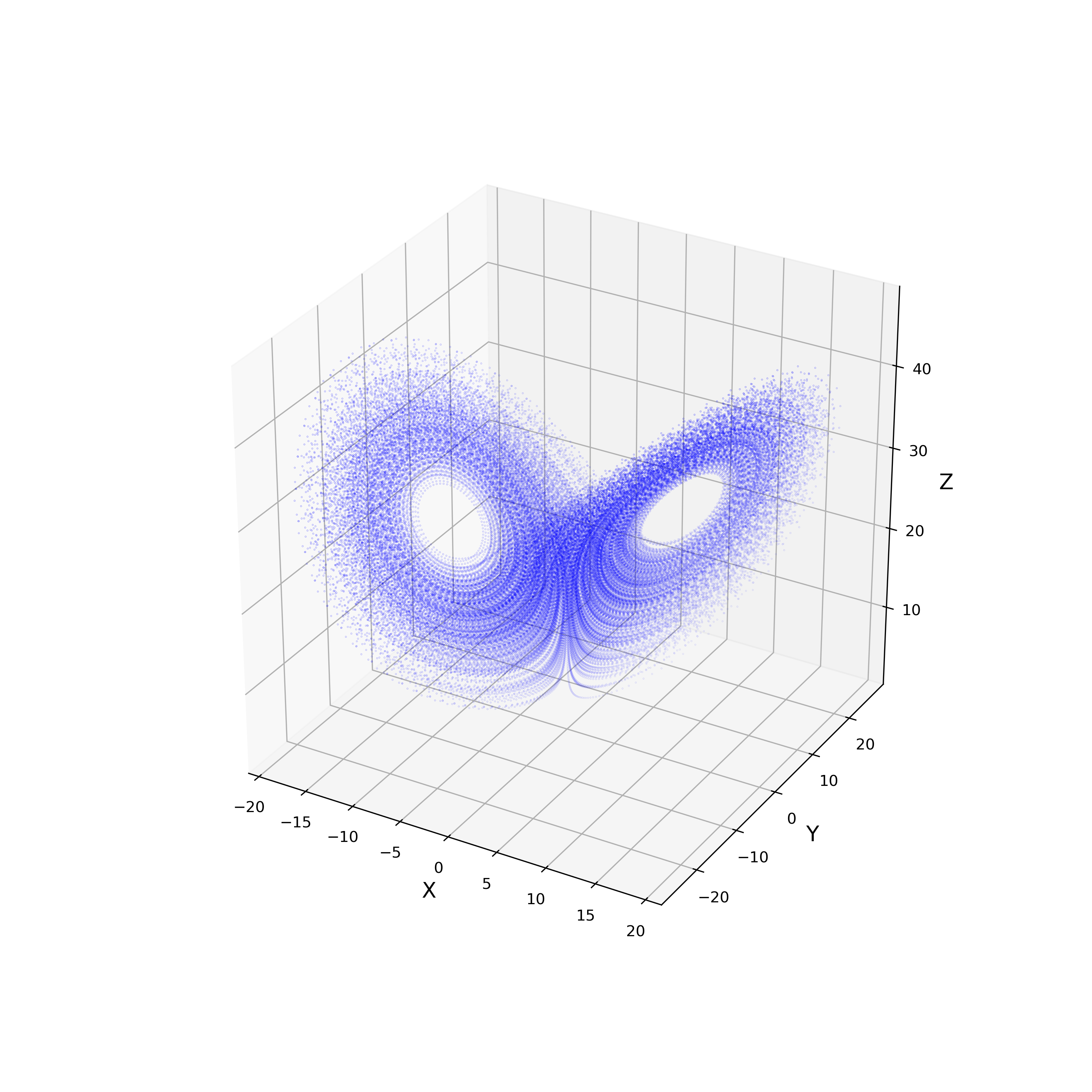}
	}\\
	\subfloat[Noisy Lorenz trajectory ($\delta = 0.01$)]{
		\includegraphics[width=0.5\linewidth]{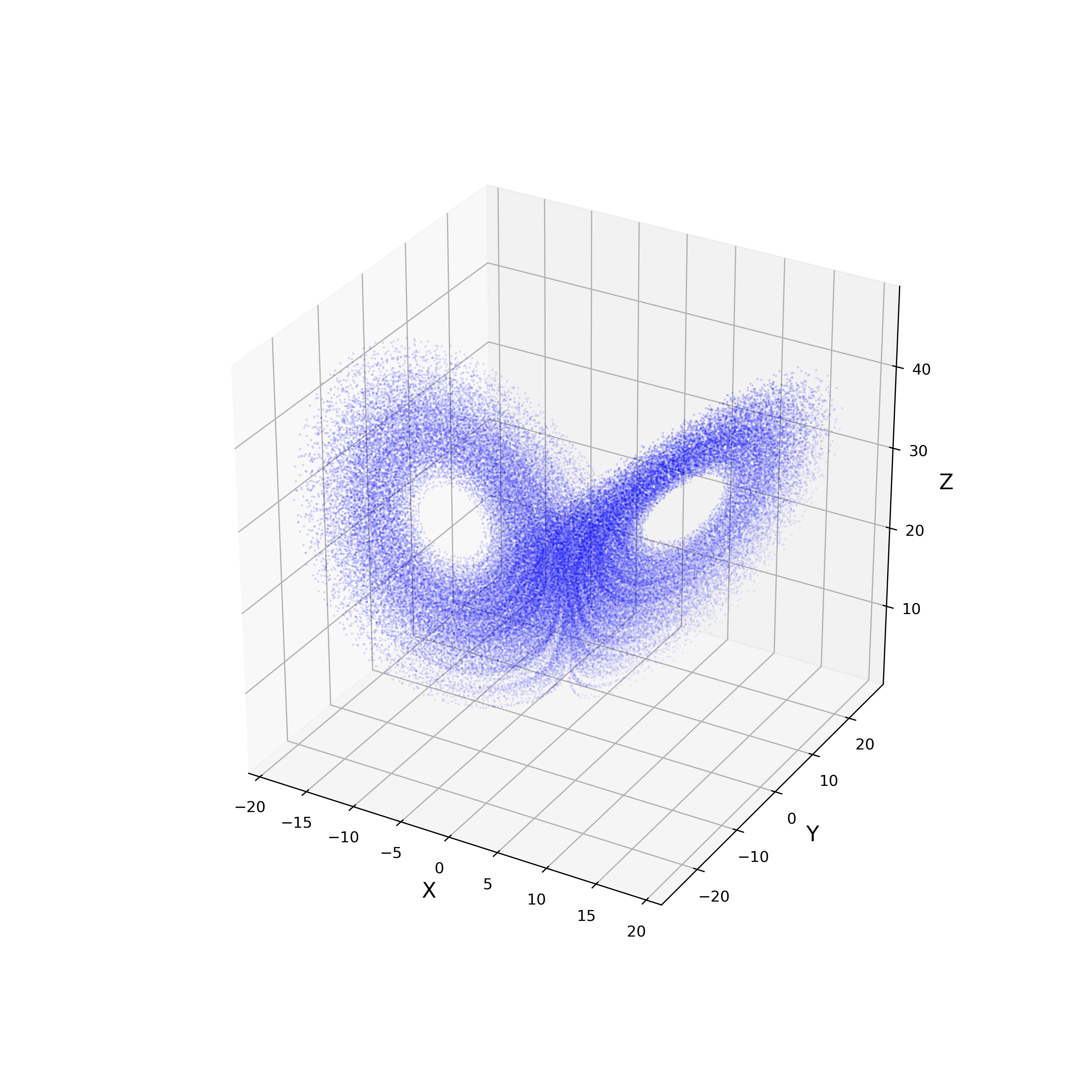} 
	}
	\subfloat[Noisy Lorenz trajectory ($\delta = 0.1$)]{
		\includegraphics[width=0.5\linewidth]{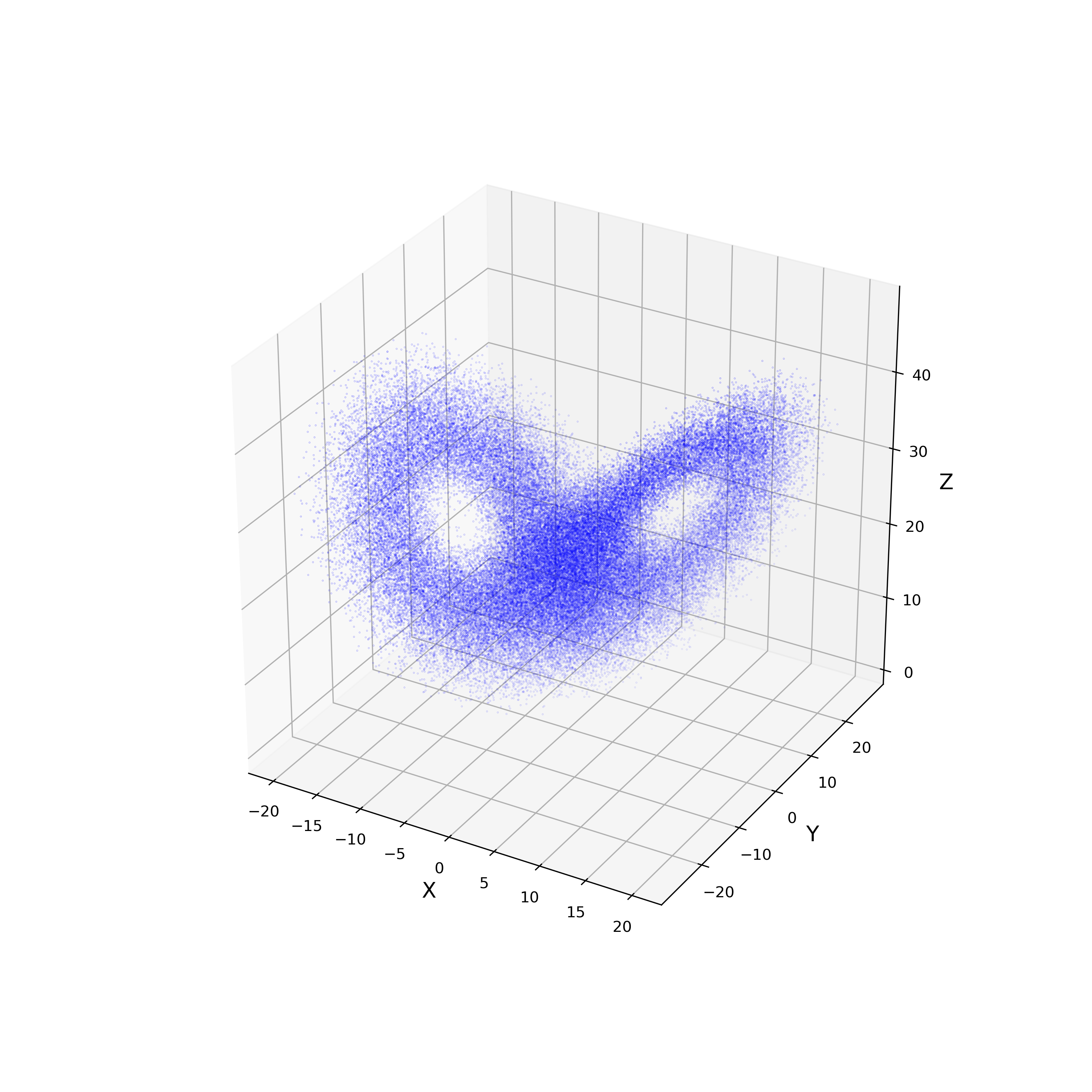}
	}
	\caption{Effect of noise on the Lorenz attractor with different magnitudes of 
		noise level $\delta$.}
	\label{fig:lorenz_noise_attractor}
\end{figure}

Fig.~\ref{fig:lorenz_noise_attractor} shows the effect of noise on the Lorenz
attractor, the resulting noisy attractor used in Section~\ref{sec:noise}. As we 
increase the noise level (as a fraction of the attractor radius) $\delta$
from 0.0 to 0.1, we see the dynamics of the attractor getting 
progressively distorted. As discussed in Section~\ref{sec:noise}, the effects of 
this distortion on the correlation dimension plots and hence the slope distributions
are clearly visible.
\section{Time Series Reconstruction: Additional Plots} 
\label{sec:appendix_reconstruction}

\begin{figure}[ht]
	\centering
	\subfloat[Weighted slope distributions ($\tau = 21$)]{
		\includegraphics[width=0.45\linewidth]{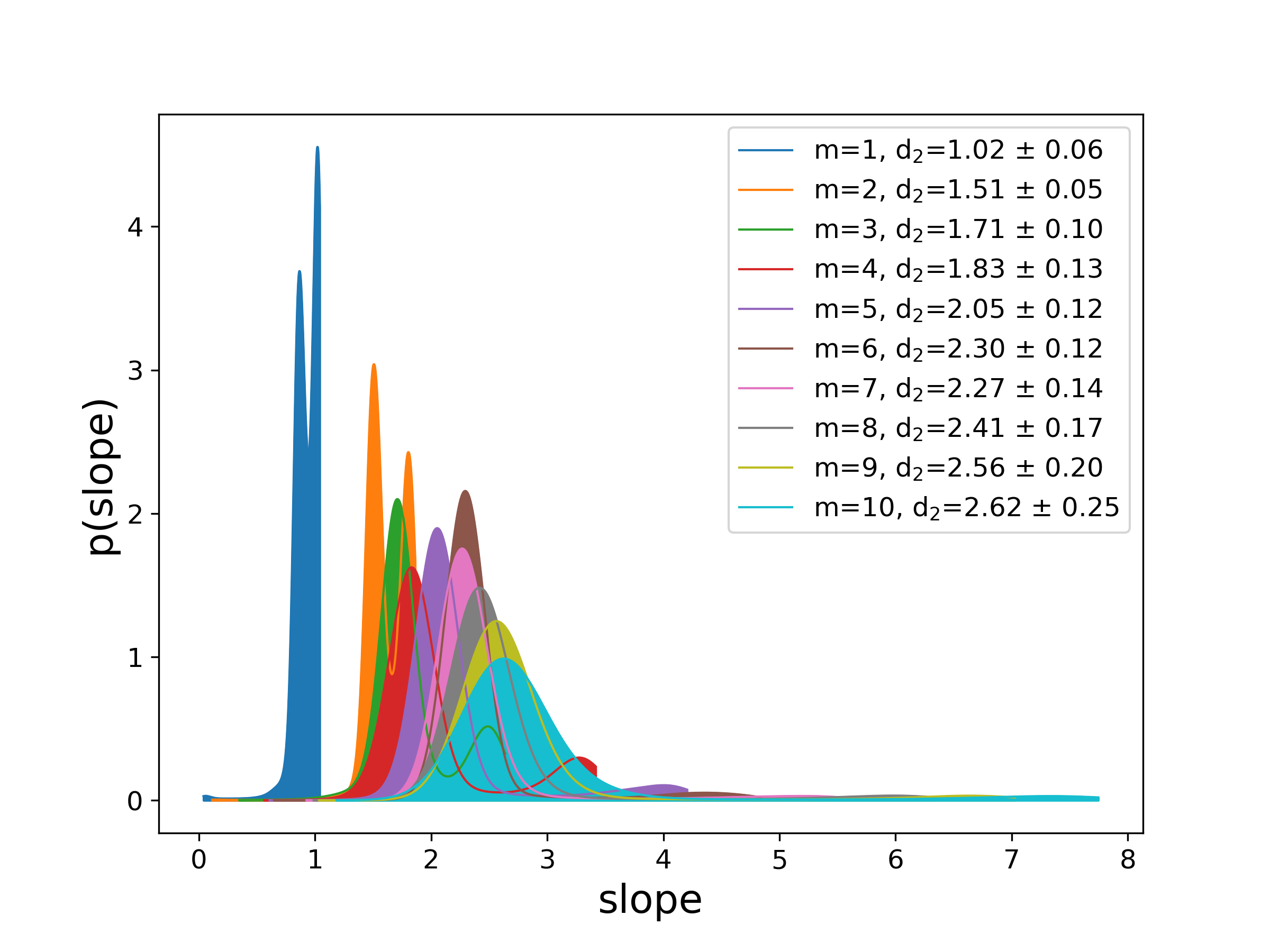}
	}
	\subfloat[Wasserstein distance ($\tau = 21$)]{
		\includegraphics[width=0.45\linewidth]{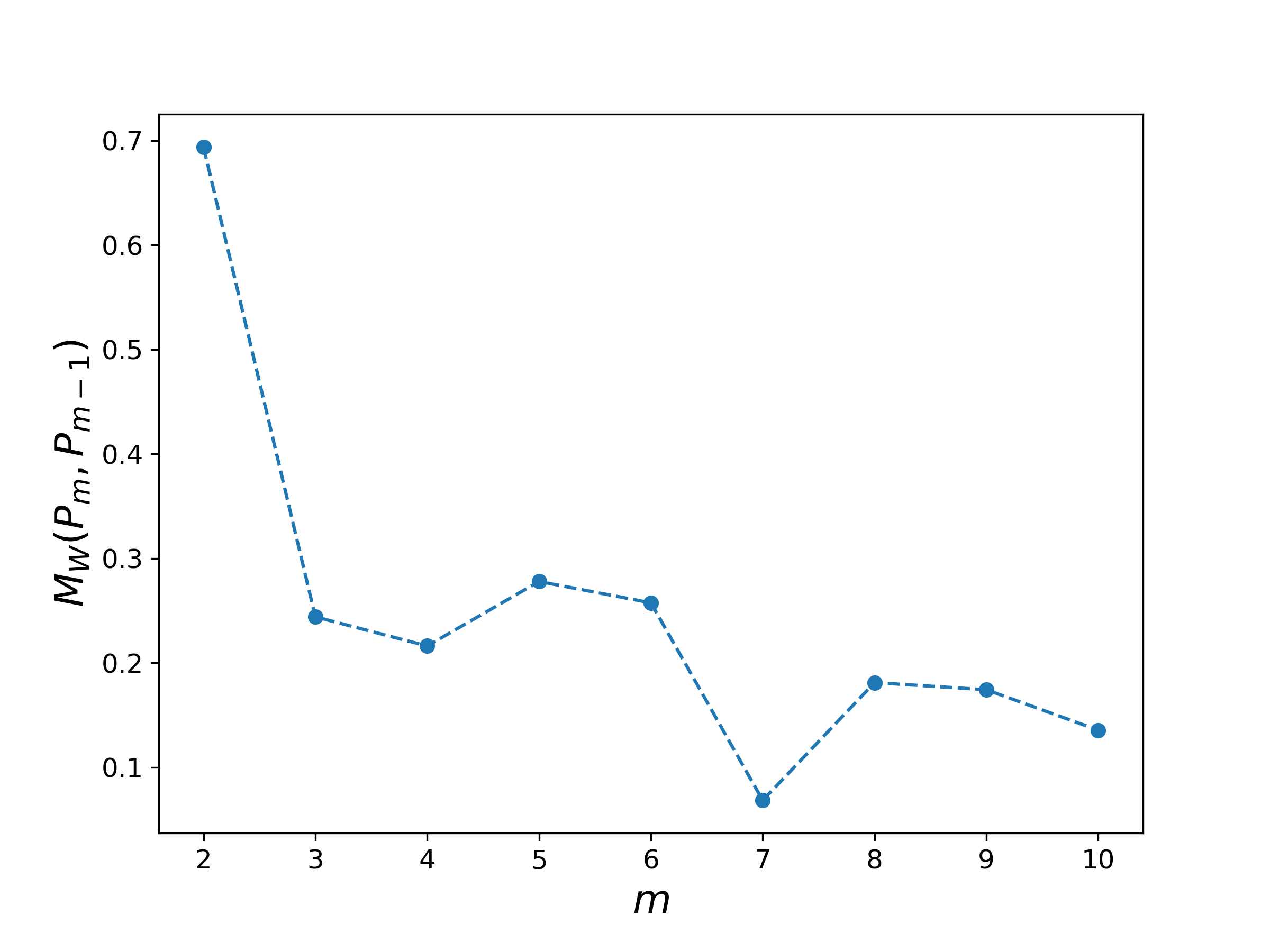} 
	}\\
	\subfloat[Weighted slope distributions ($\tau = 60$)]{
	\includegraphics[width=0.45\linewidth]{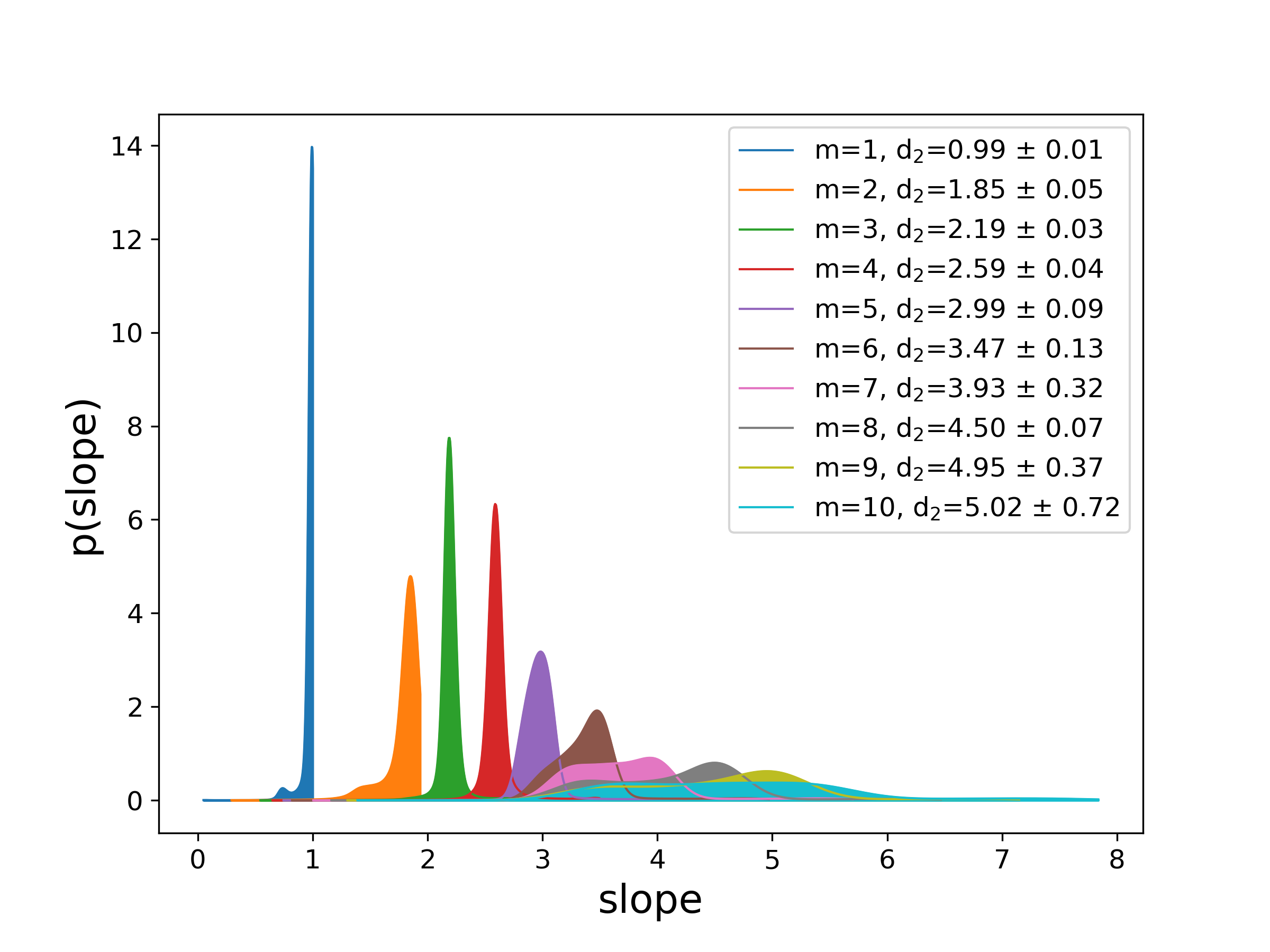}
	}
	\subfloat[Wasserstein distance ($\tau = 60$)]{
		\includegraphics[width=0.45\linewidth]{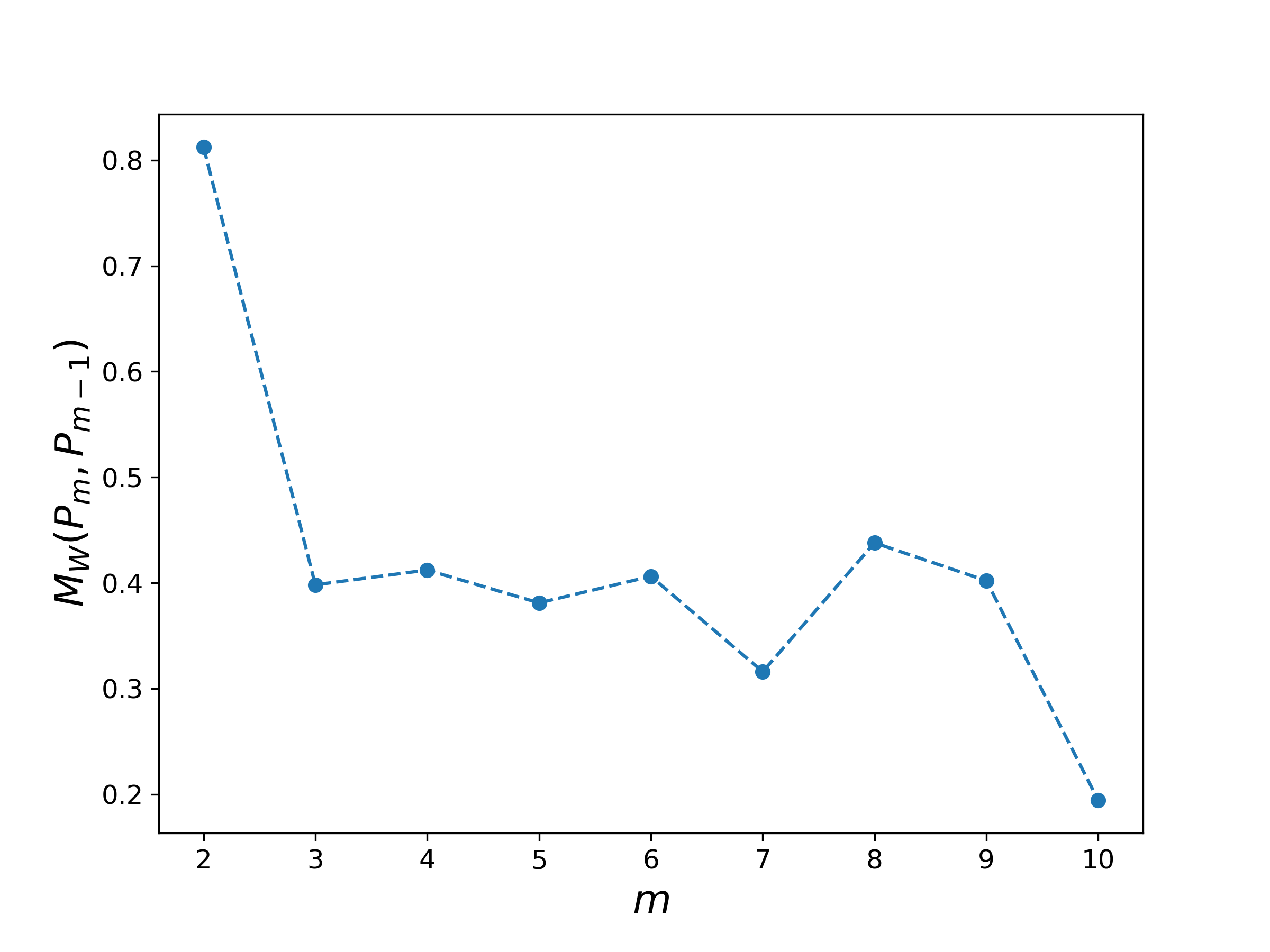} 
	}\\
	\caption{Estimating the correlation dimension for the
          reconstructed noisy Lorenz example using two different
          estimates of $\tau$.  The top row shows the slope
          distributions and the Wasserstein distance profiles for a
          value of $\tau = 21$, estimated using the heuristic of
          \citeInline{Deshmukh2020}, while the bottom row presents the
          same plots for $\tau = 60$, the estimate produced using the
          method of average mutual information \cite{fraser-swinney}.}
	\label{fig:lorenz_noise_reconstruction}
\end{figure}

\begin{figure}[ht]
	\centering \subfloat[Weighted slope distributions]{
          \includegraphics[width=0.45\linewidth]{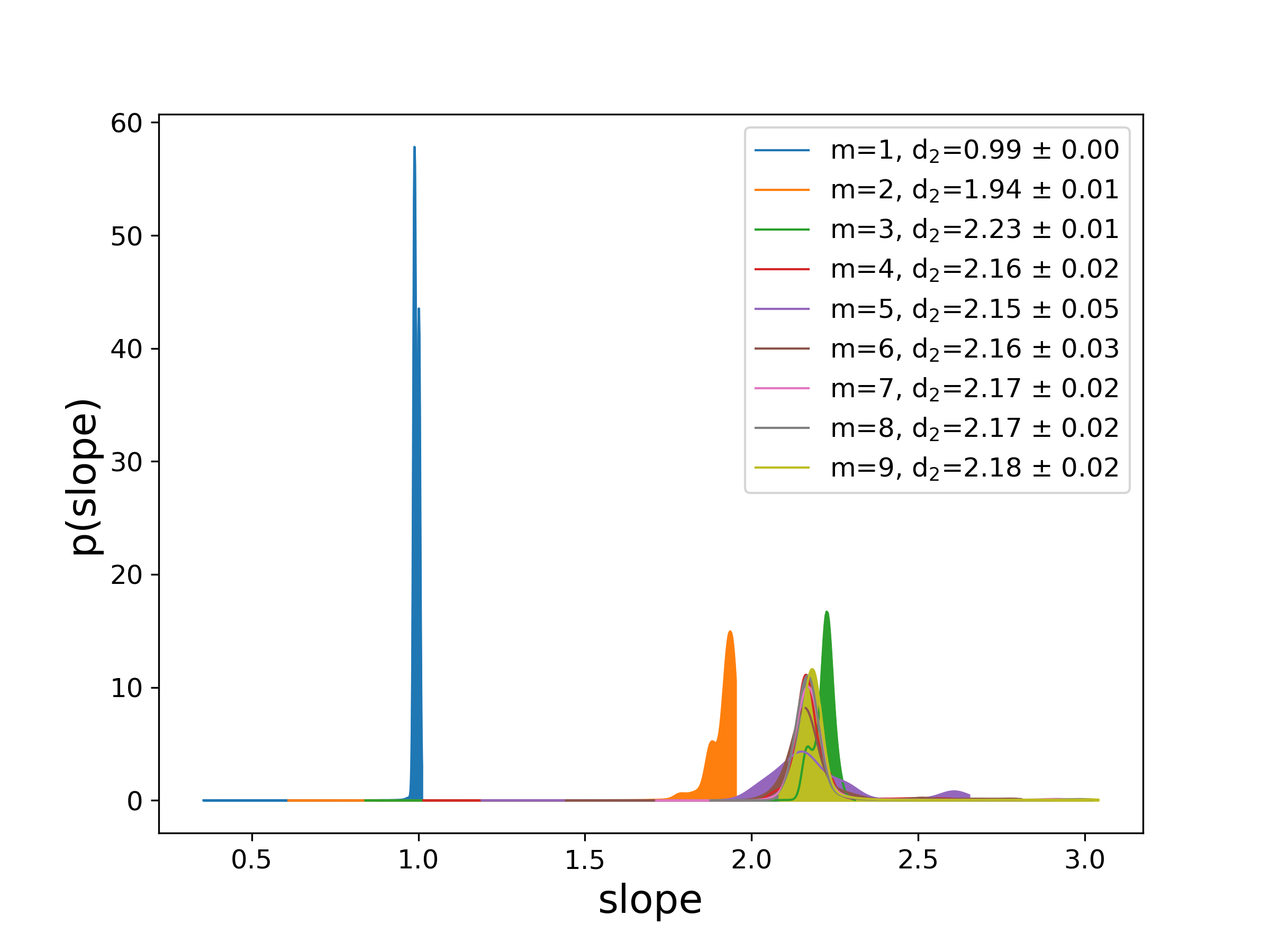}
        } \subfloat[Wasserstein distance]{
          \includegraphics[width=0.45\linewidth]{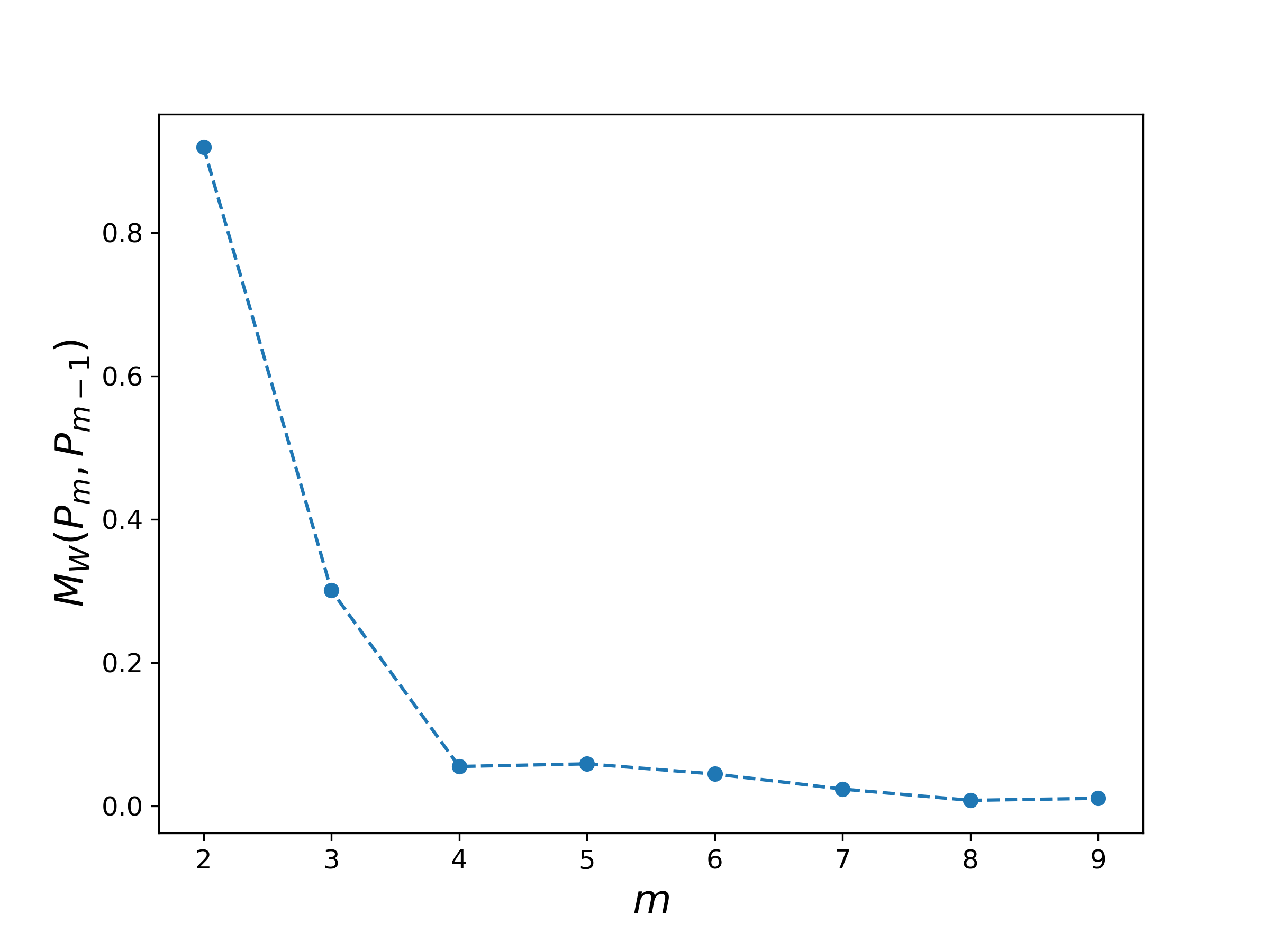}
        }
	\caption{Estimating the correlation dimension for the
          reconstructed pendulum example (1M points) using a time
          delay estimate of $\tau = 120$. }
\label{fig:pendulum_1M_reconstruction}
\end{figure}

\begin{figure}[ht]
	\centering \subfloat[Weighted slope distributions]{
          \includegraphics[width=0.45\linewidth]{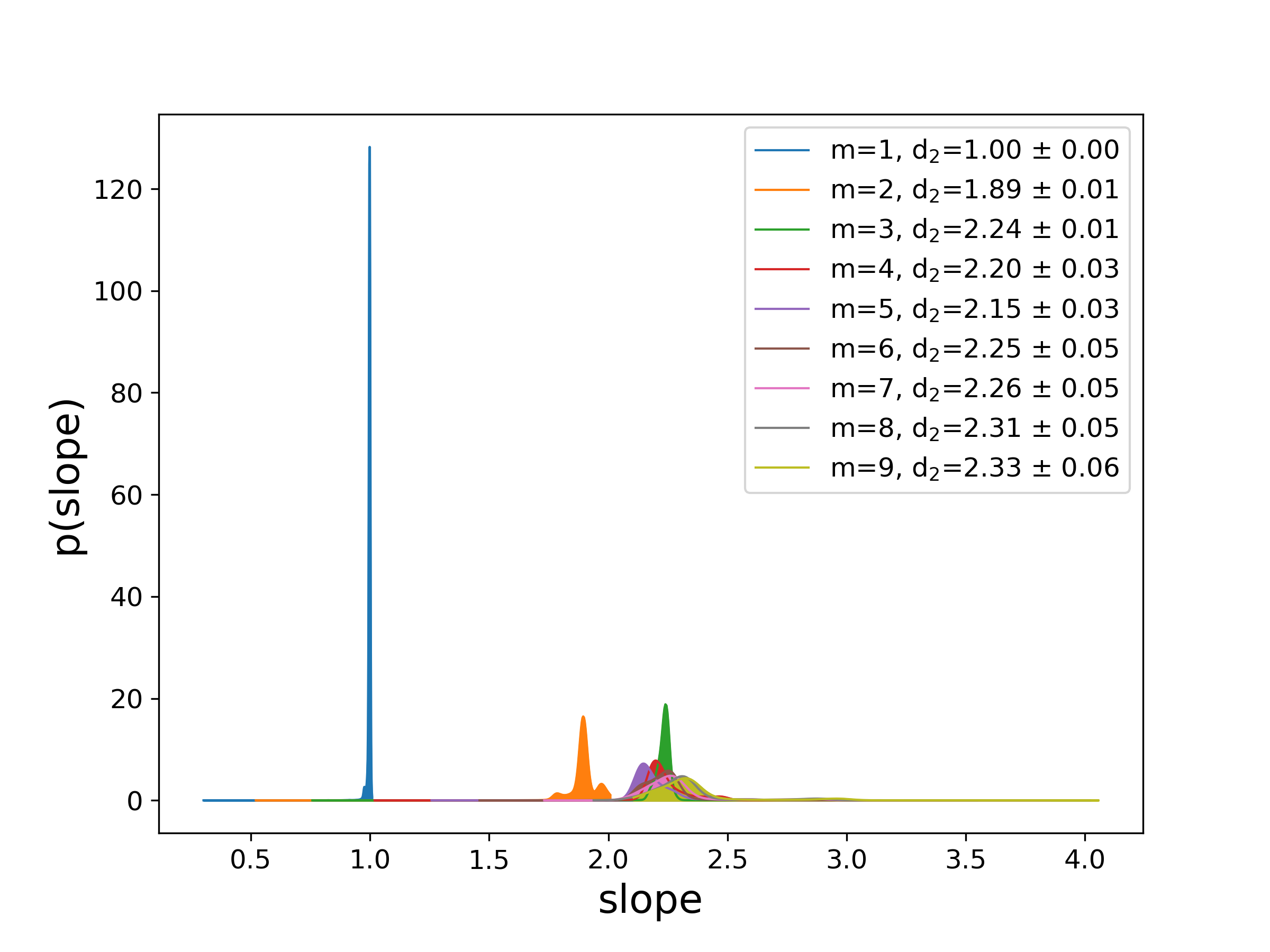}
        } \subfloat[Wasserstein distance]{
          \includegraphics[width=0.45\linewidth]{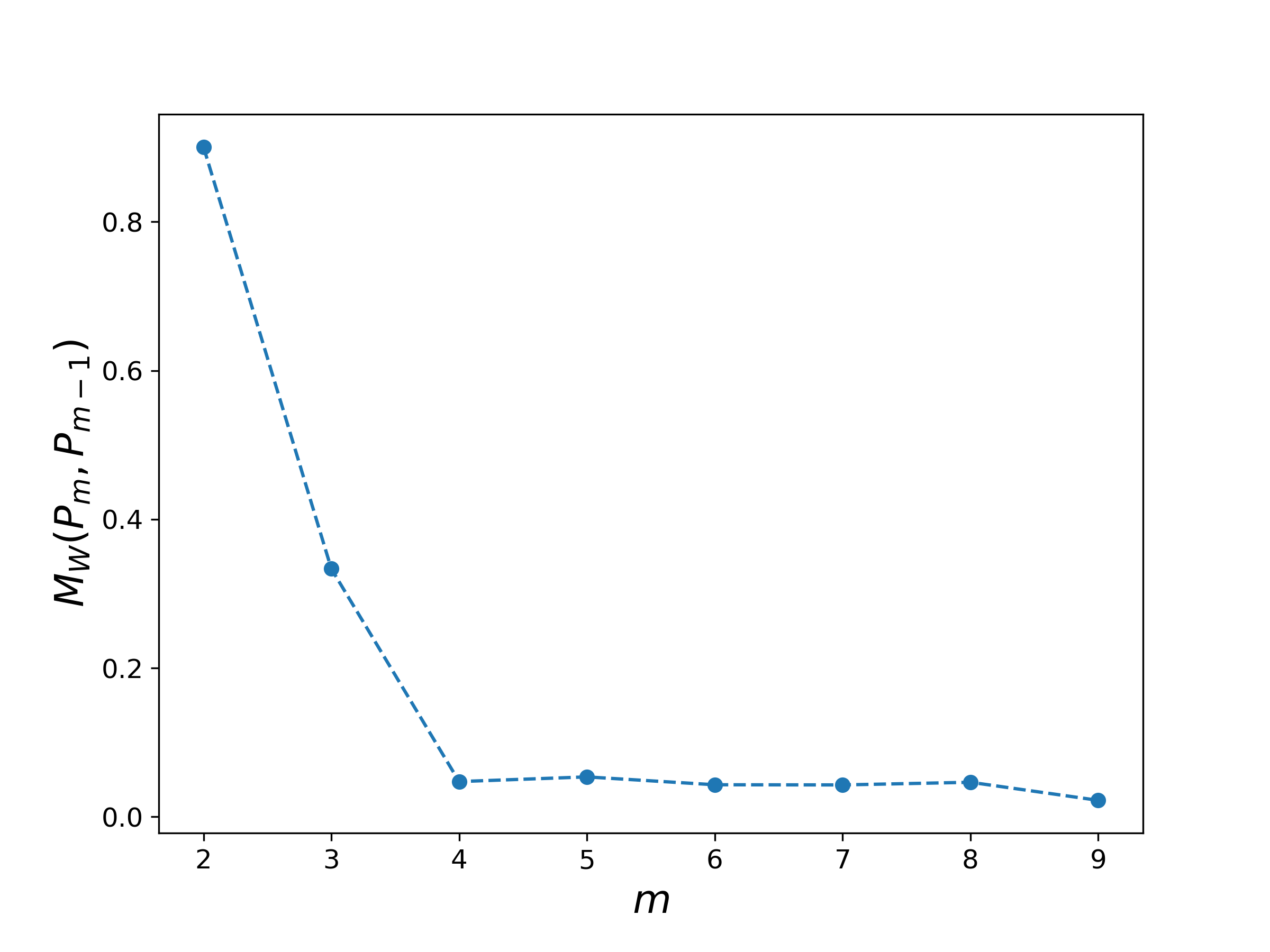}
        }
	\caption{Estimating the correlation dimension for the
          reconstructed truncated pendulum example (200,000 points)
          using a time delay estimate of $\tau = 120$. }
	\label{fig:pendulum_200K_reconstruction}
\end{figure}

\begin{figure}[ht]
	\centering
	\subfloat[Weighted slope distributions ($\tau = 24$)]{
		\includegraphics[width=0.45\linewidth]{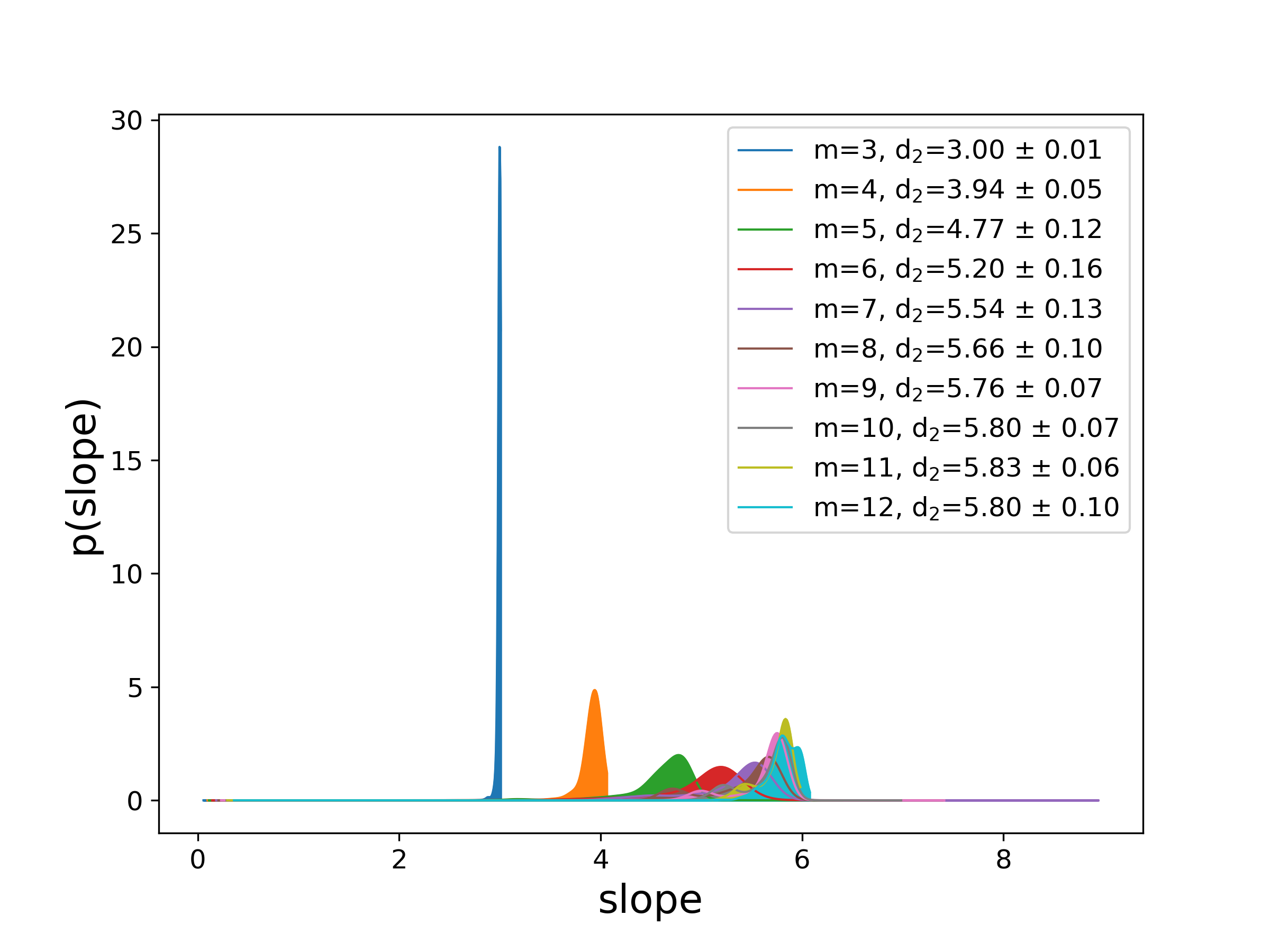}
	}
	\subfloat[Wasserstein distance ($\tau = 24$)]{
		\includegraphics[width=0.45\linewidth]{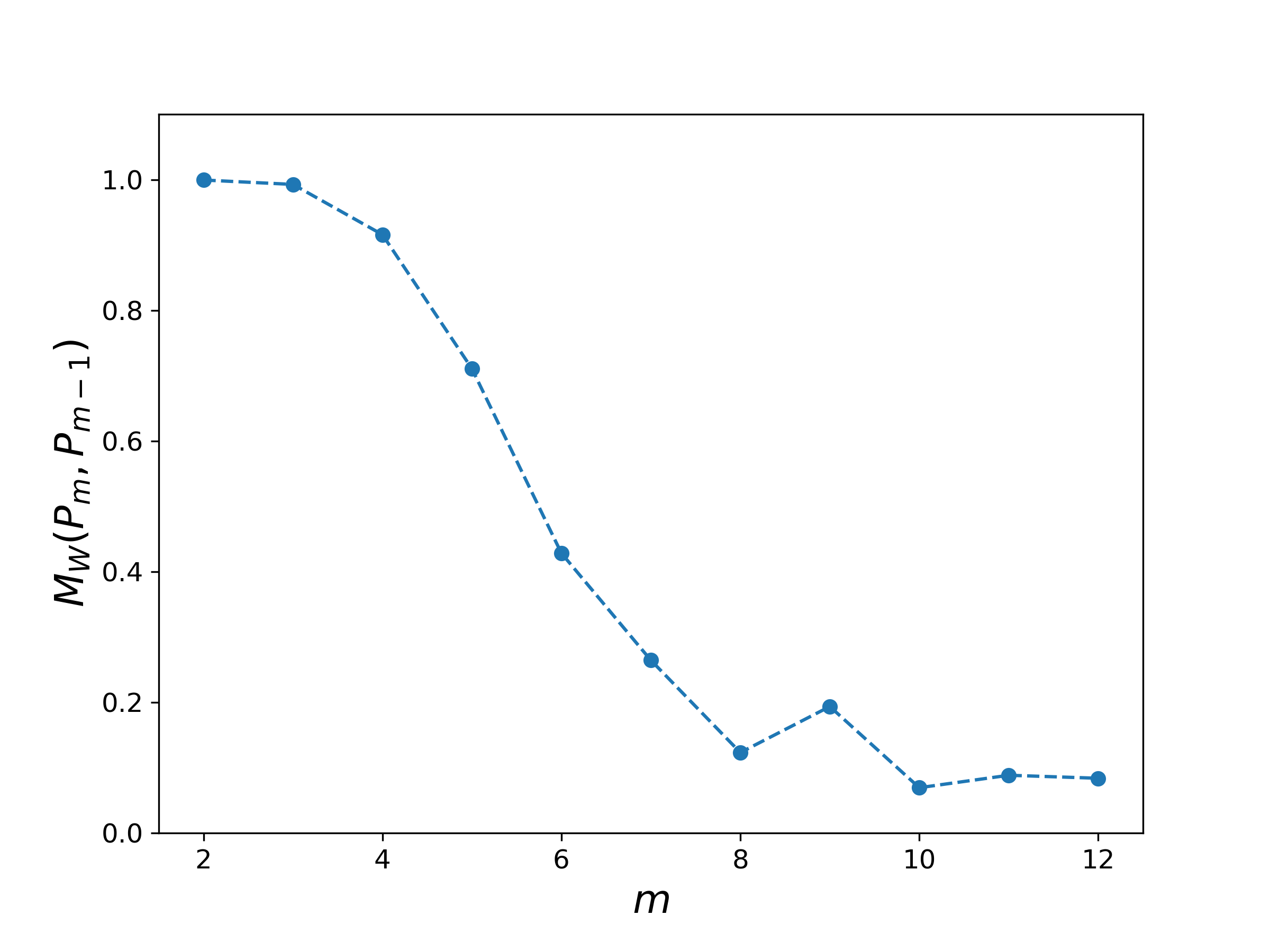} 
	}
	\caption{Estimating the correlation dimension for the
          reconstructed Lorenz-96 example using a time delay estimate
          of $\tau = 24$. }
	\label{fig:lorenz96_reconstruction}
\end{figure}

Here we present figures to support the results of Table~\ref{tab:d2_results}. 
For the noisy Lorenz data, Fig.~\ref{fig:lorenz_noise_reconstruction} shows
the slope distributions and Wasserstein distance 
between slope distributions of consecutive embedding dimensions, $M_W(P_m,P_{m-1})$.
Panels (a) and (b) use $\tau = 21$ as in Table~\ref{tab:d2_results}.
Note that $M_W(P_m,P_{m-1})$ converges more slowly for the noisy 
data than it did for the deterministic case in Fig.~\ref{fig:lorenz_reconstruction}.
The distance $M_W(P_m,P_{m-1}) <0.1$ at $m = 7$, giving $d_2= 2.27 \pm 0.14$. 

Figures~\ref{fig:lorenz_noise_reconstruction}(c) and (d) show the reconstruction results for $\tau = 60$,
the embedding delay given by the average mutual information (AMI) method.
Here $M_W(P_m,P_{m-1})$ does not reach $0.1$ for $m \le 10$.  The implication is that the embedding
dimension should be larger than $10$, contrary to the sufficient
theoretical requirement of $m=7$.
Indeed, panel (c) shows that the mode grows monotonically with $m$, reaching values much higher than the expected $d_2 = 2.59$ for the full dynamics.

Figures~\ref{fig:pendulum_1M_reconstruction} and \ref{fig:pendulum_200K_reconstruction} show the
results for the the pendulum trajectories of length $10^6$ and $2(10)^5$,
respectively, generated using a delay of $\tau = 120$. In both cases, the Wasserstein distance threshold is reached at $m = 4$,
resulting in the values in Table ~\ref{tab:d2_results}. 

Finally, Fig.~\ref{fig:lorenz96_reconstruction} shows the delay
reconstruction of the Lorenz-96 trajectory using $\tau = 24$.
Convergence occurs at $m = 10$, giving $d_2= 5.80 \pm 0.07$.

\FloatBarrier
\bibliography{master-refs}{}
\end{document}